\RequirePackage{rotating}
\documentclass[fleqn,usenatbib]{mnras}

\usepackage{newtxtext,newtxmath}

\usepackage[T1]{fontenc}
\usepackage{ae,aecompl}

\usepackage{graphicx}	
\usepackage{amsmath}	
\usepackage{amssymb}	
\usepackage[skip=1pt]{caption}
\usepackage[skip=0.5pt]{subcaption}
\usepackage[flushleft]{threeparttable}
\usepackage{rotating}
\usepackage{pdfpages}
\usepackage{longtable}

\newcommand{\msunyr}{\ensuremath{\mathit{M}_{\odot}{\rm yr}^{-1}}}   
\newcommand{\kms}{\ensuremath{{\rm km\,s^{-1}}}}                   
\newcommand{\msun}{\ensuremath{\mathit{M}_{\odot}}}   
\newcommand{\lsun}{\ensuremath{\mathit{L}_{\odot}}}                  
\newcommand{\rsun}{\ensuremath{\mathit{R}_{\odot}}}                  

\newcommand{\mdot}{\ensuremath{\dot{M}}}                             
\newcommand{\teff}{\ensuremath{\mathit{T}_{\rm eff}}}                
\newcommand{\vinf}{\ensuremath{\upsilon_{\infty}}}                          
\newcommand{\tstar}{\ensuremath{\mathit{T}_{\star}}}                 

\newcommand{\ang}{\ensuremath{\text{\AA}}}                  
\newcommand{\rin}{\ensuremath{\mathit{R}_{\rm in}}}         
\newcommand{\lsn}{\ensuremath{\mathit{L}_{\rm SN}}}         

\title[Progenitors of early-time interacting SNe]{Progenitors of early-time interacting supernovae}

\author[I. Boian and J.H. Groh]{Ioana Boian$^{1}$\thanks{Contact e-mail: \href{mailto:boiani@tcd.ie}{boiani@tcd.ie}} and
J.H.Groh$^{1}$
\\
$^{1}$Trinity College Dublin, The University of Dublin, Dublin 2, Republic of Ireland}

\date{Accepted XXX. Received YYY; in original form ZZZ}

\pubyear{2019}

\begin{document}
\label{firstpage}
\pagerange{\pageref{firstpage}--\pageref{lastpage}}
\maketitle
 
\begin{abstract}
We compute an extensive set of early-time spectra of supernovae interacting with circumstellar material using the radiative transfer code CMFGEN. Our models are applicable to events observed from 1 to a few days after explosion. Using these models, we constrain the progenitor and explosion properties of a sample of 17 observed interacting supernovae at early-times. Because massive stars have strong mass loss, these spectra provide valuable information about supernova progenitors, such as mass-loss rates, wind velocities, and surface abundances. 
We show that these events span a wide range of explosion and progenitor properties, exhibiting supernova luminosities in the $10^{8}$ to $10^{12} ~\lsun$ range, temperatures from $10\,000$ to $60\,000$ K, progenitor mass-loss rates from a few $10^{-4}$ up to $1 ~\msunyr$,  wind velocities from $100$ to $800 ~\kms$, and surface abundances from solar-like to H-depleted. Our results suggest that many progenitors of supernovae interacting with circumstellar material have significantly increased mass-loss before explosion compared to what massive stars show during the rest of their lifetimes. We also infer a lack of correlation between surface abundances and  mass-loss rates. This may point to the pre-explosion mass-loss mechanism being independent of stellar mass. We find that the majority of these events have CNO- processed  surface  abundances. In the single star scenario this points to a preference towards high-mass RSGs as progenitors of interacting SNe, while binary evolution could impact this conclusion. Our models are publicly available and readily applicable to analyze results from ongoing and future large scale surveys such as the Zwicky Transient Factory.
\end{abstract}

\begin{keywords}
interacting supernovae -- massive stars -- radiative transfer
\end{keywords}


\section{Introduction}
Core-collapse supernovae (SNe) and the massive stars ($M_{ZAMS} > 8 ~\msun$) that produce them are important objects in our Universe as they contribute a significant fraction of the radiative, kinetic, and gravitational energy, and of the heavy elements that we observe. However, the evolution of massive stars is not completely understood, especially at late stages. Additionally, the exact links between core-collapse SNe and their progenitors are not well constrained either.

It is known that massive stars lose large amounts of material during their life-times through strong winds or eruptive events ($\mdot > 10^{-6} ~\msunyr$; \citealt{smith14araa,mauron11,vink01,crowther07,groh14}). Furthermore, mass-loss might increase significantly at very late stages as suggested by recent works, both observational \citep{smith06,galyam09,ofek14b,kilpatrick17,boian18,pastorello18,pastorello19} and theoretical \citep{yooncantiello10,arnett11,moriya14a,smith14b,quataert16,fuller17}. The enhanced mass-loss can happen anytime from days before core-collapse (e.g. SN 2010mc \citealt{ofek13b}) to hundreds of years (e.g. SN 2014C \citealt{margutti14b}). The variation in mass-loss rates and possible mass-loss mechanisms in massive stars creates a variety of environments for the stars to explode into. In the cases where the circumstellar material (CSM) is dense enough, the SN lightcurves (LC) and spectra will be significantly affected. The fast SN ejecta ($v_{ej} = 10\,000 ~\kms$) will collide with the slow moving CSM ($\vinf = 10 - 3000 ~\kms$) and it will decelerate. A fraction of the kinetic energy from the SN is efficiently converted into radiation. This has a number of effects, including the brightening of the LC due to the additional source of energy, and the ionisation of the CSM. The CSM becomes optically thick, i.e. the photosphere is now in the CSM, delaying the shock-breakout and modifying the shape of the LC \citep{chevalier94,chugai01, moriya14b, forster19}. A spectrum taken at this stage will reflect the CSM properties, such as its density, velocity, and abundances \citep{dessart09,groh14,dessart15,davies19,boian19}. Therefore modelling the LC and the spectra can reveal valuable information about the SN and its progenitor, after the SN explosion. 

Due to the heterogeneity in the winds of massive stars and explosion properties, the spectral signatures of interaction between SNe and their progenitor's CSMs/winds are also diverse \citep{kiewe12,taddia13,smith17hsn}. Some events, oftentimes referred to as flash ionisation events \citep{khazov16}, show interaction signatures for only a brief period after explosion (e.g. SN 2013fs; \citealt{yaron17}). In other cases the interaction can be observed for longer periods after explosion, from months to years (e.g. SN 1988Z, SN 2005ip; \citealt{stathakis91,stritzinger12}), leading to the traditional classification of type IIn SNe. In some cases, the interaction can be so strong that the SN becomes extremely bright, earning the denomination of Superluminous SN (e.g. SN 2006gy; \citealt{ofek07,dessart15}). Deviations from a spherically symmetric CSM have also been proposed (e.g. SN 2006jc, \citealt{foley07}; PTF11iqb, \citealt{smith15}). The chemical composition of the CSM can also vary significantly, from \ion{H}{}-dominated to completely \ion{H}-depleted, as is the case in type Ibn SNe \citep{pastorello08}. We broadly refer to all of the cases described above as 'interacting supernovae', but note that other authors may reserve this term for long-term interacting type IIn supernovae alone. 

In this paper we focus on the properties of events that show brief early-time interaction signatures in their spectra. 
They may reveal information about the stellar activity right before collapse, and the ZAMS mass of the progenitor if it evolved as a single star. This work is observationally motivated by the currently increasing number of SNe observed at very early times due to the improvement in time-domain astronomy with surveys such as the Zwicky Transient Facility (ZTF; \citealt{bellm19}). Additionally, on the theoretical side, \cite{boian19} developed a grid of synthetic models exploring the effects of the SN and progenitor properties on the SN spectra 1 day post-explosion. The aim of this paper is three-fold: we model the early time optical spectra of a number of observed SNe that show interaction with their progenitors' wind/atmosphere at early-times in order to constrain their and their progenitors' properties, we introduce an extended grid of models to match SNe observed up to 4 days post-explosion, and we find diagnostic lines in the spectra that can be used for a faster analysis of the observations. All the synthetic spectra of interacting supernovae presented in this paper are publicly available and can be downloaded via WISeREP\footnote{https://wiserep.weizmann.ac.il} \citep{galyam12}.

The paper is organised as follows: Sect. \ref{sect:obs} introduces the sample of events included in the analysis; Sect. \ref{sect:emp_rel} describes the relationships that can be derived purely from the observed spectra; our methodology is outlined in Sect. \ref{sect:method}; Sect. \ref{sect:bf} contains a table with the properties of each event derived from our models and discusses in detail the best-fit models for 3 representative cases; Sect. \ref{sect:disc} explores the implications of our results; we conclude with a brief summary in Sect. \ref{sect:summary}. 

\section{Sample of Observed Supernovae}
\label{sect:obs}
\begin{figure*}
    \centering
    \includegraphics[width=0.99\textwidth]{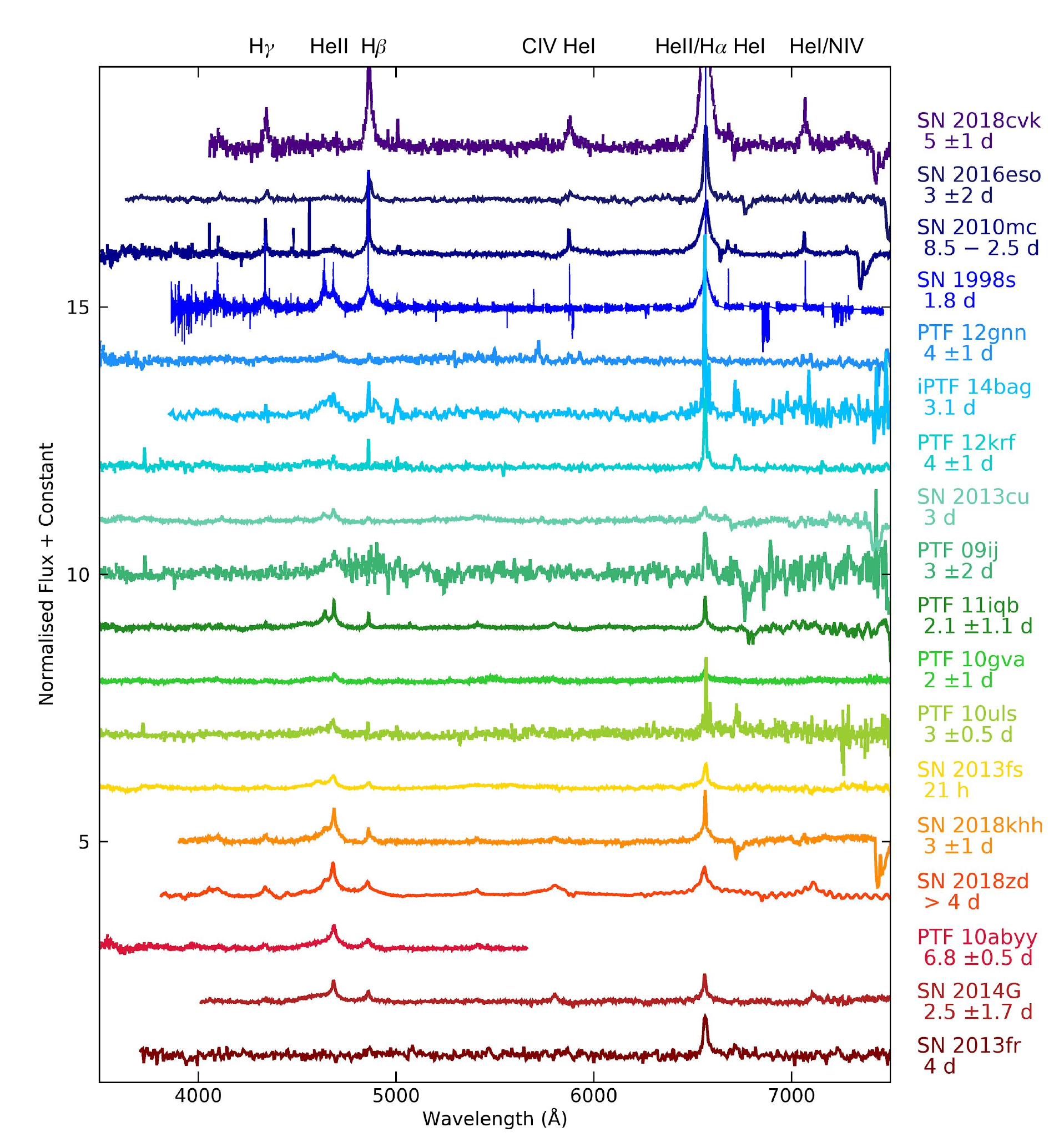}
    \caption{Normalised optical spectra of the observed events analysed in this paper. We also included a spectrum of SN 1998S for comparison. The spectra have been shifted for clarity, the labels on the right indicate the names of the events and the estimated post-explosion age of the spectrum, and the locations of the strongest lines are marked at the top.}
    \label{fig:snsample}
\end{figure*}

We have selected 17 type II SNe that show interaction signatures in their early spectra. We looked for events with at least 1 publicly available spectrum in the optical range ($\sim 3500 - 8000 ~\ang$) with a good signal to noise ratio and relatively well constrained explosion times. For each event, we have chosen the earliest spectrum available, with the exception of iPTF13dqy and iPTF13ast, for which the earliest spectra were taken less than 1 day after explosion. Because of the rapid temperature evolution during the first day, a different set of models would be needed to analyse those events \citep{groh14,yaron17}. For these two events we fit later spectra, and we also include the results from the previous works in our analysis. All the spectra have been downloaded via WISeREP \citep{galyam12} and are shown in Fig. \ref{fig:snsample}. For most of the events, the explosion times are taken as the mean between the last non-detection and the first detection. For a small number of events, namely SN 2010mc, PTF 11iqb, SN 2013fs, iPTF14bag, the explosion times have been estimated in previous works by fitting the lightcurves, and we use these values. The distances and peak magnitudes were taken from The Open Supernova Catalog \citep{guillochon17} unless specified otherwise. The spectral observations were taken using various instruments and therefore vary in resolution. In some cases, this information is not available and the average full-width half-maximum (FWHM) of the flash-ionisation lines is taken as the upper limit for the spectral resolution in velocity space. 

The quality of the results obtained from the spectral modelling relies on a number of observable parameters, such as the distance to the event, the age of the selected spectra relative to the explosion time, the observed brightness of the event at the time of the spectrum, and the resolution of the spectra. Therefore we describe below the properties of each SN and the observational setup employed in each case, compiled from a number of publicly available resources. This information is summarised in Table \ref{tab:sne}.

\begin{sidewaystable*}
    \small
	\centering
	\caption{Observational properties of the SNe included in our sample.}
	\label{tab:sne}
	\begin{tabular}{ccccccccccc} 
		\hline
		Name & Host Galaxy & Redshift & Distance & Discovery & Last & Spectrum & Time after & Instrument & Resolving Power & References \\
		 & & & (Mpc) & Date & non-detection & Date & explosion & (Telescope) & (Resolution) & \\
		\hline
		\hline
		PTF 09ij & \ldots & 0.124 & 598.7 & 2009-05-20 & 2009-05-16 & 2009-05-21 & $3 \pm 2$ d & DBSP & 1200 & 1, 2, 3, 4 \\
		 & & & & & & & & (Palomar 5.1m) & 250 \kms &\\
		PTF 10abyy & \ldots & 0.0297 & 134.4 & 2010-12-03 &  2010-12-02 & 2012-12-09 & $6.8 \pm 0.5$ d & LRIS & 1000 & 1, 2, 5\\
		 & & & & & & & & (Keck I) & 300 \kms & \\
		PTF 10gva & SDSS J122355.39+ & 0.0276 & 124.3 & 2010-05-05 & 2010-05-03 & 2010-05-06 & $2 \pm 1$ d & LRIS & 1090 & 1 ,2 ,5 ,6, 7 \\
		 & 103448.9 &  & & & &  & &  (Keck I) & 275 \kms & \\
		SN 2010mc & A172130+ & 0.035 & 153 & 2010-08-20 & 2010-08-17 & 2010-08-26 & $8.5 - 2.5$ d & GMOS & 3000 & 1, 8, 9  \\
		(PTF 10tel) & 4807 & & & & & & &  (Gemini) & 100 \kms & \\
		PTF 10uls & \ldots & 0.0479 & 201 & 2010-09-07 &  2010-09-06 & 2010-09-10 & $3 \pm 0.5$ d & Mayall RC Spec & 1000 & 1, 2, 10 \\ 
		 & & & & & & & & (Kitt Peak 4m) & 300 \kms & \\
		PTF 11iqb & NGC 151 & 0.0125 & 50.4 & 2011-07-23 & 2011-07-22 & 2011-07-24 &  $2.1 - 1.1$ d & GMOS & 1000 & 1, 9, 11, 12 \\
		& & & & & & & &  (Gemini) & 300 \kms & \\
		PTF 12gnn & \ldots & 0.0308 & 139.5 & 2012-07-09 & 2012-07-07 & 2012-07-12 &  $4 \pm 1$ d & Kast & 1200 & 1, 2, 13 \\
		 & & & & & & & & (Lick 3m) & 250 \kms & \\
		PTF 12krf & \ldots & 0.0625 & 289.6 & 2012-11-04 &  2012-11-02 & 2012-11-07 & $4 \pm 1$ d & DBSP & 1500 & 1, 2, 4 \\
		 & & & & & & & & (Palomar 5.1m) & 200 \kms & \\
		 SN 2013cu & UGC 9379 & 0.0253 & 108 & 2013-05-03 & 2012-05-03 & 2013-05-06 &  3 d & ALFOSC &  750 & 1, 14, 15, 16 \\
		 (iPTF 13ast) & & & & & & & & (NOT) & 400 \kms & \\
		 SN 2013fs & NGC 7610 & 0.0118 & 50.9 & 2013-10-06 & 2013-10-05 & 2013-10-07 & $21$ h & ALFOSC &  600 & 1, 16, 17\\
		(iPTF 13dqy) & & & & & & & & (NOT) & 500 \kms & \\
		SN 2013fr & MCG+04-10-24 & 0.021 & $87.0 \pm 1.6$ & 2013-09-28 & & 2013-10-03 & $4$ d & EFOSC &  355 & 18, 19 \\
		 & & & & & & & & (NTT) & 845 \kms & \\
		iPTF 14bag & \ldots & 0.116 & 557.2 & 2014-05-18 & 2014-05-18 & 2014-05-21 & $3.1$ d &  DIS & 1000 & 1, 2, 20 \\
		 & & & & & & & & (APO-3.5m) & 300 \kms & \\
		SN 2014G & NGC 3448 & 0.0045 & $24.4 \pm 9$ & 2014-01-14 & 2014-01-10 & 2014-01-14 & $2.5 \pm 1.7$ d & Andor iDus DU440 & 1000 & 21, 22, 23, 24  \\
		 & & & & & & & & (Galileo 1.22m) & 300 \kms & \\
		SN 2016eso & ESO 422-G19 & 0.016 & 71.9 &  2016-08-08 & 2016-08-04 & 2016-08-09 &  $3 \pm 2$ d & EFOSC2 & 355 & 2, 19, 25, 26\\
		 & & & & & & & & (NTT) & 845 \kms & \\
		SN 2018cvk & ESO 233-G & 0.0247 & 111.5 & 2018-06-25 & 2018-06-23 & 2018-06-29 & $5 \pm 1$ d & Goodman & 2800 & 2, 27, 28, 29  \\
		& & & & & & & & (SOAR) & 107 \kms & \\
		SN 2018khh & 2MASX J22031497- & 0.0229 & 103.1 & 2018-12-20 & 2018-12-17 & 2018-12-21 & $3 \pm 1$ d  & Goodman & 2800 & 2, 29, 30 \\
		& 5558516 & & & & & & & (SOAR) & 107 \kms & \\
		SN 2018zd & \ldots & 0.0029 & 13.2 & 2018-03-02 & & 2018-03-06 &  $>4$ d  & FLOYDS-N & 370 & 2, 31, 32  \\
		& & & & & & & & (FTN) & 810 \kms & \\
	\end{tabular}	
	\begin{tablenotes}
     \item {\bf References:} (1) \cite{khazov16}, (2) \cite{guillochon17}, (3) \cite{kasliwal09}, (4) \cite{oke82} , (5) \cite{Oke95}, (6) \cite{galyam10}, (7) \cite{chenko10}, (8) \cite{ofek13b}, (9)  \cite{hook04}, (10) \cite{mayallRC}, (11) \cite{Parrent11}, (12) \cite{smith15}, (13) \cite{miller94}, (14) \cite{galyam14}, (15) \cite{groh14a}, (16) \cite{ALFOSC}, (17) \cite{yaron17}, (18) \cite{bullivant18}, (19) \cite{smartt15b}, (20) \cite{DIS}, (21) \cite{nakano14}, (22) \cite{bose16}, (23) \cite{terreran16}, (24) \cite{rafanelli12}, (25) \cite{brimacombe16}, (26) \cite{stanek16}, (27) \cite{brimacombe18}, (28) \cite{nicholls18}, (29) \cite{clemens04}, (30) \cite{brimacombe18b}, (31) \cite{itagaki18}, (32) \cite{sand14}.
    \end{tablenotes}
\end{sidewaystable*}

--{\it PTF 09ij} was discovered on 2009-05-20 at a redshift of $z=0.124$ \citep{kasliwal09}, which corresponds to a luminosity distance of $d_{L} = 598.7$ Mpc. The spectrum used in this paper was taken on 2009-05-21, $3 \pm 2$ days post-explosion (given the last non-detection was on 2009-05-16), with the low-to-medium resolution Double Spectrograph (DBSP; \citealt{oke82}) on Palomar 5.1m \citep{khazov16}. The only available photometric observation was taken 1 day prior to the spectrum and has $M_{R} = -18.46$ mag.

--{\it PTF 10abyy:} Discovered on 2010-12-03 at $z = 0.0297$ ($d_{L} = 134.4$ Mpc), PTF10abyy has one of the oldest spectra in our sample, obtained with the Low Resolution Imaging Spectrograph (LRIS; $R \simeq 1000$; \citealt{Oke95}) at Keck I on 2010-12-09, $6.8 \pm 0.5$ days post-explosion \citep{khazov16}. Photometrically it is well sampled in the r-band and it resembles a type II-P LC.

--{\it PTF 10gva} was discovered on 2010-05-05 in the galaxy SDSS J122355.39+103448.9, at redshift of $z = 0.0276$ \citep{galyam10}, corresponding to a distance of $d_{L} = 124.3$ Mpc. The galaxy has a reddening of $E(B-V) = 0.0263 \pm 0.0008$ \citep{schlafly11}. The SN reached a maximum luminosity on 2010-05-12, having $m_{max,R} = 16.77$ mag and $M_{max,R} = -18.67$ mag \citep{galyam10,rubin16}. The SN was also detected in the UV with Swift on 2010-05-07, and fitting the Swift photometric measurements revealed a black-body temperature of $20000$ K \citep{chenko10}. In this paper, we analyse the only public spectrum, obtained on 2010-05-06 with the LRIS instrument on Keck I, $2 \pm 1 $ days post-explosion. 

--{\it SN 2010mc} or PTF 10tel is a well known event made famous by its pre-SN outburst observed 40 days prior to explosion, outburst during which the star lost $ \simeq 10^{-2} ~\msun$ at $2000 ~\kms$ \citep{ofek13b}. The SN is located at $z=0.035$ ($d_{L} = 153$ Mpc) and was detected on 2010-08-20 \citep{ofek13b}. The first spectrum, which we analyse in this paper, was obtained on 2010-08-26, $ \simeq 8.5^{+0}_{-2.5}$ days post-discovery with the Gemini Multi-Object Spectrograph South (GMOS-S; $R \simeq 3000$; \citealt{hook04}) on Gemini. Having shown narrow lines from the SN-CSM interaction for a prolonged amount of time, SN 2010mc is classified as SN~IIn. 

--{\it PTF 10uls} was discovered on 2010-09-07 at $z = 0.0479$ ($d_{L} = 201$ Mpc). While it is well sampled photometrically (only r-band), with a LC that closely resembles those of PTF12gnn and iPTF13dqy, it has only one publicly available spectral observation, obtained on 2010-09-10 with the Mayall RC Spectrograph on Kitt Peak 4m ($R \simeq 1000$, \citealt{mayallRC}), $3 \pm 0.5$ days post-explosion (last non-detection on 2010-09-06;  \citealt{khazov16}).

--{\it PTF 11iqb} was discovered on 2011-07-23 \citep{Parrent11} by PTF in the galaxy NGC 151, at a distance of $50.4$ Mpc (redshift $z = 0.0125$; \citealt{smith15}) and has a reddening of $E(B-V) = 0.0284$ mag \citep{Schlegel98}. While PTF11iqb is observationally well sampled both in photometry and spectroscopy \citep{smith15}, we analyse the earliest spectrum of this event, obtained on 2011-07-24, $2.1^{+0}_{-1.1}$ days post-explosion, with GMOS. The narrow lines resulting from the interaction of the SN ejecta with the CSM are not resolved, but late-time spectra of higher resolution place a constraint of $\vinf \leq 100 ~\kms$ on the terminal wind velocity \citep{smith15}. Additionally, \cite{smith15} also derive from LC fitting a mass-loss rate of $1.5 \times 10^{-4} ~\msunyr$ lost in about 8 years prior to collapse in a disk-like geometry, and a blackbody temperature of $25\,000$ K. 

--{\it PTF 12gnn} was discovered on 2012-07-09 at a redshift of $z = 0.0308$, which corresponds to a luminosity distance of $d_{L} = 139.5$ Mpc. The last non-detection was on 2012-07-07. A spectrum was obtained on 2012-07-12 with the Kast instrument \citep{miller94} at the Lick Observatory, $4 \pm 1$ days post-explosion \citep{khazov16}. At peak, this SN had $M_{R} = -17.9$, and its LC resembled iPTF13dqy.

--{\it PTF 12krf} was first detected on 2012-11-04 at $z = 0.0625$ \citep{khazov16}, i.e. $d_{L}= 289.6$ Mpc. A spectrum was obtained $4 \pm 1$ days post-explosion (last-non detection 2012-11-02), on 2012-11-07 using the DBSP. Photometrically it is sparsely observed, and mainly around maximum light ($M_{R} = -18.86$ mag).

--{\it SN 2013cu:} Also known as iPTF 13ast, this type IIb SN has been well studied due to its very early-time spectrum obtained 15.5h post-explosion \citep{galyam14,groh14a}. This SN resides in the UGC 9379 galaxy, at a distance of 108 Mpc ($z = 0.02534$), and was discovered on 2013-06-03 \citep{galyam14}. \cite{groh14a} modelled the 15.5 h spectrum in a similar fashion to the work performed in this paper and obtained $L = 10^{10} ~\lsun$, $\mdot \simeq 3 \times 10^{-3} ~\msunyr$, $\vinf \simeq 100 ~\kms$, and $46$\% H, $52$ \% He, and N-enhanced/C-depleted abundances, implying an LBV/YHG progenitor. These values are in relative agreement with the values analytically derived by \cite{galyam14}. In this work we analyse the spectrum obtained 3 days post-explosion, on 2013-05-06, which still shows interaction signatures, covers a wider range of wavelengths allowing us to probe more species and better constrain the abundances, and helps us build a temporal evolution. The spectrum modelled in this paper was obtained with the Alhambra Faint Object Spectrograph and Camera (ALFOSC; \citealt{ALFOSC}) on the Nordic Optical Telescope (NOT). 

--{\it SN 2013fr:} SN 2013fr (SNhunt213) was discovered on 2013-09-28 by the Catalina Sky Survey in the galaxy MCG+04-10-24 at $z=0.021$ ($d = 87.0 \pm 1.6$ Mpc; \citealt{bullivant18}).  A spectrum was obtained on day 4 with EFOSC \citep{smartt15b} at a resolution of  $R \simeq 355 $. By day 7 the narrow lines have subsided. The LC and spectral evolution suggest a type II-L SN of low velocity and possibly low explosion energy, with late-time IR excess suggesting strong pre-explosion mass-loss rates \citep{bullivant18}.

--{\it SN 2013fs:} is also known as iPTF 13dqy and it is one of the best studied SNe in our sample due to its very early discovery, only 3 hours after explosion. It resides in a nearby galaxy, NGC7610, at a distance of $d=50.95$ Mpc ($z=0.011855$; \citealt{yaron17}). iPTF13dqy is also currently the SN with the earliest observed post-explosion spectrum. \cite{yaron17} obtained a spectrum only 6 hours post-explosion, and using radiative transfer modelling they suggested the SN had a luminosity of $2.0-3.5 \times 10^{10} ~\lsun$, and was produced by a Red Supergiant (RSG) with $\mdot = 2.0-4.0 \times 10^{-3} ~\msunyr$, $\vinf = 100 ~\kms$, and solar surface abundances. \cite{morozova17} fits the multi-band LCs of this event and suggests the progenitor was a RSG with $M_{ZAMS}= 13.5 ~\msun$, $\mdot = 0.15-1.5 ~\msunyr$ and $\vinf=10 - 100 ~\kms$.  \cite{dessart17} place a constraint on the amount and extension of the material ejected pre-SN ($\simeq 0.01 ~\msun$ and $\simeq 2 \times 10^{14}$ cm) using multi-gourp radiation hydrodynamics and radiative transfer models. \cite{Moriya18} explore the effects of wind acceleration on the light curve of this event and derive a mass-loss rate of $\simeq 10^{-3} ~\msunyr$ and $\vinf=10 ~\kms$. \cite{bullivant18} present additional observations during the first 100 days, including photometry, spectroscopy and spectropolarimetry. While we include the fit to the earliest spectrum computed by \cite{yaron17} in the discussion, additionally we analyse a later spectrum, obtained 21 hours post-explosion, on 2013-10-07 with ALFOSC on the NOT. The narrow lines disappear after $\simeq 2$ days, after which the SN resembles a typical type II-P.

--{\it iPTF 14bag:} Discovered on 2014-05-18 at a redshift of $z = 0.116$ ($d_L = 557.2$ Mpc), PTF14bag has only one publicly available spectral observation, on 2014-05-21 with the Dual Imaging Spectrograph (DIS; \citealt{DIS}) at the Apache Point Observatory (APO-3.5m; \citealt{khazov16}) at a medium resolution of $R \simeq 1000$. The epoch of the spectral observation is estimated as $3.1$ days post-explosion \citep{khazov16}. The flux at discovery was $m_R = 20.48 $ mag (WISeREP), but no other photometric observations are publicly available.

--{\it SN 2014G:} Also known as iPTF14fe, this SN was discovered on 2014-01-14 in the host galaxy NGC 3448 at $z = 0.004503$ \citep{nakano14}. The host of SN 2014G is placed at $d = 24.4 \pm 9.0$ Mpc, and has a total extinction coefficient of $E(B-V)_{tot} = 0.254 \pm 0.072$ \citep{bose16} or $E(B-V)_{tot} = 0.21 \pm 0.11$ \citep{terreran16}. The SN reached a maximum luminosity of $m_{max,U} = 13.65$ mag and $M_{max,U} = -17.85$ mag on 2014-01-18 and the LC evolution followed that of a typical type II-L. This SN was monitored for over 1 year having multiple photometric measurements from the UV to the NIR and several spectra over the optical range \citep{terreran16,bose16}. The spectra show narrow emission lines from the SN-CSM interaction for the first 9 days post-explosion. We analyse the first spectrum, which was taken on 2014-01-14, $2.5 \pm 1.7$ days post-explosion with the Andor iDus DU440 ($300 ~\kms$, \citealt{terreran16,rafanelli12}). Late time spectroscopic features suggest a possible bipolar CSM \citep{terreran16}, but polarimetry measurements were inconclusive \citep{bose16}. The strength of the [OI] $\lambda \lambda 6300,6363$ lines point to a progenitor of $M_{ZAMS} = 15-19 ~\msun$ \citep{terreran16}, while semi-analytic models of the LC, place the progenitor around $9~\msun$ and $630 ~\rsun$.  

--{\it SN 2016eso:} Also known as ASASSN-16in or PS16ejl, this SN was detected by the All Sky Automated Survey for Supernovae (ASAS-SN) on 2016-08-08 in the galaxy ESO 422-G19 at $z=0.016$ ($d_{L} = 71.9656$ Mpc; \citealt{brimacombe16}). The last non-detection was on 2016-08-04 \citep{stanek16}. A spectrum was obtained on 2016-08-09, $3 \pm 2$ days post-explosion with the EFOSC2 instrument ($R = 355$,  \citealt{smartt15b}), showing signatures of interaction between the SN and the CSM. The photometric observations are sparse and there is a large gap around the peak. 

--{\it SN 2018cvk:} SN 2018cvk (ASASSN-18nx, Gaia18dii) was discovered on 2018-06-25 in the galaxy ESO 233-G7 at $z=0.02473$ ($d_{L} = 111.51$ Mpc; \citealt{brimacombe18}). A spectrum was obtained on 2018-06-29 with the Goodman Spectrograph \citep{clemens04} at the Southern Astrophysical Research (SOAR) Telescope at a resolution of $R \simeq 2800$. Given the last non-detection was on 2018-06-23 \citep{nicholls18}, this spectrum is $5 \pm 1$ days-old. Photometrically it only has a few post-peak G-band detections. 

--{\it SN 2018khh} (ASASSN-18abz, Gaia19aup) was discovered on 2018-12-20 \citep{brimacombe18b} in 2MASX J22031497-5558516 ($z = 0.0229$, $d_{L} = 103.1$ Mpc) around its peak and has a few photometric points in the G-band and 1 spectrum obtained with the Goodman spectrograph at SOAR on 2018-12-21. The last non-detection was on 2018-12-17 \citep{brimacombe18b}, making the spectrum $3 \pm 1$ days-old. 

--{\it SN 2018zd} (Gaia18anr, ATLAS18mix) was discovered on 2018-03-02 \citep{itagaki18} and has been well sampled photometrically in multiple filters \citep{mikolajczyk18}. Its host is at a redshift of $z = 0.002979$, i.e. $d_{L} = 13.214$ Mpc. In this work, we analyse the spectrum obtained with the Folded Low Order whYte-pupil Double-dispersed Spectrograph (FLOYDS-N, $R \simeq 300-500$, \citealt{sand14}) at the Faulkes Telescope North (FTN) on 2018-03-06, at least 4 days after the explosion. 

\section{Empirical Relationships}
\label{sect:emp_rel}

We expect the early-time spectra of supernovae interacting with their progenitors to fall into three main classes based on the ionisation level of the species present in the events, i.e low ionisation, medium ionisation, and high ionisation \citep{boian19}. A high ionisation spectrum may show emission from highly ionised species such as \ion{He}{ii}, \ion{N}{v}, \ion{O}{v}, \ion{O}{vi}, while the low ionisation end will develop strong \ion{He}{i}, \ion{N}{ii}, and/or \ion{Fe}{ii}, with \ion{N}{iii}, \ion{N}{iv}, \ion{C}{iii}, and/or \ion{C}{iv} appearing in between depending also on the abundances. These classes can then be mapped back to the SN properties, such as the temperature and luminosity. Additionally, the strengths of the lines can also be related to the mass-loss rate of the progenitor and/or the surface abundances. The observed spectra are shown in Fig. \ref{fig:snsample}, broadly ordered from high ionisation (red) at the bottom of the figure to low ionisation (purple) at the top. 

In order to have a more quantitative classification of our sample of supernovae we have measured the equivalent width, W, of several relevant lines in the optical range. We have chosen the ratio of the \ion{He}{ii} $\lambda 4686$ to \ion{He}{i} $\lambda 5876$ as a proxy for the temperature. Other ratios such as \ion{C}{iii} to \ion{C}{iv} or \ion{N}{iii} to \ion{N}{iv} would also reflect the temperature of the SN, but these lines may not be present in the spectra depending on the surface abundances of the progenitor. The \ion{He}{i} $\lambda 5876$ line was chosen due to its strength and isolation, i.e. it is not usually blended with other lines. While the \ion{He}{ii} $\lambda 4686$ introduces several uncertainties in the calculation of the equivalent width due to blending from the adjacent \ion{C}{iii} and/or \ion{N}{iii} lines, it is the strongest \ion{He}{ii} line in the optical range. As a proxy for $\mdot/\vinf$ we have chosen the \ion{H}{$\beta$} line, again due to the guaranteed presence of \ion{H}{} in the spectra of type II SNe, the proximity in wavelength to the other lines used in this analysis, and the fact that it is less sensitive to time-dependent effects than \ion{H}{$\alpha$}. A caveat of using \ion{H}{i} and \ion{He}{i} lines would be the possible contamination from the interstellar medium (ISM), but higher resolution observations should circumvent this issue in cases where \vinf\ is higher than the velocity of the nebula. Note that measuring the equivalent widths using different methods will also introduce extra uncertainties.

To measure the equivalent widths of the lines listed above, the observed spectra were first normalised. We have approximated the continuum emission in the optical range by fitting a polynomial function specific to each of the observed spectra and divided the observed spectrum by that function. We estimate the normalisation method introduces and error of 5 \% in the W values. The equivalent widths were then measured in the 4841 - 4881 \ang~range for the \ion{H}{$\beta$} $\lambda 4861$ line, 5856 - 5896 \ang~for the \ion{He}{i} $\lambda 5876$ line, and 4666 - 4706 \ang~for the \ion{He}{ii} $\lambda 4686$ line. The errors are taken as the averaged standard deviation of the flux measured over several wavelength ranges where no obvious emission lines are present. 

\begin{figure*}
    \begin{subfigure}[t]{0.49\textwidth}
        \includegraphics[width=1.0\linewidth]{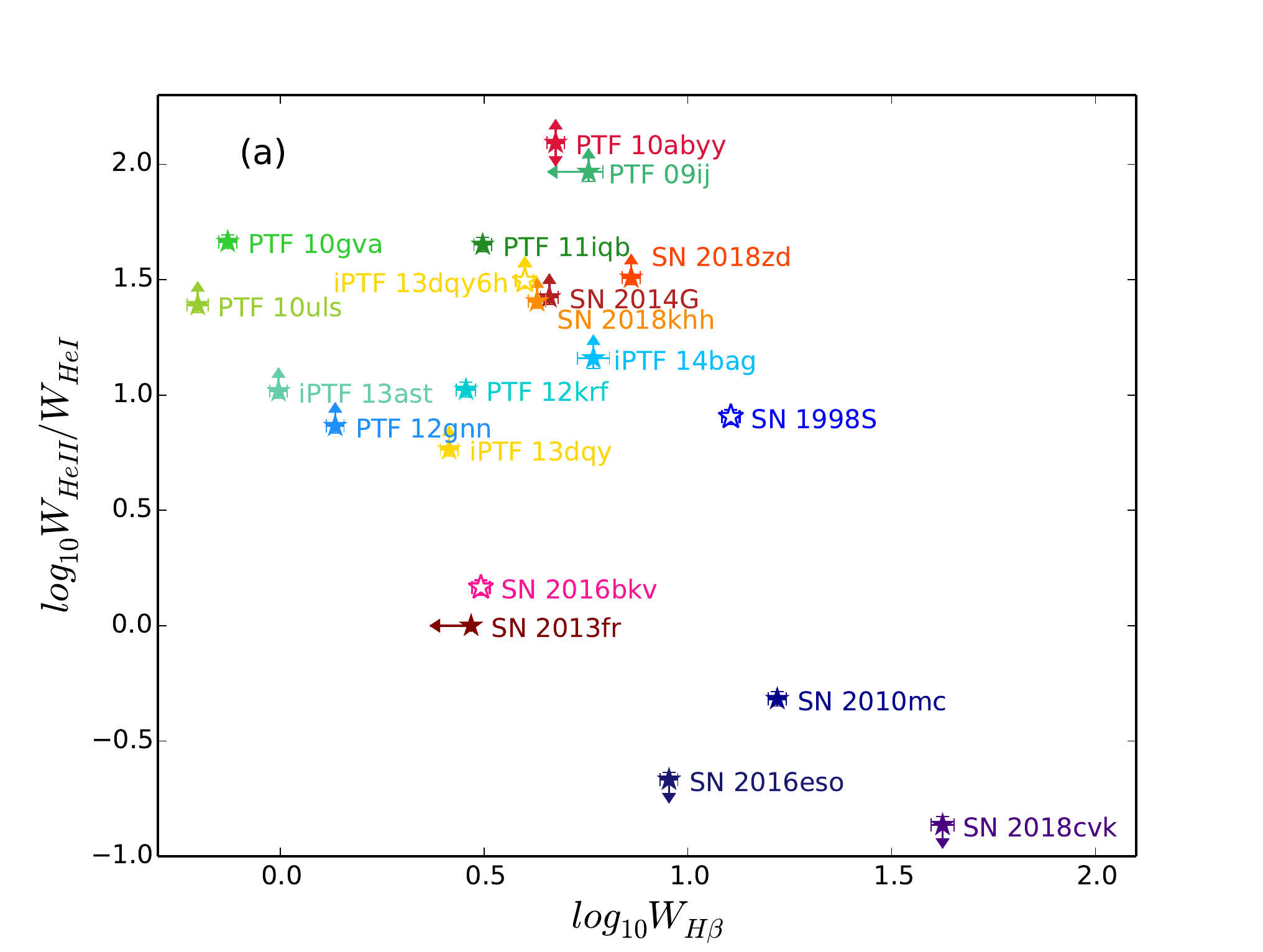}
    \end{subfigure}\hfill
    \begin{subfigure}[t]{0.49\textwidth}
        \includegraphics[width=1.0\linewidth]{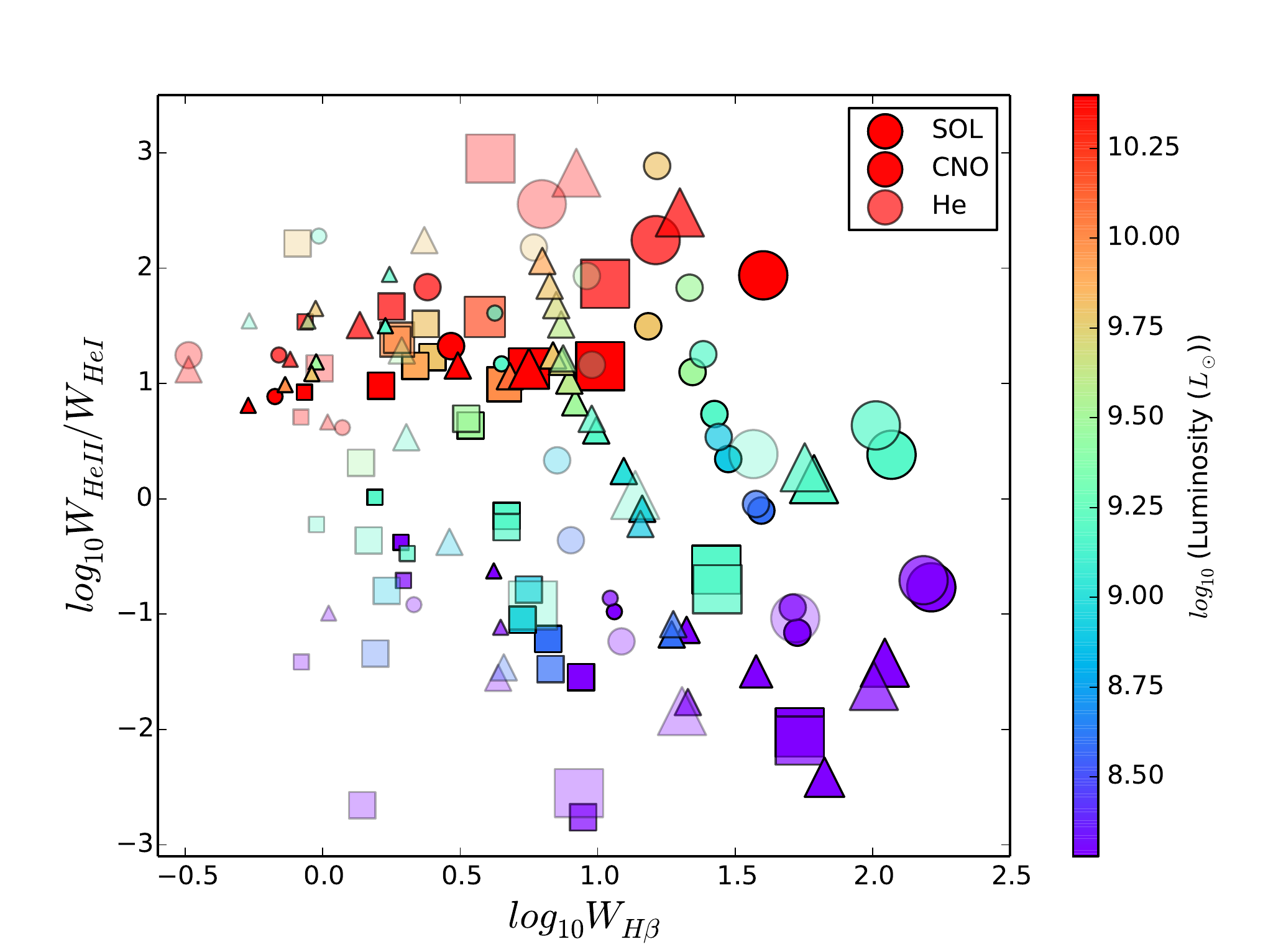}
    \end{subfigure}
    
    \caption{The equivalent width of the \ion{H}{$\beta$}$\lambda 4861$ line vs the \ion{He}{ii}$\lambda 4686$ to \ion{He}{i}$\lambda 5876$ ratio for the observed SNe (left) and the models (right). We have added some previously studied SNe (empty stars): SN 1998S at 1.8 days post-explosion (blue), SN 2016bkv at 4 days post-explosion (pink), and PTF13dqy (SN 2013fs) at 0.6 hours post-explosion (gold). On the right, the symbols correspond to models at 1 day (circles), 2 days (triangles), and 3 days (squares) post-explosion for different luminosities (colour) and mass-loss rates (size, $10^{-3}$ to $10^{-2}$ \msunyr). The different transparency levels represent the different surface abundances, the solid symbols having solar-like surface abundances, the semi-transparent have CNO-processed abundances, while the most transparent symbols are the \ion{He}{}-rich models.}
    \label{fig:ew}
\end{figure*}

Figure \ref{fig:ew}a shows the equivalent widths of the selected lines for all the supernovae in our sample. We can see that some of events are clearly low ionisation SNe (e.g. SN 2010mc), clustering in the bottom right corner, i.e. low \ion{He}{ii}/\ion{He}{i} ratios. They also show distinctively stronger \ion{H}{i} emission than the higher ionisation events, which could be due to two factors. Firstly the lower temperatures lead to higher recombination rates and therefore stronger \ion{H}{i} emission lines. Secondly these SNe could have higher \ion{H}{} abundances in the CSM to begin with. The majority of the observations, however, seem to tend towards the top half of the plot, but show no other obvious trend. The exact locations on the plot are also unknown for some of these events, since they might not show clear emission in one of the lines used in the analysis, most commonly the \ion{He}{i} emission for the high ionisation cases or \ion{He}{ii} for the low ionisation cases. We have instead placed limits on the maximum W value of a line that cannot be detected in each spectrum by calculating the area under a Gaussian with a peak equal to the standard deviation of the flux of each SN over an area with no evident lines and assuming a FWHM equal to that of a Gaussian fit to the \ion{H}{$\beta$} line of the corresponding SN. SN 2013fr does not clearly show any of the lines used for this analysis, so the noise limit is used for all of its lines. PTF10abyy is also truncated at 5600 \ang, thus lacking the \ion{He}{i} line. Based on the similarity of the available spectrum to other events we have placed it in the higher ionisation end, but its position could also be much lower than presented in Fig. \ref{fig:ew}a. We have also added to this plot the measurements of 3 other spectra of events we do not analyse in detail in this paper, but which have constraints on their explosions and progenitors from previous works. We have added the 2 day spectrum of SN 1998S \citep{shivvers15}, the 4 day spectrum of SN 2016bkv \citep{hosseinzadeh18, nakaoka18}, and the 6 h spectrum of iPTF13dqy \citep{yaron17}. \cite{deckers19} show that SN 2016bkv is fit by parameters corresponding to the medium to low ionisation range and has a low mass-loss rate ($\mdot = 6 \times 10^{-4} ~\msunyr$ and $L = 5.5 \times 10^{8} ~\lsun$), which matches its placement in the lower left side of Fig. \ref{fig:ew}a. Modelling the spectrum of SN 1998S revealed $L=1.5 \times 10^{10} ~\lsun$ and $\mdot= 6 \times 10^{-3} ~\msunyr$ \citep{shivvers15}, confirming the high-ionisation, high mass-loss rate prediction given by its top right  placement in Fig. \ref{fig:ew}a. For PTF13dqy, \cite{yaron17} modelled the earliest spectrum, obtained 6 hours after explosion and found $\mdot = 2-4 \times 10^{-3} ~\msunyr$ and $L = 2.0-3.5 \times 10^{10} ~\lsun$, values which match its position in the equivalent widths figure, in between the two previously discussed SNe on the x-axis, and in the top half of the figure. This spectrum shows no \ion{He}{i} emission so an upper limit was calculated as previously discussed, therefore its position could be much higher.

Overall Fig. \ref{fig:ew}a provides a good indication of the type of spectrum a SN interacting with its progenitor shows at early times and could be used as an initial guide in determining \teff~ and \mdot. Care must be taken when using this method as the time of the observation, the SN velocity, and the surface abundances of the progenitor can also influence the strength of the employed emission lines. We will further discuss trends and some sources of scatter in the following section, where we show a similar figure for synthetic spectra.

\section{Radiative Transfer Modelling Using CMFGEN}
\label{sect:method}
Spectroscopic modelling allows a reliable determination of the properties of progenitors of interacting SNe, such as abundances, mass-loss rates, and CSM velocities. Here we use the radiative transfer code CMFGEN \citep{hm98}, with the implementations for modelling interacting SNe as described in \citet{groh14a} and \citet{boian19}. In short, CMFGEN computes the transport of radiative energy through stationary expanding atmospheres in spherical symmetry, in the non-local thermodynamic equilibrium regime. 

In our setup we do not need to specify the source of energy, although it is widely accepted to come from the conversion of kinetic energy as the SN ejecta shocks the wind or CSM of the progenitor. Therefore our main input physical parameters are the luminosity of the event $L$ at the inner boundary, the progenitor's mass-loss rate \mdot, terminal wind velocity \vinf, the surface abundances, and the location of the inner boundary \rin, which relates to the ejecta velocity and explosion time. Due to the dense CSM these SNe are embedded in a pseudo-photosphere. Therefore we define two temperatures, \tstar\ at the inner boundary ($\tau \simeq 10.0$), and \teff\ at $\tau = 2/3$. A detailed description of our setup and models can be found in \cite{boian19}. 

In our models the SN luminosities range from $1.9 \times 10^{8} ~\lsun$ to $2.5 \times 10^{9} ~\lsun$, the progenitor mass-loss rates are in between $10^{-3} ~\msunyr$ and  $10^{-2} ~\msunyr$, and we have three different radii ($8 \times 10^{13}$ cm, $16 \times 10^{13}$ cm, and $32 \times 10^{13}$ cm) corresponding to 1.0, 1.8 and 3.7 days post-explosion if we assume a constant SN velocity of $10000 ~\kms$. We keep a constant wind velocity of $\vinf = 150 ~\kms$. We also explore three different surface abundance scenarios solar-like, CNO-processed, and He-rich. The relative abundances in mass fractions for each set are as follows: the solar-like case has $\ion{H}{} = 0.70$, $\ion{He}{}=0.28$, $\ion{C}{} = 3.02 \times 10^{-3}$, $\ion{N}{} = 1.18 \times 10^{-3}$, and $\ion{O}{} = 9.63 \times 10^{-3}$; the CNO-processed case has $\ion{H}{} = 0.70$, $\ion{He}{} = 0.28$, $\ion{C}{} = 5.58 \times 10^{-5}$, $\ion{N}{} = 8.17 \times 10^{-3}$, and $\ion{O}{} = 1.32 \times 10^{-4}$; and the  He-rich case has $\ion{H}{} = 0.18$, $\ion{He}{} = 0.80$, $\ion{C}{} = 5.58 \times 10^{-5}$, $ \ion{N}{} = 8.17 \times 10^{-3}$, and $\ion{O}{} = 1.32 \times 10^{-4}$.

However, the models can be extrapolated/interpolated to different parameter spaces using several scaling relations. Firstly the terminal wind velocities of massive stars cover a wide range of values. However, what we are truly fitting in our models and what can be typically constrained from observations in cases where the lines are not resolved is $\mdot/\vinf$ (but see \citealt{grafener16}). Therefore \mdot\ can be scaled for different values of \vinf\ as needed. Secondly, the ionisation level of the species present in a spectrum is in essence dictated by the temperature structure. In our models we input the luminosity of the supernova, which relates to \tstar\ by the Stefan-Boltzmann law, $L = 4\pi \rin^2\sigma \tstar^{4}$ \citep{boian19}. In the process of finding a best-fit model for an observed event, we fit the emission and absorption lines first, i.e. we find a good fit for the temperature, and then we match the luminosity to the observed absolute flux. Oftentimes this requires an adjustment in luminosity. When this happens, in order to keep the same temperature, \rin\ is also adjusted according to the Stefan-Boltzmann law. In addition, if the luminosity changes, the mass-loss rate needs to be scaled in order to preserve the optical depth scales. We adopt the relation $\mdot \propto \lsn^{3/4}$ from \cite{grafener16} (note Equation 7 in \cite{grafener16} contains a typo in the \lsn\ exponent, using '4/3' rather than '3/4'). Lastly, since we have not accounted for dust extiction in the processing of the observed spectra, we redden the synthetic fluxes using the \cite{fitzpatrick99} parameterization and determine the best-fit values for the colour excess, E(B-V) and the parameter of relative visibility, R(V). We employ these relations when fitting the observed spectra as will be shown in Sect. \ref{sect:bf}.  

\subsection{Empirical relationships explained by CMFGEN models}
\begin{sidewaysfigure*}
    \centering
    \includegraphics[width=1.0\textheight]{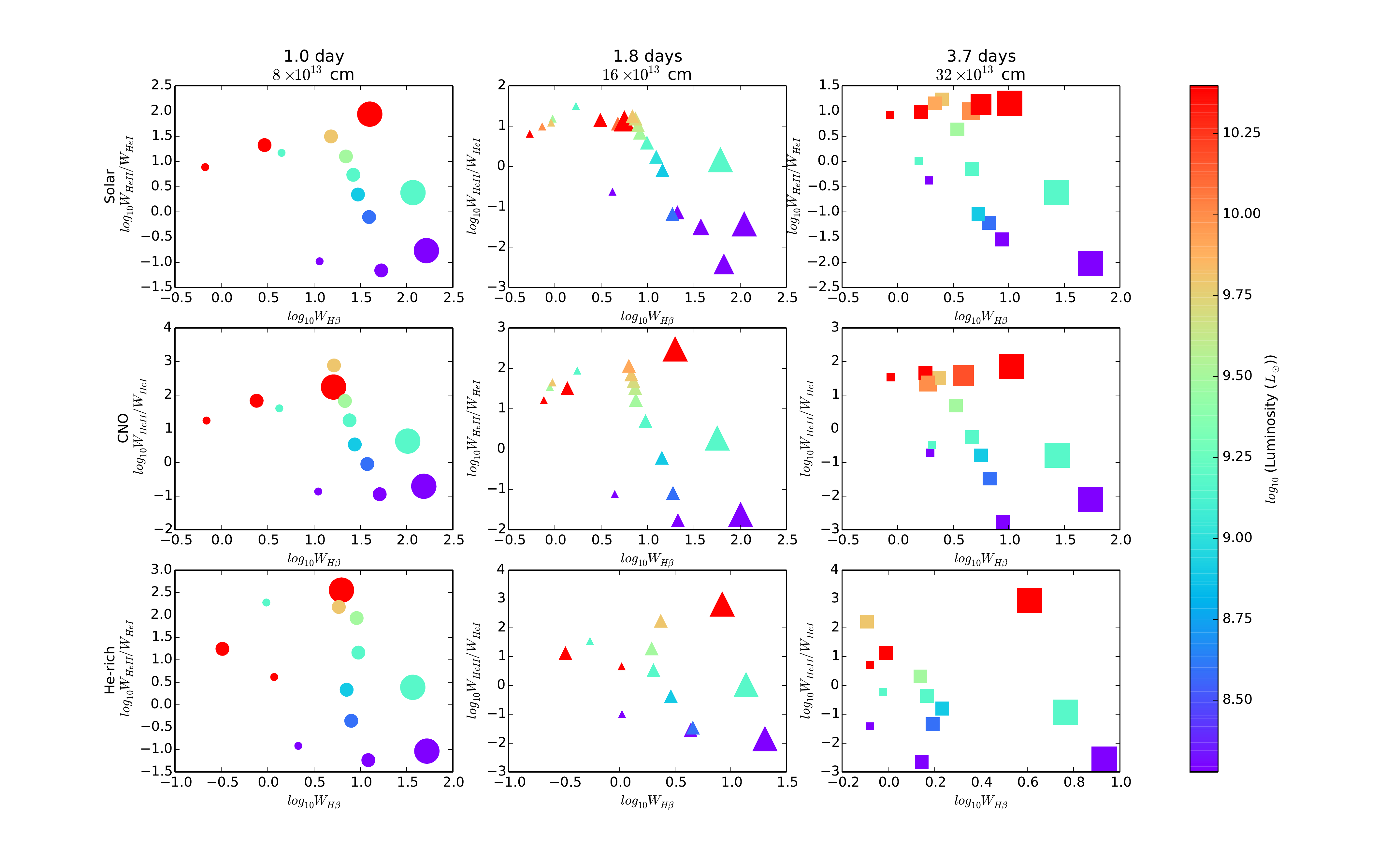}
    \caption{The equivalent width of the \ion{H}{$\beta$}$\lambda 4861$ line vs the \ion{He}{ii}$\lambda 4686$ to \ion{He}{i}$\lambda 5876$ ratio for a set of CMFGEN synthetic spectra. We explore 3 post-explosion times, with models at 1 day post-explosion on the first column, models at 1.8 days on the second column, and models at 3.7 days on the third column. The explosion times correspond to the inner radii of $8 \times 10^{13}$ cm (first column), $16 \times 10^{13}$ cm (second column), and $32 \times 10^{13}$ cm (third column). There are also 3 sets of surface abundances, solar-like (top row), CNO-processed (middle row), and He-rich (bottom row). The colours relate to SN luminosities, while the sizes correspond to the progenitors' mass-loss rates, from $10^{-3}$ for the smallest symbol, to $10^{-2}$ for the largest.}
    \label{fig:ew_models_large}
\end{sidewaysfigure*}

The empirical relationships observed in our sample of SNe (Sect. \ref{sect:emp_rel}) are reproduced well by our models. Similarly to Fig. \ref{fig:ew} a, in Figs. \ref{fig:ew}b and \ref{fig:ew_models_large} we show the equivalent widths of the \ion{H}{$\beta$} line vs the ratio of the \ion{He}{ii} to the \ion{He}{i} equivalent width, for a set of synthetic spectra from our library of models. The equivalent width values have been obtained in a similar fashion to those from the observed spectra and they are given in Appendix \ref{apx:models}. By analysing the strengths of these lines as previously discussed for our sample of observed events, we aim to identify possible trends and relate them to the properties of the SN and its progenitor. We summarise our findings below in terms of trends in mass-loss rate (represented by the size of the symbols in Fig. \ref{fig:ew_models_large}), luminosity of the SN (colour coded in Fig. \ref{fig:ew_models_large}), the surface abundances (each row of panels in Fig. \ref{fig:ew_models_large} corresponds to a certain set of abundances), and inner radius (each column in Fig. \ref{fig:ew_models_large} follows one radius).

 Mass-loss rate effects:
 The \ion{H}{$\beta$} line was chosen as a proxy for \mdot. Figure \ref{fig:ew_models_large} shows that a lower \mdot\ leads to lower \ion{H}{$\beta$} emission at all times, luminosities, and abundances considered. Another effect of an increase in \mdot\ is the decrease in \teff\ for the same $L$ and \rin. This can be seen especially in the models with $1.5 \times 10^{9} ~\lsun$, for which the \ion{He}{ii}/\ion{He}{i} ratio decreases for increasing \mdot. 
 
 Luminosity effects: 
 An increase in luminosity for models with otherwise identical properties leads to an increase in temperature. This affects the spectral lines in Fig. \ref{fig:ew_models_large} in two main ways: the \ion{He}{ii} to \ion{He}{i} ratio increases, and the \ion{H}{$\beta$} luminosity decreases. Therefore, following the models with different \lsn but the same \mdot\ in  Fig. \ref{fig:ew_models_large}, for most sets of abundances and radii we see that the models tend towards higher \ion{He}{ii} to \ion{He}{i} rations and slightly towards lower \ion{H}{$\beta$} emission. This breaks for the highest \lsn\ model, since that model is too hot for \ion{He}{ii} as well. Additionally,  Fig. \ref{fig:ew_models_large} shows that the He-rich models increase in \ion{H}{$\beta$} flux after $L=1.5 \times 10^{9} ~\lsun$. However, this is due to contamination from the close-by \ion{He}{ii} $\lambda 4859$ line, which increases at higher \lsn.
 
Abundance effects:
The abundance effects are slightly more complex than other parameters. The models with solar-like and CNO-processed surface abundances are quite similar to each other, but not identical. The models at lower \lsn\ do not change much between the two cases, but the models at higher \lsn\ show much stronger \ion{He}{ii}. While in this case neither the \ion{H}{} nor the \ion{He}{} abundances change, the opacity changes as a result of the difference in CNO ratios. Another reason for this discrepancy could be the contamination of the \ion{He}{ii} line from the near-by \ion{N}{iii/iv} lines. 
Decreasing the H abundance/increasing the He abundance shifts the models strongly toward the left side of the plot due to the expected lower H emission. Additionally, at low \lsn\ the \ion{He}{ii} to \ion{He}{i} ratio decreases slightly. This is due to the fact that there is little or no \ion{He}{ii} emission due to the low \tstar, but the \ion{He}{} abundance is increasing, therefore the \ion{He}{i} line increases while the \ion{He}{ii} does not. The opposite is true at the high \lsn\ end, where \ion{He}{ii} increases due to the increased amount of \ion{He}{}, but \ion{He}{i} does not due to the high \tstar. At medium \lsn\ however, for high \mdot\ the ratio is lower in the He-rich models, but at low \mdot\ the ratio is higher, which is probably an effect of \teff, which is higher for lower \mdot.

Inner radius effects:
Increasing \rin\ will lead to a decrease in \tstar\ for a model with the same luminosity. Increasing \rin\ and keeping $\mdot/\vinf$ constant produces a lower density in the wind. Therefore in Fig. \ref{fig:ew_models_large}, the general trend between the models at different times (i.e. different radii) is that the larger \rin\ models move downwards because a lower \tstar\ implies a lower \ion{He}{ii}/\ion{He}{i} ratio, and to the left side of the plot due to the decrease in density. 

Overall, many qualitative trends can be identified in Fig. \ref{fig:ew_models_large}, however many degeneracies also appear and detailed modelling paired with high quality observations are needed to break them. Comparing the figure showing the equivalent widths measured in the observed events (Fig. \ref{fig:ew}a) to the figure of the equivalent widths measured in our models (Fig. \ref{fig:ew}b) we can see that our grid samples the entire observed parameter space. These types of metrics could be useful in quickly constraining the properties of SNe interacting with their progenitors. They can also provide a starting guide when computing detailed spectral models, as we have in the following section.

\section{Constraining progenitor and explosion properties}
\label{sect:bf}

Here we estimate SN and progenitor properties such as \lsn, \tstar, \mdot, and surface abundances for the events in our sample by comparing the observed spectra to our models. For the sake of brevity, we discuss in this section the detailed modelling of three SNe in our sample, illustrating the low-, medium-, and high-ionisation cases. A summary of the results can be found in Table \ref{tab:bf_param} and detailed discussions on the best-fits of the remaining SNe in our sample can be found in Appendix \ref{apx:bfs}. The values quoted in this section and in Table \ref{tab:bf_param} have been scaled following the relations described in Sect. \ref{sect:method}.

\begin{figure*}
    \begin{subfigure}[t]{0.99\textwidth}
        \includegraphics[width=0.96\textwidth]{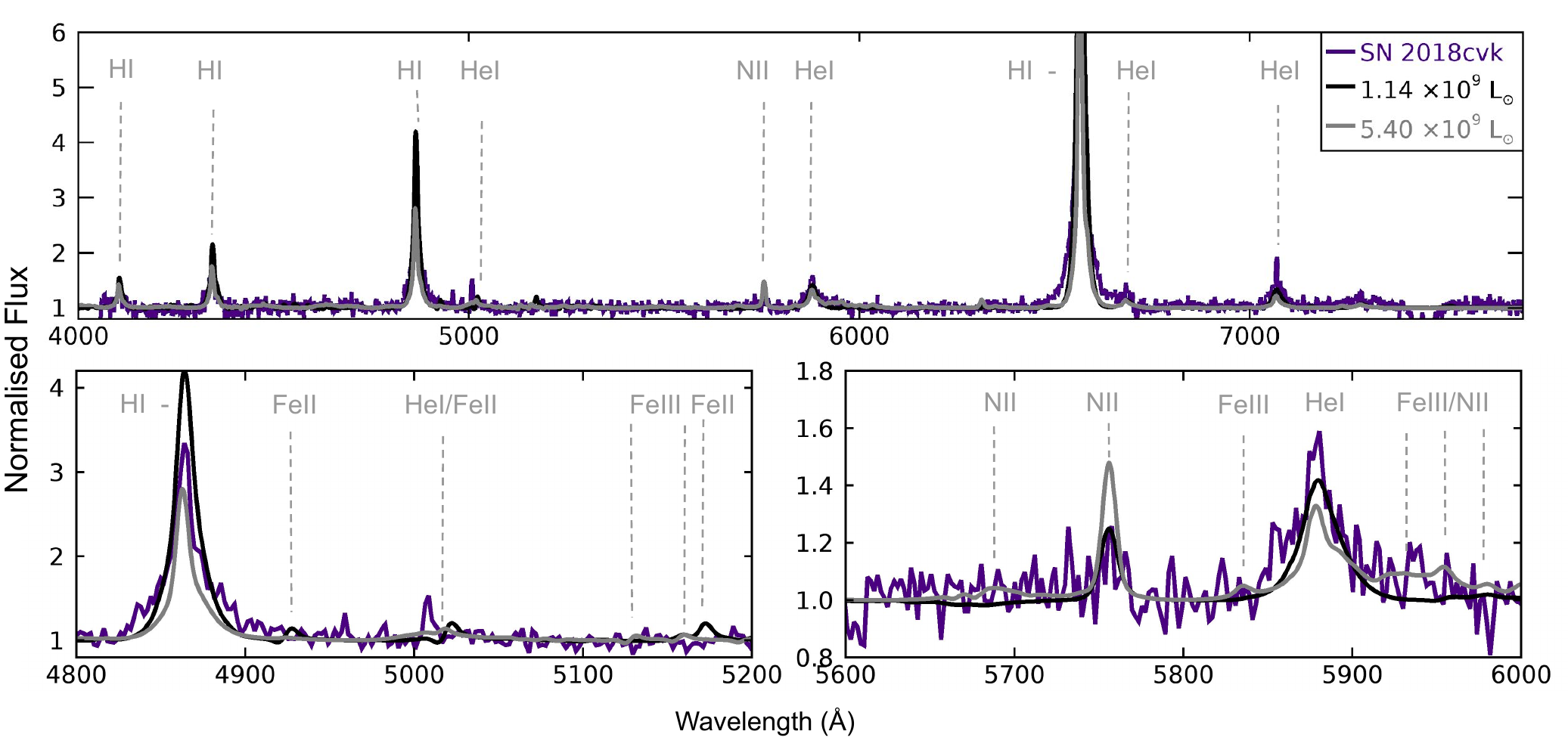}
        \caption{The models have $L = 1.14 \times 10^{9} ~\lsun$, $\rin = 78.0 \times 10^{13}$ cm, $\mdot = 7.49 \times 10^{-2} ~\msunyr$, $\vinf = 500 ~\kms$, $\tstar = 9919$ K, and CNO-processed surface abundances (black), and $L = 5.4 \times 10^{9} ~\lsun$, $\rin = 60.7 \times 10^{13}$ cm, $\mdot = 3.33 \times 10^{-2} ~\msunyr$, $\vinf = 500 ~\kms$, $\tstar = 16620$ K, and CNO-processed surface abundances (grey), respectively. The bottom panels are zoomed-in regions of the top panel.}
        \label{fig:18cvkL}
    \end{subfigure}

    \begin{subfigure}[b]{0.99\textwidth}
        \includegraphics[width=0.96\textwidth]{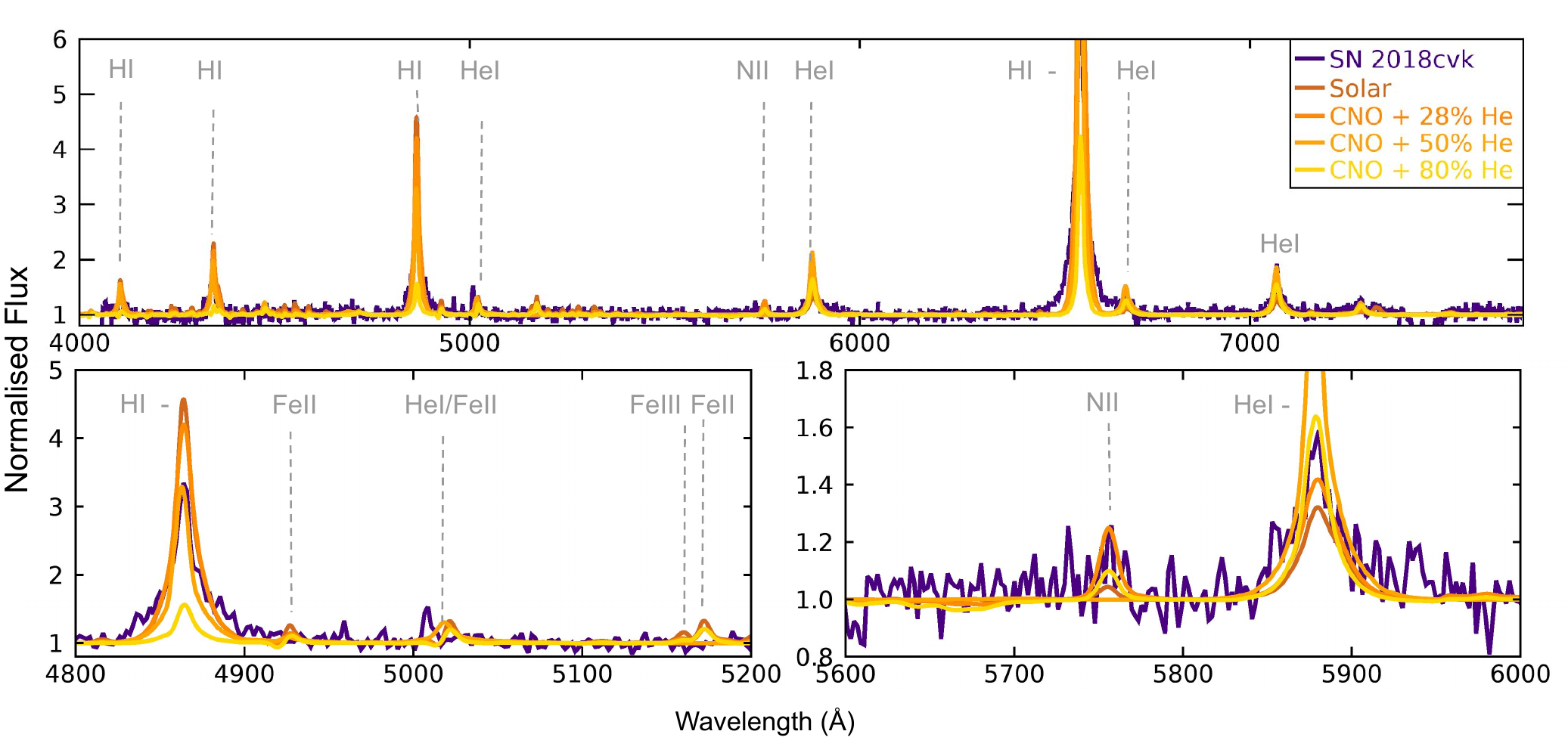}
        \caption{The models with solar (brown), CNO-processed abundances and 28\% \ion{He}{} (dark orange), and 80\% \ion{He}{} (gold) have the same properties as the black model in panel (a), except for the surface abundances (and a slight difference in \tstar due to the difference in opacity). The model with 50\% \ion{He}{} (orange) is closer in properties to the grey model in panel (a), except for the surface abundances. }
        \label{fig:18cvkZ}
    \end{subfigure}
    
    \begin{subfigure}[b]{0.99\textwidth}
        \includegraphics[width=0.96\textwidth]{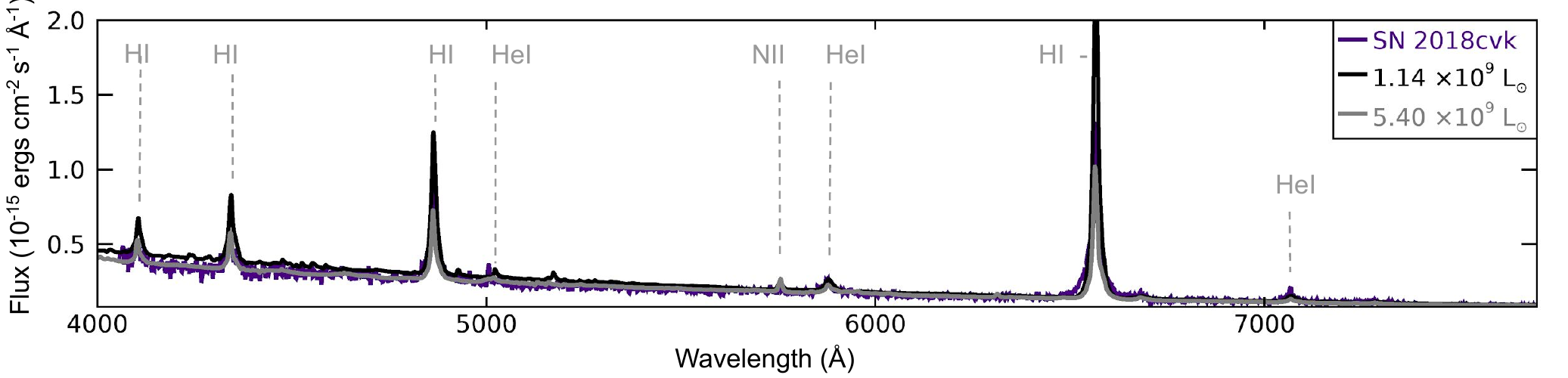}
    \caption{Absolute flux of SN 2018cvk and the models from panel (a). Assuming a distance of $d = 111.51$ Mpc, the best-fit is given by a colour excess of E(B-V) = 0.1 and a relative visibility R(V) = 3.1 for the black model, and E(B-V) = 0.2 and R(V) =3.1 for the grey model. }
    \label{fig:18cvkAbs}
    \end{subfigure}
    
    \vspace{-0.2cm}
    \label{fig:18cvkBestFits}
    \caption{The observed spectrum of SN 2018cvk (dark purple) and the closest fitting models.}
\end{figure*}

SN 2018cvk is a clear example of a low-ionisation interacting SN, exhibiting mainly \ion{H}{i} and \ion{He}{i} lines. Since its spectrum was taken $5 \pm 1$ days post-explosion, we compare it to our set of models computed at the latest times post explosion (at 3.7 days). This is not a major issue since the explosion time in our models is estimated assuming a constant $\mathrm{v_{ej}} = 10\,000 ~\kms$, which may vary in this case, and also the models have to be scaled up as explained in Sect. \ref{sect:method} eventually leading to a higher \rin\ and hence larger $\mathrm{t_{exp}}$. Our models show that SN 2018cvk has $L = 1.1 - 5.4 \times 10^{9} ~\lsun$, $\mdot = 3.33-7.49 \times 10^{-2} ~\msunyr$ ($\vinf = 500 ~\kms$), $\tstar = 9900 - 16600$ K, and $\rin = 60.7 - 78.0 \times 10^{13}$ cm (Fig. \ref{fig:18cvkL}). The luminosity was adjusted by a factor of $6.0$ and $3.6$ respectively, in order to fit the absolute flux of the observations (assuming a distance of 111.51 Mpc) and the spectral energy distribution (SED) is best-fit by a colour excess of $E(B-V)=0.05$ ($R_V = 3.1$), and $E(B-V)=0.2$ for the higher \lsn\ model (Fig. \ref{fig:18cvkAbs}). Our best-fit model slightly overestimates the \ion{H}{i} lines and underestimates the \ion{He}{i} lines. The \ion{He}{i} have a clear narrower component which might be due to contamination from the ISM. It may also be that the \ion{He}{} to \ion{H}{} ratio is slightly underestimated in our model, but it would not be as high as in the \ion{He}{}-rich models. The abundances at the stellar surface could be either CNO-processed or solar-like. The two models differ by the presence of a \ion{N}{ii}$\lambda 5754$ line in the CNO case (due to the increased \ion{N}{} abundnace) and stronger \ion{Fe}{ii} lines in the $4900 - 5400 ~\ang$ region in the solar-like case. The different strengths of the \ion{Fe}{ii} lines are due to slight differences in the T structure in the outer wind caused by differences in the cooling rates, where coincidentally the \ion{Fe}{ii} lines are also formed. However none of these lines are above the SNR of the spectrum (Fig. \ref{fig:18cvkZ}) making the distinction between the two abundance scenarios difficult. Fig. \ref{fig:18cvkZ} also shows that the \ion{He}{} abundance falls in between 28\% and 50\%.
SN 2018cvk is comparable in luminosity and mass-loss rate to the other low-ionisation spectra in our sample (SN 2016eso and SN 2010mc; Appendix \ref{apx:bfs}). Due to the high mass-loss rates and relatively low \lsn\ we believe SN 2018cvk might exhibit interaction signatures for a prolonged period of time, perhaps belonging to the classical SN IIn category.  

\begin{figure*}
    \begin{subfigure}[t]{0.99\textwidth}
        \includegraphics[width=0.96\textwidth]{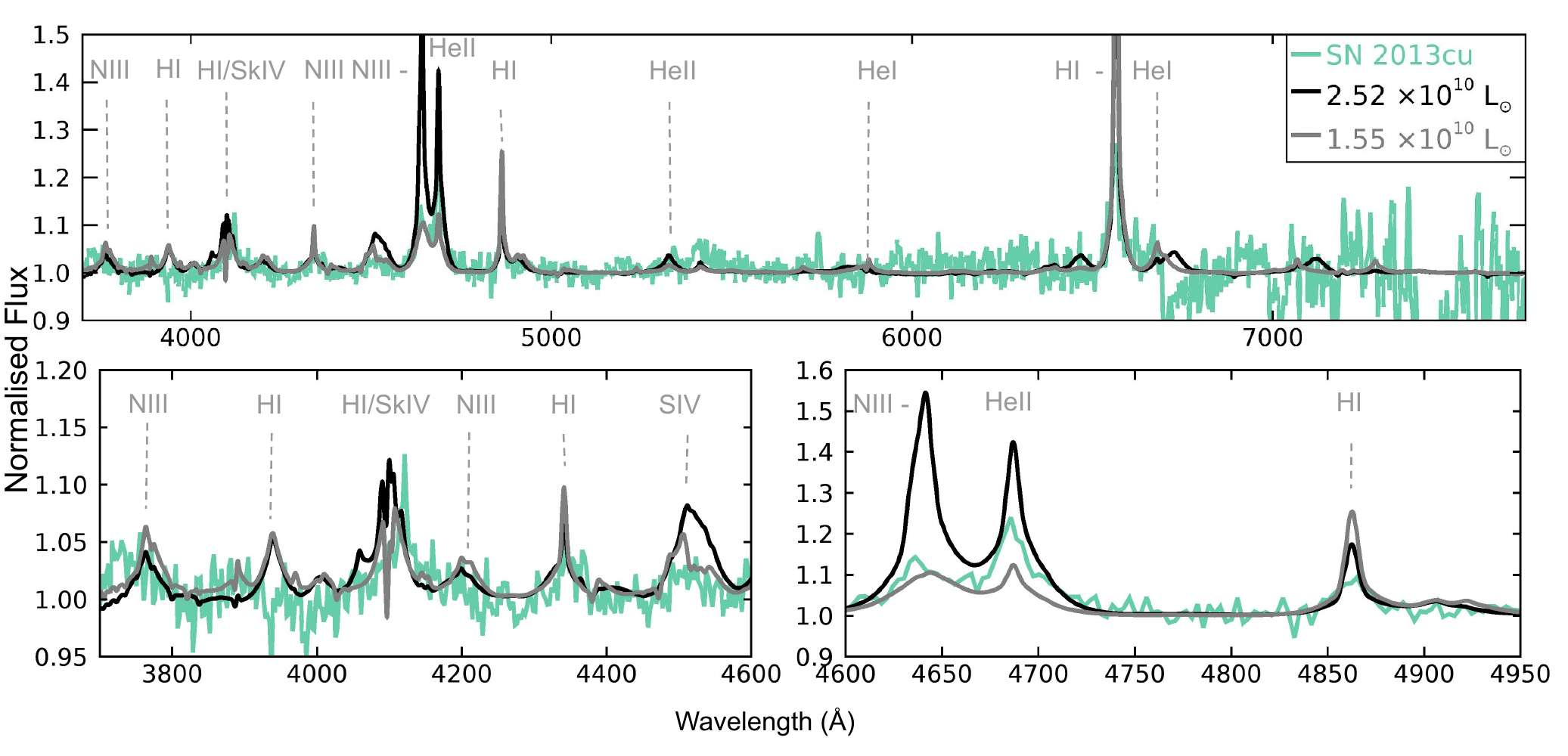}
        \caption{The models have $L = 2.5 \times 10^{10} ~\lsun$, $\rin = 64.0 \times 10^{13}$ cm, $\mdot = 5.7 \times 10^{-3} ~\msunyr$, $\vinf = 100 ~\kms$, $\tstar = 23\,800$ K, and CNO-processed surface abundances (black), and $L = 1.5 \times 10^{10} ~\lsun$, $\rin = 71.5 \times 10^{13}$ cm, $\mdot = 6.7 \times 10^{-3} ~\msunyr$, $\vinf = 100 ~\kms$, $\tstar = 19\,900$ K, and CNO-processed surface abundances (grey), respectively. The bottom panels are zoomed-in regions of the top panel.}
        \label{fig:13astL}
    \end{subfigure}
    
    \begin{subfigure}[b]{0.99\textwidth}
        \includegraphics[width=0.96\textwidth]{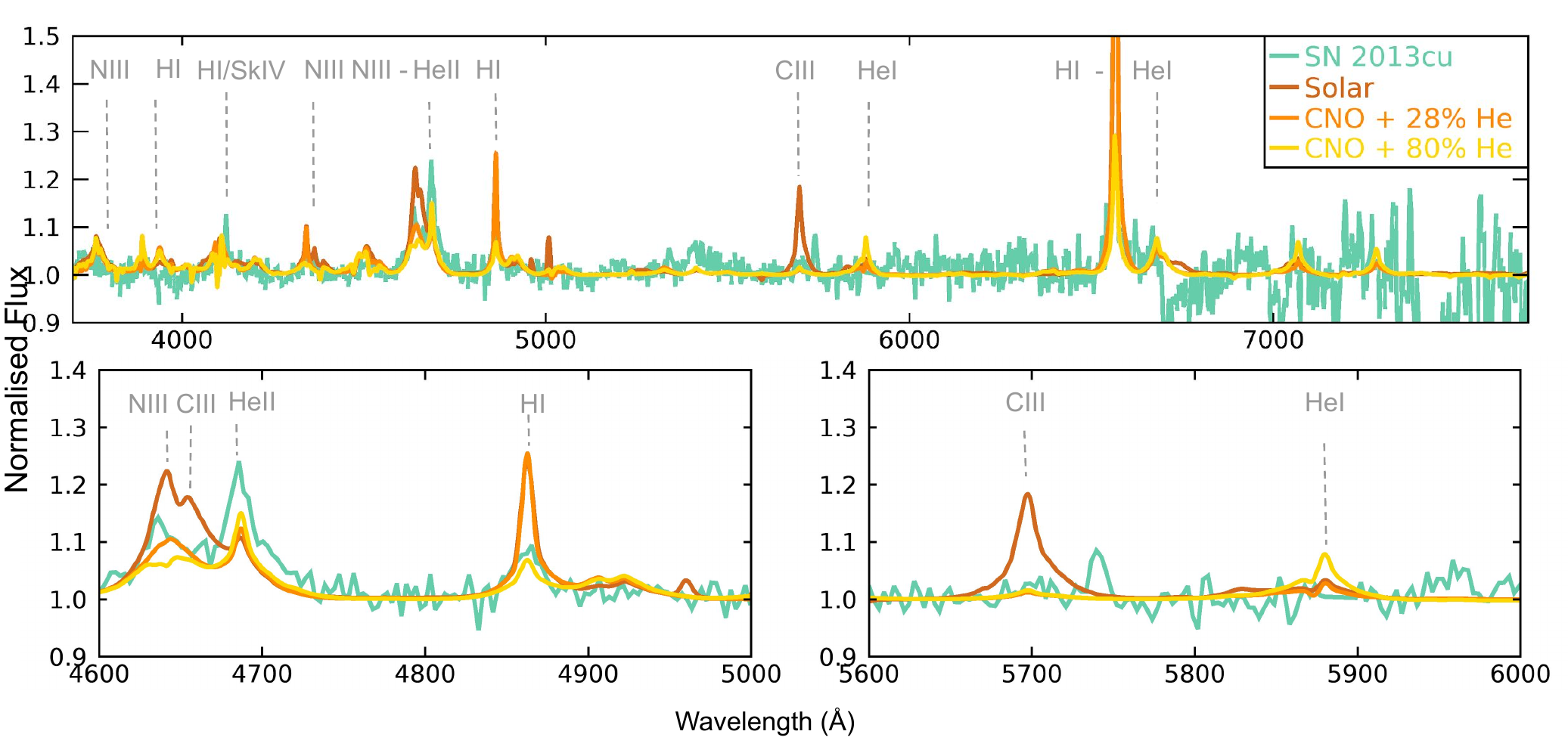}
        \caption{The models here have similar properties to the grey model from panel (a) and solar-like (brown), CNO-processed with 28\% \ion{He}{} (dark orange), and with 80\% \ion{He}{} (gold) surface abundances.}
        \label{fig:13astZ}
    \end{subfigure}
    
    \begin{subfigure}[b]{0.99\textwidth}
        \includegraphics[width=0.96\textwidth]{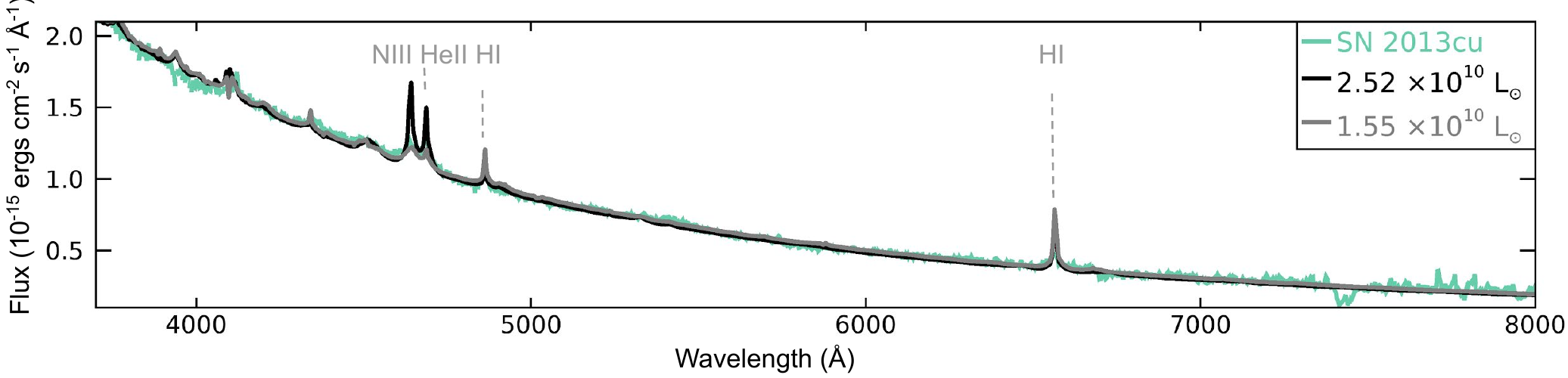}
        \caption{Observed absolute flux of SN 2013cu and the models from panel (a). Assuming a distance of $d = 108$ Mpc, the best-fit is given by a colour excess of E(B-V) = 0.1 and a relative visibility R(V) = 3.1 for both models.}
        \label{fig:13astAbs}
    \end{subfigure}
    
    \label{fig:13astBestFits}
    \caption{The 3 day-old observed spectrum of SN 2013cu (green) and the closest fitting models.}
\end{figure*}

iPTF13ast/SN 2013cu at 3 days post-explosion presents emission lines characteristic of a medium-ionised spectrum, such as \ion{N}{iii}, \ion{C}{iii} and \ion{S}{iv}. The best-fit model for this spectrum falls in between a model with $L = 1.5 \times 10^{10} ~\lsun$, $\rin = 71.5 \times 10^{13}$ cm, $\tstar = 19\,900$ K,  $\mdot = 6.7 \times 10^{-3} \msunyr$ and a model with $L = 2.5 \times 10^{10} ~\lsun$, $\rin = 64.0 \times 10^{13}$ cm, $\tstar = 23\,800$ K,  $\mdot = 5.7 \times 10^{-3} \msunyr$ (Fig. \ref{fig:13astL}). The terminal wind velocity cannot be well constrained using this spectrum alone since the lines are not resolved, therefore we adopt the velocity from \cite{groh14a} of $\vinf=100 ~\kms$. It can been seen in Fig. \ref{fig:13astL} that the lower \tstar\ model slightly underestimates the \ion{He}{ii} lines, while the higher \tstar\ model overestimates both \ion{He}{ii} and \ion{N}{iii}. Both models in Fig. \ref{fig:13astL} have CNO-processed surface abundances and $28\% $ \ion{He}{} and they clearly overestimate \ion{H}{i} emission. Fig. \ref{fig:13astZ} shows the $L = 1.55 \times 10^{10} ~\lsun$ model for three different surface abundance, and we can see that the \ion{He}{} abundance should be higher. This supports the findings from previous works and the II-b classification of this event. The CNO lines are well fitted by the models with CNO-processed abundance. A solar surface abundance would strongly overestimate the \ion{C}{iii} lines. Fitting the SED assuming a distance of $108$ Mpc reveals weak reddening, with a colour excess of $E(B-V)=0.01$ and $R_V =3.1$ (Fig. \ref{fig:13astAbs}). The scaling factors required for the \lsn\ to match the absolute flux were 5.0 and 4.0 respectively. Comparing to previous results which model the spectrum at 15.5 hours \citep{groh14a, grafener16,galyam14}, our analysis reveals slightly higher \lsn\ and \mdot, and similar abundances. The \lsn\ is expected to be higher since both spectra are pre-peak, with the latter spectrum analysed here being taken at a brighter stage.

\begin{figure*}
    \begin{subfigure}[t]{0.99\textwidth}
        \includegraphics[width=0.96\textwidth]{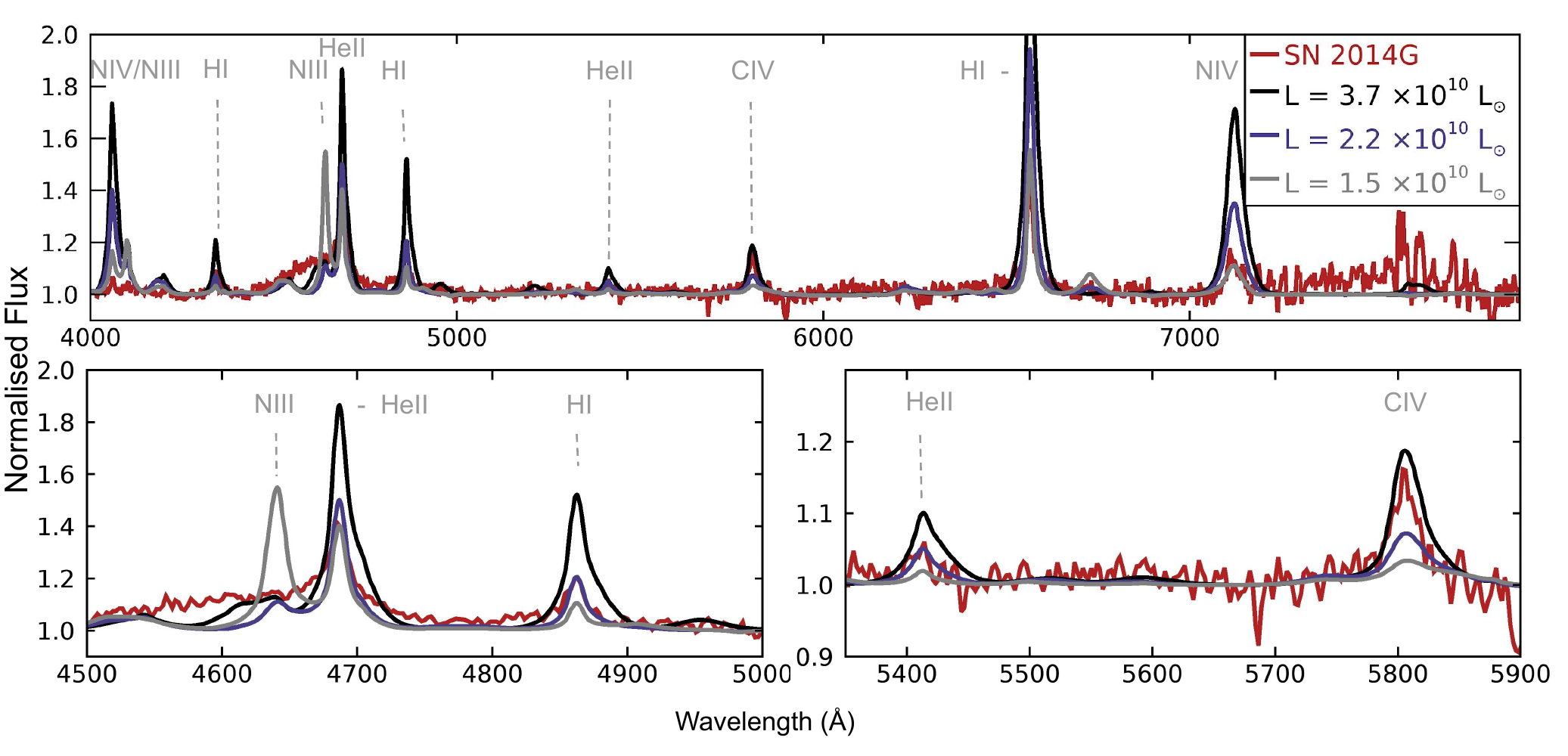}
        \caption{The models have $L = 3.7 \times 10^{10} ~\lsun$, $\rin = 39.2 \times 10^{13}$ cm, $\mdot = 49.6 \times 10^{-3} ~\msunyr$, $\vinf = 500 ~\kms$, $\tstar = 33\,500$ K, and CNO-processed surface abundances (black), $L = 2.2 \times 10^{10} ~\lsun$, $\rin = 39.2 \times 10^{13}$ cm, $\mdot = 33.9 \times 10^{-3} ~\msunyr$, $\vinf = 500 ~\kms$, $\tstar = 29\,300$ K, and CNO-processed surface abundances (purple), and $L = 1.5 \times 10^{10} ~\lsun$, $\rin = 39.2 \times 10^{13}$ cm, $\mdot = 45.2 \times 10^{-3} ~\msunyr$, $\vinf = 500 ~\kms$, $\tstar = 26\,700$ K, and CNO-processed surface abundances (grey) respectively. The bottom panels are zoomed-in regions of the top panel.}
        \label{fig:14gL}
    \end{subfigure}

    \begin{subfigure}[b]{0.99\textwidth}
        \includegraphics[width=0.96\textwidth]{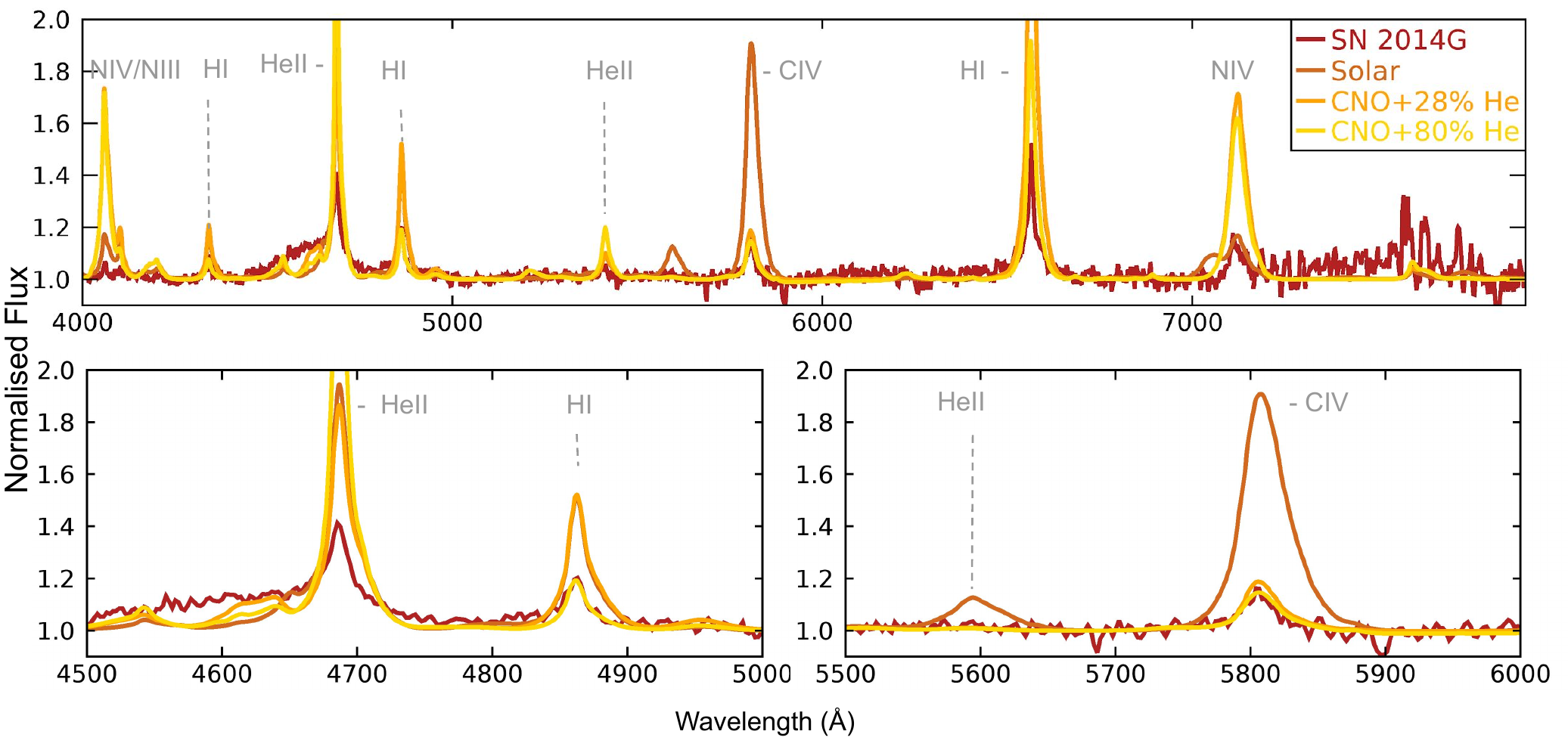}
        \caption{The models here have similar properties to the black model from panel (a) and solar-like (brown), CNO-processed with 28\% \ion{He}{} (dark orange), and with 80\% \ion{He}{} (gold) surface abundances.}
        \label{fig:14gZ}
    \end{subfigure}

    \begin{subfigure}[b]{0.99\textwidth}
        \includegraphics[width=0.96\textwidth]{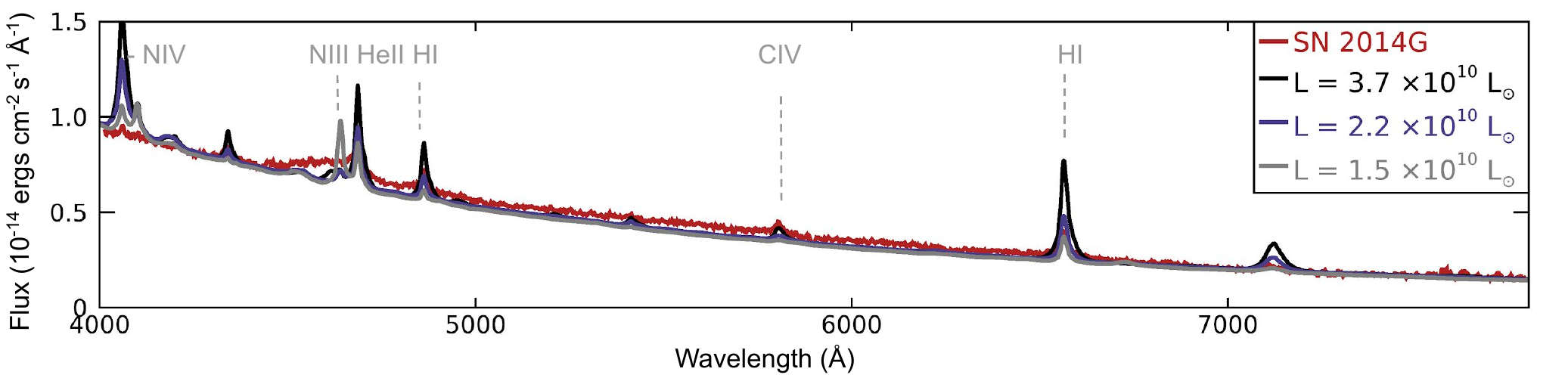}
        \caption{Observed absolute flux of SN 2014g and the models from panel (a). Assuming a distance of $d = 24.5$ Mpc, the best-fit is given by E(B-V) = 0.15, R(V) = 3.1 for the black model, E(B-V)=0.13, R(V) = 3.1 for the purple model, and E(B-V)=0.11, R(V) = 3.1 for the grey model.}
        \label{fig:14gAbs}
    \end{subfigure}
    
    \caption{The observed spectrum of SN 2014g (dark red) and the closest fitting models.}
    \label{fig:14gBestFits}
\end{figure*}

To discuss the spectroscopic modelling and diagnostic lines of high-ionisation events we use SN 2014G, which shows strong \ion{He}{ii}, \ion{C}{iv}, \ion{N}{iv}, and \ion{N}{v} features in its early spectra. Many spectra in our sample are very similar to that of SN 2014G, such as those of PTF 10uls, PTF 10abyy, PTF 11iqb, or SN 2018khh. Our modelling shows that SN 2014G has $L = 2.2 - 3.7 \times 10^{10} ~\lsun$, $\mdot = 33.9 - 49.6 \times 10^{-3} ~\msunyr$, $\vinf = 500 ~\kms$, $\rin = 39.2 \times 10^{13}$ cm, $\tstar = 29300 - 33500$ K and CNO-processed surface abundances, with $Y$ larger than $0.28$, but not as high as $0.80$ (Figs. \ref{fig:14gL} \& \ref{fig:14gZ}). Fitting the SED and assuming a distance of $24.5$ Mpc reveals an extinction coefficient of $E(B-V) = 0.17 \pm 0.03$ and $R_V = 3.1$ (Fig. \ref{fig:14gAbs}), matching previous works. \cite{terreran16} place a constraint of $M_{ZAMS} = 17 \pm 2 ~\msun$ and suggest a RSG or a Yellow Hypergiant (YHG) with enhanced mass-loss since \cite{bose16} place a constraint of $9 ~\msun$ at the pre-explosion stage. They also suggest a possible bipolar outflow. Our results support the RSG/YHG hypothesis. The CNO-processed surface abundances would be expected from a more massive ($\sim 20$ \msun) RSG, a YHG or a BSG star. The mass-loss rate and wind velocities are however atypical of quiescent RSGs thus implying enhanced mass-loss at the pre-explosion stages. If we take into consideration that the narrow lines were only visible for 9 days, and naively assume a constant ejecta velocity of $v_{sn} = 10\,000~\kms$, then the dense CSM extends up to $77.7 \times 10^{8}$ km. Further assuming a stellar radius of $R = 630 ~\rsun = 4.38 \times 10^{8}$ km \citep{bose16} and a constant wind velocity of $\vinf = 500 ~\kms$ then the wind as we observe it has only blown for $0.46$ years before explosion. With a mass-loss rate of $(1.4 - 5.0) \times 10^{-2} ~\msunyr$ we get a CSM mass of $(0.6-2.3) \times 10^{-2} ~\msun$. This implies that most of the mass was lost prior to this pre-SN episode, which is in agreement with stellar evolution models . These values would be underestimated if the geometry of the CSM deviates from spherical symmetry. \cite{hillier19} indeed derive a CSM mass of $\sim 1~\msun$ from the LC fitting. The inability to simultaneously fit all the lines observed in the flash ionisation spectrum could point to an aspehrical geometry. Interestingly, PTF11iqb which has a similar spectrum to SN 2014G has also shown signs of asphericity \citep{smith15}.

\begin{sidewaystable*}
    \footnotesize
	\noindent
	\caption{Best-fit parameters for our sample of observed SNe, derived using CMFGEN models. Here, \rin\ is the inner radius, \lsn, the luminosity of the SN, \tstar, the temperature at an optical depth of $10.0$, \vinf, the terminal wind velocity, \mdot\ the mass-loss rate of the progenitor, and $D$, the density parameter. These quantities are described in detail in \ref{sect:method}. In the cases where the widths of the narrow CSM lines correspond to the resolution of the spectra, we use the resolution as the upper limit of \vinf, which also propagates to the corresponding \mdot. The mass-loss rates for a fixed wind velocity of $150 ~\kms$ have also been included. The abundances reflect the progenitor's surface/CSM abundances, and the fraction of each element for the three cases are given in Sect. \ref{sect:method}. We also include the colour excess, E(B-V), required to fit the observations, assuming R(V)=3.1. The parameters for the last 4 events included in this table are taken from the referenced works. The SN types are taken from the current literature where available, and estimated from the public light curves if no previous classification was found.}
	\label{tab:bf_param}
	\hskip-0.5cm
	\begin{threeparttable}
	\begin{tabular}{lccccccccccc} 
		\hline
		SN Name & \rin & \lsn & Scaling & E(B-V) & $\mdot$ & $\vinf$ & $\mdot (\vinf/150 ~\kms)$ & $D$ & \tstar\ ($\tau = 10$ ) & Abundances & SN Type \\
		 & $10^{13}$ cm & $10^9 ~\lsun$ & factor & & $10^{-3} \msunyr$ & \kms & $10^{-3} \msunyr$ & $10^{15}$ g $\mathrm{cm^{-1}}$ & $10^{3} $ K & & \\
		\hline
		PTF09ij & $62.4-66.4$ & $43.0-57.0$ & $3.8-4.3$ & $0.22$ & $< 22.7$ & $<250$ & $8.94-13.62$ & $3.0-4.6$ & $26.7-29.3$ & CNO & II$^{[5]}$  \\
		PTF10abyy & $230-240$ & $510.0-1375.0$ & $51-55$ & $0.37-0.44$ & $506.7-1066.6$ & $800$ & $95-199.98$ & $31.8-67.0$ & $26.7-33.5$ & SOL/CNO & II-P* \\
		PTF10gva & $45.3-48.0$ & $56.7 - 200.0$ & $8.0-9.0$ & $0.01-0.02$ & $ < 9.5$ & $ < 275$ & $4.75 - 5.18$ & $1.6-1.7$ & $33.6 - 47.5$ & SOL/CNO  & II-P*\\
		SN 2010mc\tnote{\dag} & $22.6-25.3$ &  $0.98-1.56 $ & $2.0-2.5$ & $0.07-0.1$ & $10.1-11.9$ & $300$ & $5.05-5.95$ & $1.7 - 2.0$ & $16.7-19.8$ &  CNO & II-n$^{[6]}$/L* \\
		PTF10uls & $59.9-64.0$  & $60.0- 87.5$ & $3.5-4.0$ & $0.3$ & $15.4-28.3$ & $300$ & $7.7-14.15$ & $2.6 - 4.7$ & $29.3 - 33.4$ & CNO & II-P$^{[7]}$ \\
		PTF11iqb\tnote{\dag} & $15.2-15.6$ & $2.9-3.6$ & $0.9-0.95$ & $0.3$ & $<2.0$ &$<100$ & $1.0-3.0$ & $0.3 -1.0$ & $28.2-29.9$ & CNO & II-P$^{[8]}$ \\
		PTF12gnn\tnote{\dag} & $61.6-67.9$ & $13.9-23.3$ & $3.7-4.5$ & $0.22$ & $<15.4$ & $<250$ & $4.98-9.24$ & $1.7 - 3.1$ & $19.9-23.8$ & CNO & II-P*\\
		PTF12krf & $139.5-153.5$ & $71.3-119.7$ & $19-23$ & $0.47$ & $< 42.0$ & $<200$ & $27.3-31.5$ & $9.1 - 10.5$ & $19.9-23.8$ & CNO & II-P/b$^{[7]}$ \\ 
		SN 2013cu & $64.0-71.6$ & $15.5-25.2$ & $4.0-5.0$ & $0.01$ & $5.7-6.7$ & $100$ & $8.55- 10.05$ & $2.8 - 3.3$ & $19.9 - 23.8$ & CNO/He & II-b$^{[9]}$ \\
		SN 2013fs & $25.3-26.8$ & $35.0-62.5$ & $2.5-2.8$ & $0.1$ & $4.0-4.3$ & $100$ & $6.0-6.45$ & $2.0-2.2$ & $39.8-47.5$ & SOL & II-P$^{[2]}$ \\
		SN 2013fr & $55.4-63.2$ & $5.9 - 9.4$ & $3.9-12$ & $0.33-0.36$ & $<108.9$  & $<845$ & $8.3-19.33$ & $2.8 - 6.5$ & $16.8-19.8$ & SOL/CNO & II-L $^{[10]}$ \\
		iPTF14bag & $160-189.3$ & $108.5 - 150.0$ & $35-100$ & $0.45$ & $<189.7$ & $<300$ & $43.2-94.85$ & $14.5 - 31.8$ & $19.9- 23.6$ & CNO & II$^{[5]}$ \\
		SN 2014G & $39.2$ & $22.5 - 37.5$ & $1.5$ & $0.14-0.2$ & $33.9-49.6$ & $500$ & $10.17-14.88$ & $3.4- 4.9$ & $29.3-33.5$ & CNO & II-L$^{[11]}$ \\
		SN 2016eso & $69.2-70.1$ & $3.7- 7.0$ & $4.68-4.8$ & $0.04-0.12$ & $<177.1$ & $<845$ & $9.73-31.44$ & $3.3 - 10.5$ & $14.1 - 16.6$ & SOL/CNO & II \\
		SN 2018cvk & $60.7-78.0$ & $1.14-5.4$ & $3.6-6.0$ & $0.05-0.2$ & $33.3-74.9$ & $500$ & $10.0-22.47$ & $3.3 -7.5$ & $9.9-16.6$ & SOL/CNO & II-n/L* \\ 
		SN 2018khh & $78.4$ & $60.0-150.0$ & $6.0$ & $0.02-0.06$ & $38.3-127.8$ & $500$ & $11.5-38.34$ & $3.8 - 12.8$ & $26.7 - 33.5$ & CNO & II-n/L* \\
		SN 2018zd & $78.4-84.7$ & $70.0-150.0$ & $6.0-7.0$ & $0.19$ & $<207.0$ & $<810$ & $21.37-38.33$ & $7.2 - 12.8$ & $26.6-33.4$ & SOL & II-P* \\
		\hline
		SN 1998S$^{[1]}$ & $60$  & $15 \pm 5$ & & $0.05$ & $6.0 \pm 1 $ & $ 40 \pm 5 $ & & $7.2-7.8$ & $\sim 19-23$ & CNO & II-n/L\\
		 & & & & & & & & & & ($X=0.49$) & \\
		SN 2013fs$^{[2]}$ & $13-14$ & $20 - 35$ & & $0.05$ & $2.0-4.0$ & $ 100$ & & $1.0 - 2.0$ & $48.0-58.0$ & SOL & II-P \\
		SN 2013cu$^{[3]}$ & $15.0$ & $10$ & & $A_V = 0.1$ & $3.0$ & $100$ & & $1.5$ & $\sim 39$ & CNO & II-b\\
		 & & & & & & & & & & ($X=0.46$) & \\
		SN 2016bkv$^{[4]}$ & $7.5-8.5$ & $0.55$ & & $0.01$ & $0.6$ & $150 - 300$ & & $0.1 - 0.2$ & $25.9$ & SOL & II-P\\
	\end{tabular}
	\begin{tablenotes}
        \item[\dag] The flux for these events has been scaled in the available files by an unknown factor.
        \item[*] The II-P vs II-L classification is arbitrarily based on a visual inspection of the light curve. These events have rather ambiguous LC properties, or short-lasting plateaus.
        \item[]  References: (1) \cite{shivvers15}, (2) \cite{yaron17}, (3) \cite{groh14a}, (4) \cite{deckers19} , (5) \cite{khazov16}, (6) \cite{ofek13b}, (7) \cite{rubin16}, (8) \cite{smith15}, (9) \cite{galyam14}, (10) \cite{bullivant18}, (11) \cite{terreran16},
    \end{tablenotes}
    \end{threeparttable}
\end{sidewaystable*}

\begin{figure*}
    \centering
    \includegraphics[width=0.73\textheight]{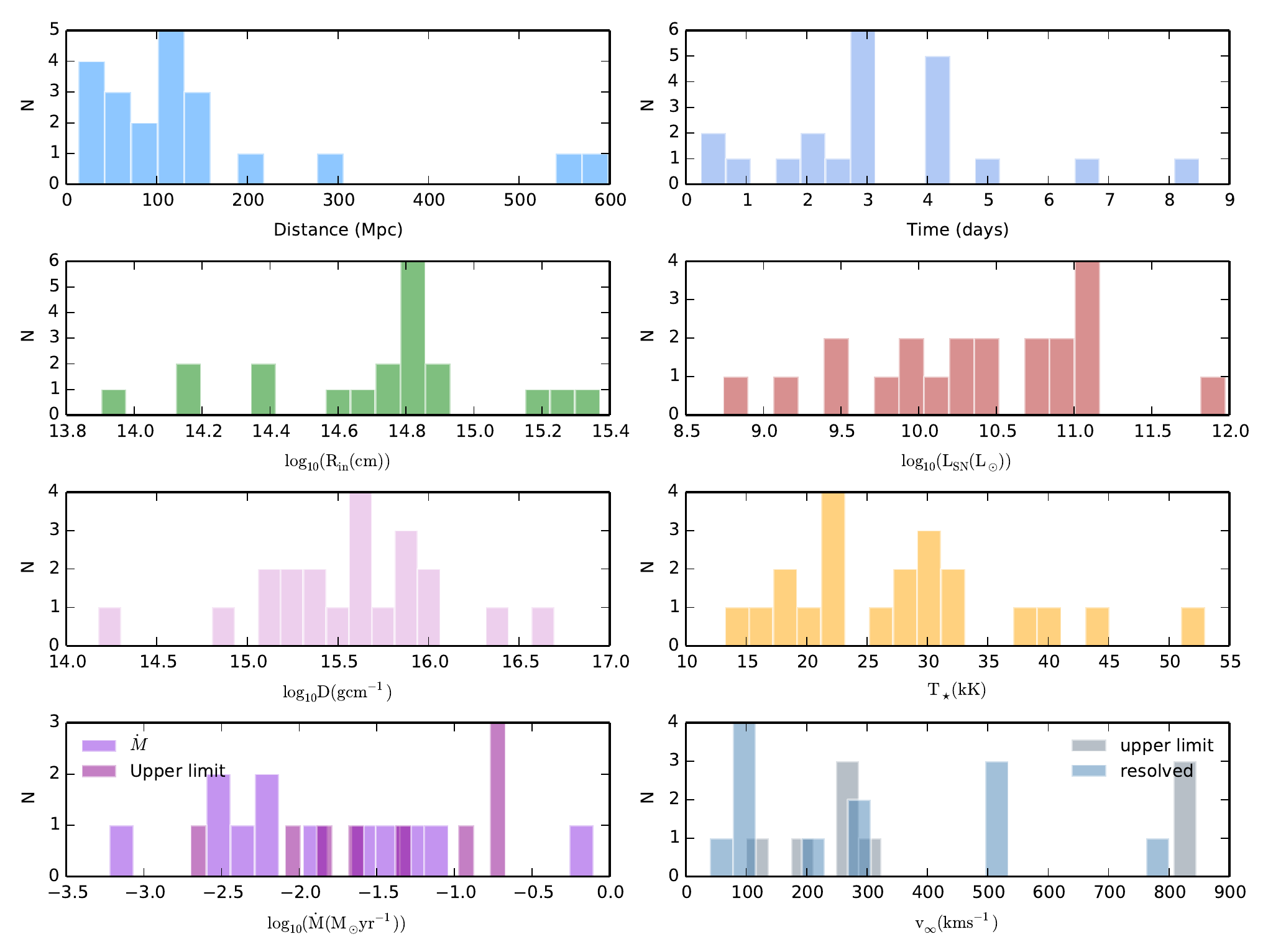}
    \caption{The distributions of the observed distances to the supernovae in our sample and of the post-explosion ages of the spectra analysed in this paper (top row), and the distributions of the following best-fit parameters: inner radii (\rin, left column, second row), luminosities (\lsn, right column, second row), density parameters ($D$, left column, third row), temperatures (\tstar, right column, third row), mass-loss rates (\mdot, left column, last row), and terminal wind velocities (\vinf, right column, last row). Note that these are based on the mean values, and that for some events due to the resolution of the spectra we could place only an upper limit to \vinf and therefore also to \mdot. }
    \label{fig:dist}
\end{figure*}

\begin{figure*}
    \centering
    \begin{subfigure}{0.49\textwidth}
        \includegraphics[width=1.0\textwidth]{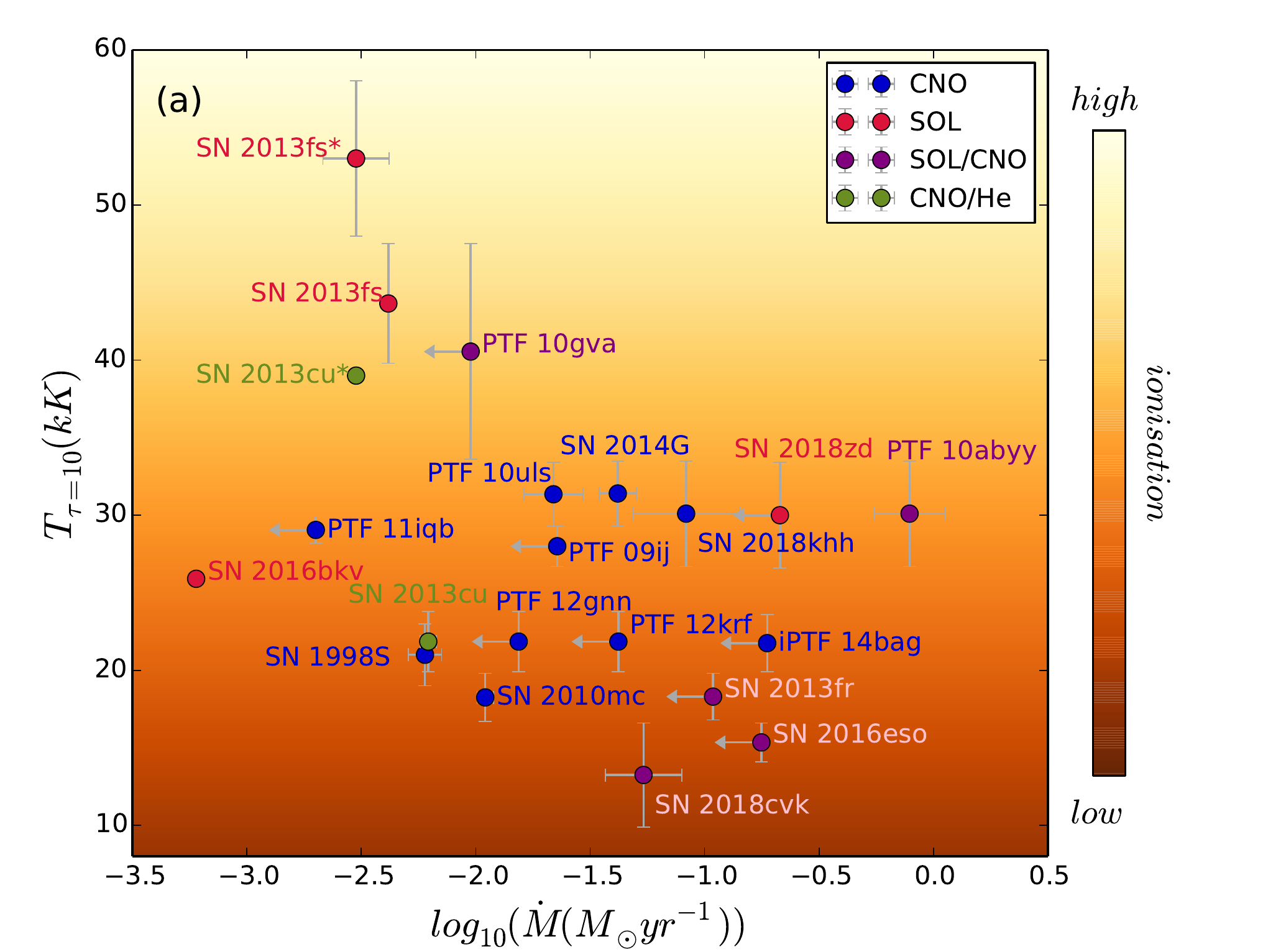}
    \end{subfigure}
    \begin{subfigure}{0.49\textwidth}
        \includegraphics[width=1.0\textwidth]{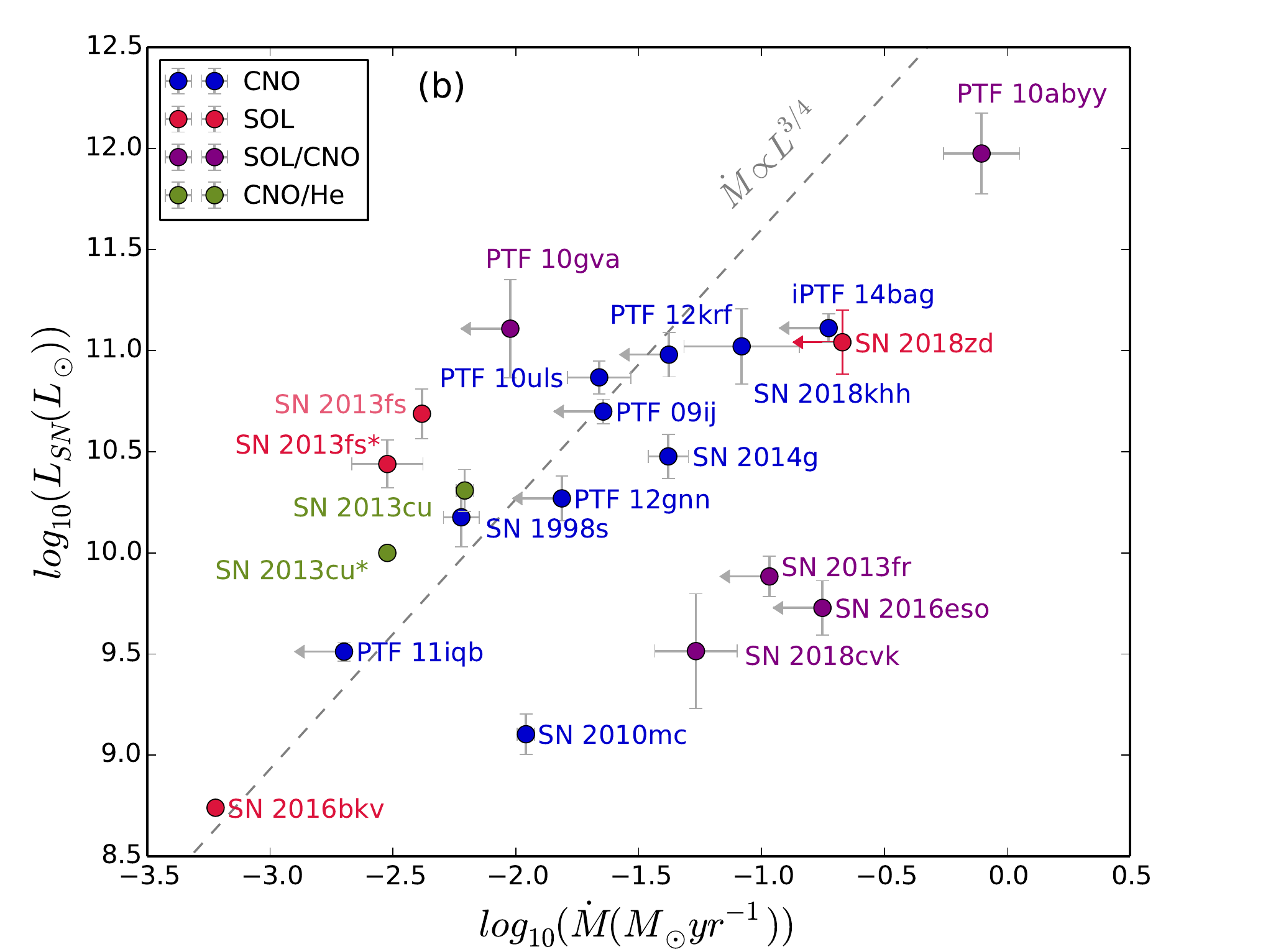}
    \end{subfigure}
    
    \begin{subfigure}{0.49\textwidth}
    \includegraphics[width=1.0\textwidth]{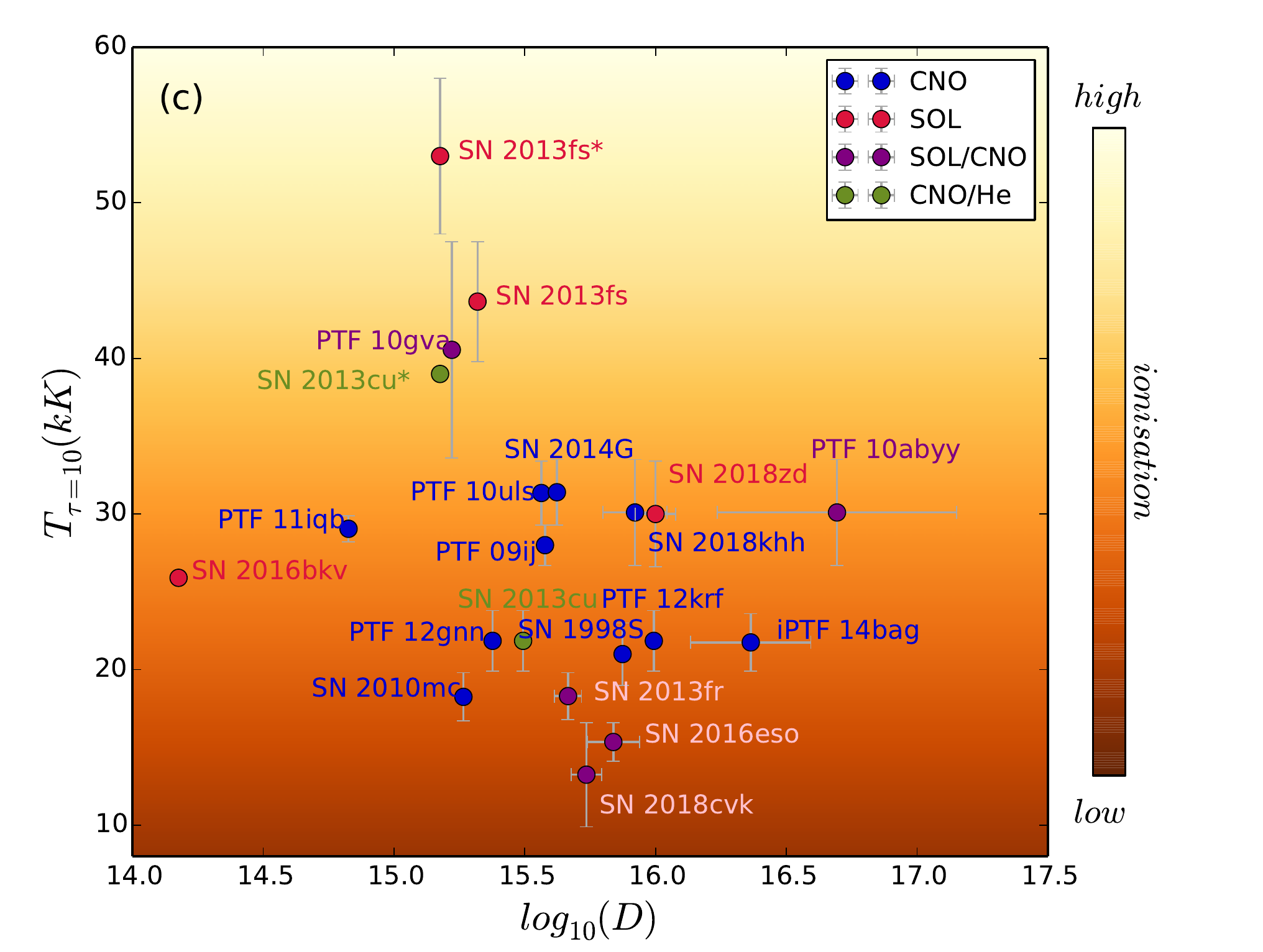}
    \end{subfigure}
    \begin{subfigure}{0.49\textwidth}
        \includegraphics[width=1.0\textwidth]{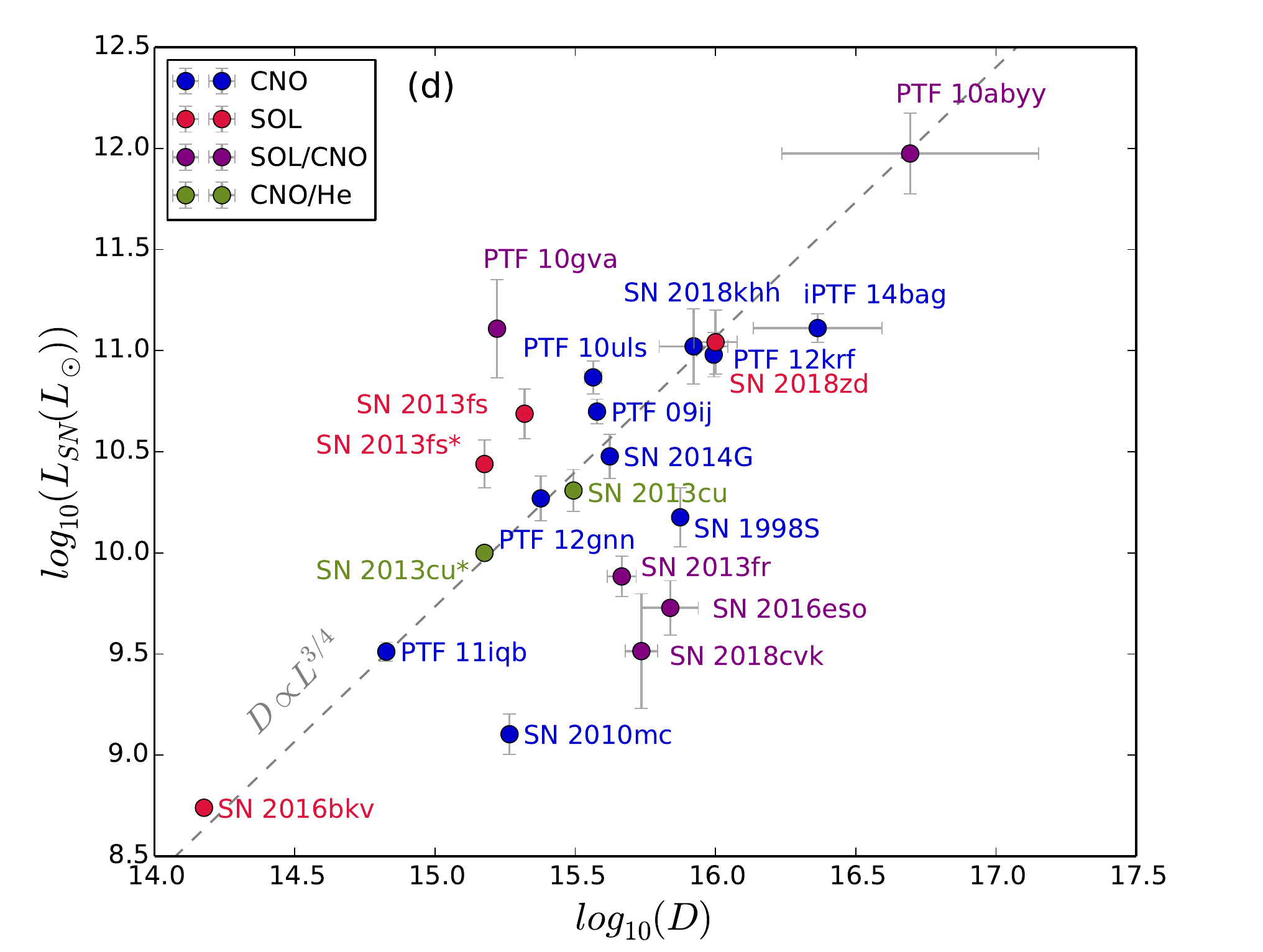}
    \end{subfigure}
    
    \caption{Temperature, luminosity, mass-loss rate, and density relations for our SN sample. The colour code refers to the surface abundances of the progenitor. In the right side panels we include for illustration purposes the expected scaling between mass-loss rate and luminosity (top), and density and luminosity respectively (bottom). All the values here are derived from CMFGEN modelling of the observed SNe. We have included four previously studied SNe, i.e. SN 2016bkv, SN 1998S, SN 2013cu at 1 day post-explosion (SN 2013cu*), and SN 2013fr at 6 h post-explosion (SN 2013fr*) as described in Sect. \ref{sect:emp_rel}. All the values can be found in Table \ref{tab:bf_param}.}
    \label{fig:TLMD}
\end{figure*}

\begin{figure}
   \centering
   \includegraphics[width=1.0\columnwidth]{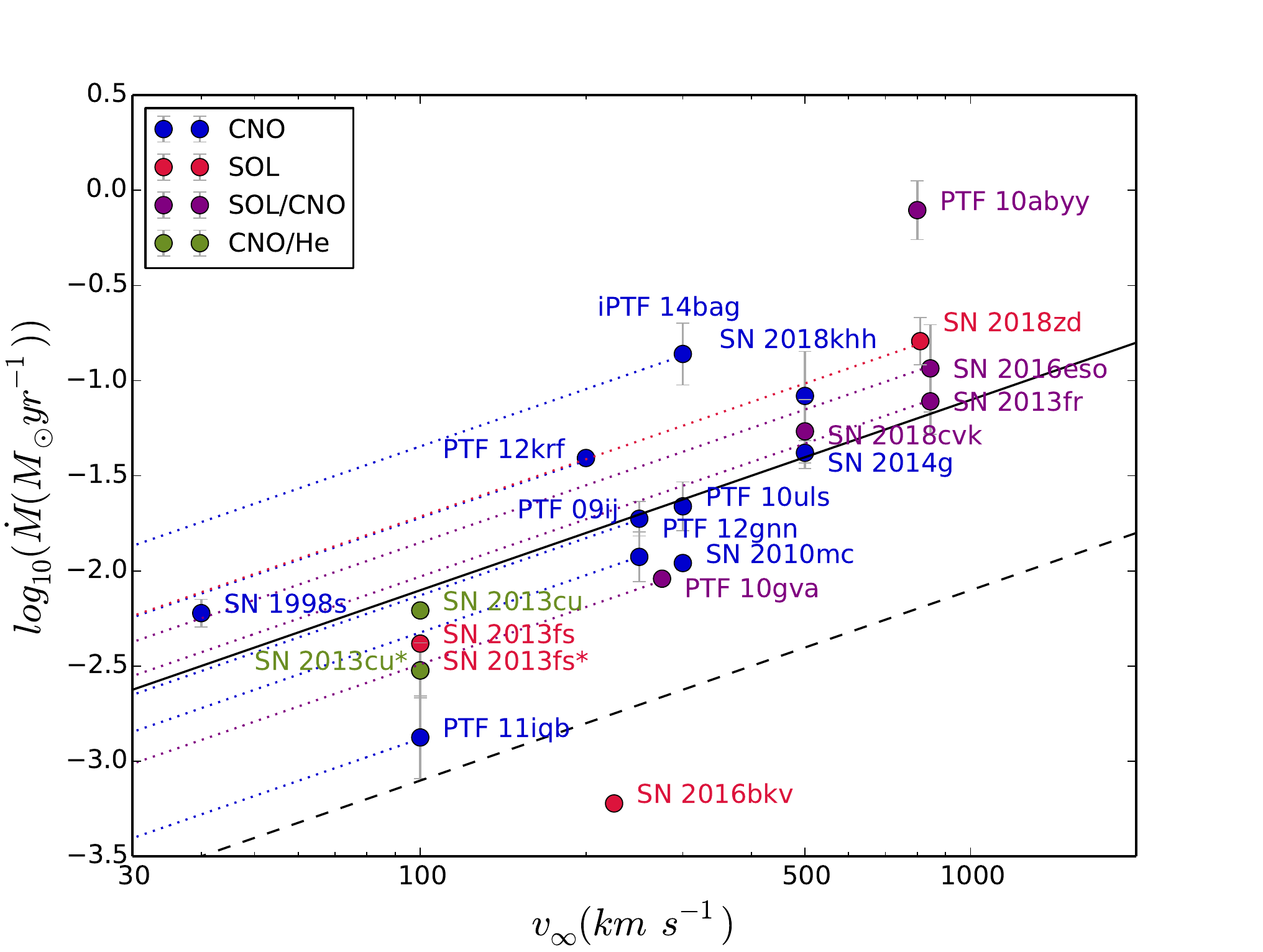}
    \caption{Mass-loss rate vs terminal wind velocity. The solid and dashed lines correspond to a wind density parameter of $5 \times 10^{16} \mathrm{g\, cm^{-1}}$ and $5 \times 10^{15} \mathrm{g\, cm^{-1}}$ respectively. The dotted lines represent the possible locations of the events for which we only have an upper limit on the wind velocity. Colour codes as in Figure 8.}
   \label{fig:Mdotvsv}
\end{figure}

The three spectra discussed above are representative of three distinct regimes, namely low-, medium-, and high-ionisation. However the different classes are not separated by a clear cut in the parameter space, but there is a rather continuous distribution of properties. We have compiled our results in Figures \ref{fig:dist}, \ref{fig:TLMD}, and \ref{fig:Mdotvsv}, which show the distributions of the observed distances to the events, the ages of the spectra, and the modelled inner radii, luminosities, density parameters, temperatures, mass-loss rates and terminal wind velocities, and we discuss them below. Overall our results show, as Fig. \ref{fig:ew} previously implied, that the SNe in our sample span a large portion of the interacting SNe parameter space, despite the limited number of events. The inner radii derived from spectral modelling range from $7 \times 10^{13}$ cm to $2.5 \times 10^{15}$ cm, with most events having $5 \times 10^{14} < \rin < 10^{15}$ cm, most likely mirroring the spectral ages distribution which peaks around 3 to 4 days (Fig. \ref{fig:dist}). Our sample covers evenly an extended range of luminosities, from a few $10^{8} ~\lsun$ to $10^{12} ~\lsun$ (Figs. \ref{fig:dist} and \ref{fig:TLMD}). In terms of temperature, the SNe go from as cool as $9\, 900$ K (corresponding to the low-ionisation cases)  to as hot as $58\, 000$  K (highly ionised spectra), but unlike the luminosities, which are more evenly spread, most of the T are in the $20\, 000$ to $35\,000$ range, with most events exhibiting a medium-ionisation spectrum(Figs. \ref{fig:dist} \& \ref{fig:TLMD}). We have included the density parameter, D, defined similarly to \cite{chevalier11} as $D = \frac{\mdot}{4 \pi \vinf}$, because for a significant portion of the events in our sample, we were only able to obtain an upper limit for the wind velocity, and hence only an upper limit for the mass-loss rate. Most events fall around the $5 \times 10^{16} g~cm^{-1}$ value (Figs. \ref{fig:dist} and \ref{fig:Mdotvsv}), which is considered a typical density for type IIn SNe \citep{chevalier11,smith17hsn}. The universally high densities in our sample suggest that the main difference between flash ionisation events and longer lived type IIn SNe is given by the extension of the CSM rather than its density. The mass-loss rates are also spread evenly, from $6 \times 10^{-4} ~\msunyr$ to $\simeq 1.0 ~\msunyr$. The wind velocities are mostly around $100 - 300 ~\kms$, but also reach values as high as $800 ~\kms$ (Fig. \ref{fig:dist}). 

Considering that the strength of the \ion{H}{$\beta$} line and the ratio of the \ion{He}{ii} to \ion{He}{i} line were chosen as proxies for the mass-loss rate and the temperature, Figs. \ref{fig:TLMD}a and c should qualitatively mimic Fig. \ref{fig:ew}a. Overall this holds, with some discrepancies being introduced by the low resolution of the observations (for many events only an upper limit for the mass-loss rate could be obtained), by the lack of certain lines in the spectrum (in Fig. \ref{fig:ew}a many events have only upper limits for the equivalent widths), by the insufficient sampling of the models, or by differences in surface abundances. For example, SN 2016bkv has a higher T than its \ion{He}{ii} to \ion{He}{i} ratio would suggest. This is a result of a slight mismatch between the \ion{He}{i} strength in the best-fit model \citep{deckers19} and that measured in the observations. SN 2013fr also seems to have a lower T than its W values would indicate, but it's placement at the 0 value in Fig. \ref{fig:ew}a is due to the noise limit being adopted for both of the \ion{He}{} lines used, since its spectrum does not show any clear emission line other than \ion{H}{$\alpha$}. Figs. \ref{fig:TLMD} a and c also point to the expected trend that higher wind densities lead to lower T. This is of course degenerate with \rin\ or the age of the spectrum.   

In the case of interacting SNe, a large percentage of the luminosity is provided by converting the SN ejecta kinetic energy into radiation due to the deceleration in the CSM. Therefore the CSM density is proportional to the observed luminosity, described by the $\mdot \propto \lsn^{3/4}$ relation. Our results show that this relation holds both in the \mdot-L space (Fig. \ref{fig:TLMD}b) and the D-L space (Fig. \ref{fig:TLMD}d), with some scatter given by our uncertainties and slightly different post-explosion times.

In terms of surface abundances, the majority of our events are fit well by the assumption that the progenitor had CNO-processed surface abundances, which is to be expected for evolved massive stars. A small number of events (15.7 \%) show solar-like surface abundances, with another 26.3 \% having either solar-like or CNO-processed abundances. According to our modelling, the only event that could have a CSM significantly depleted in \ion{H}{} is SN 2013cu, which is a reasonable result since this event has been classified as a type IIb SN. Our models show no clear correlation between the surface abundances and other SN or progenitor properties (Fig. \ref{fig:TLMD}b). This has broader implications on massive star evolution which will be discussed in detail in Sect. \ref{sect:disc}. 

\begin{figure}
    \centering
    \includegraphics[width=1.0\columnwidth]{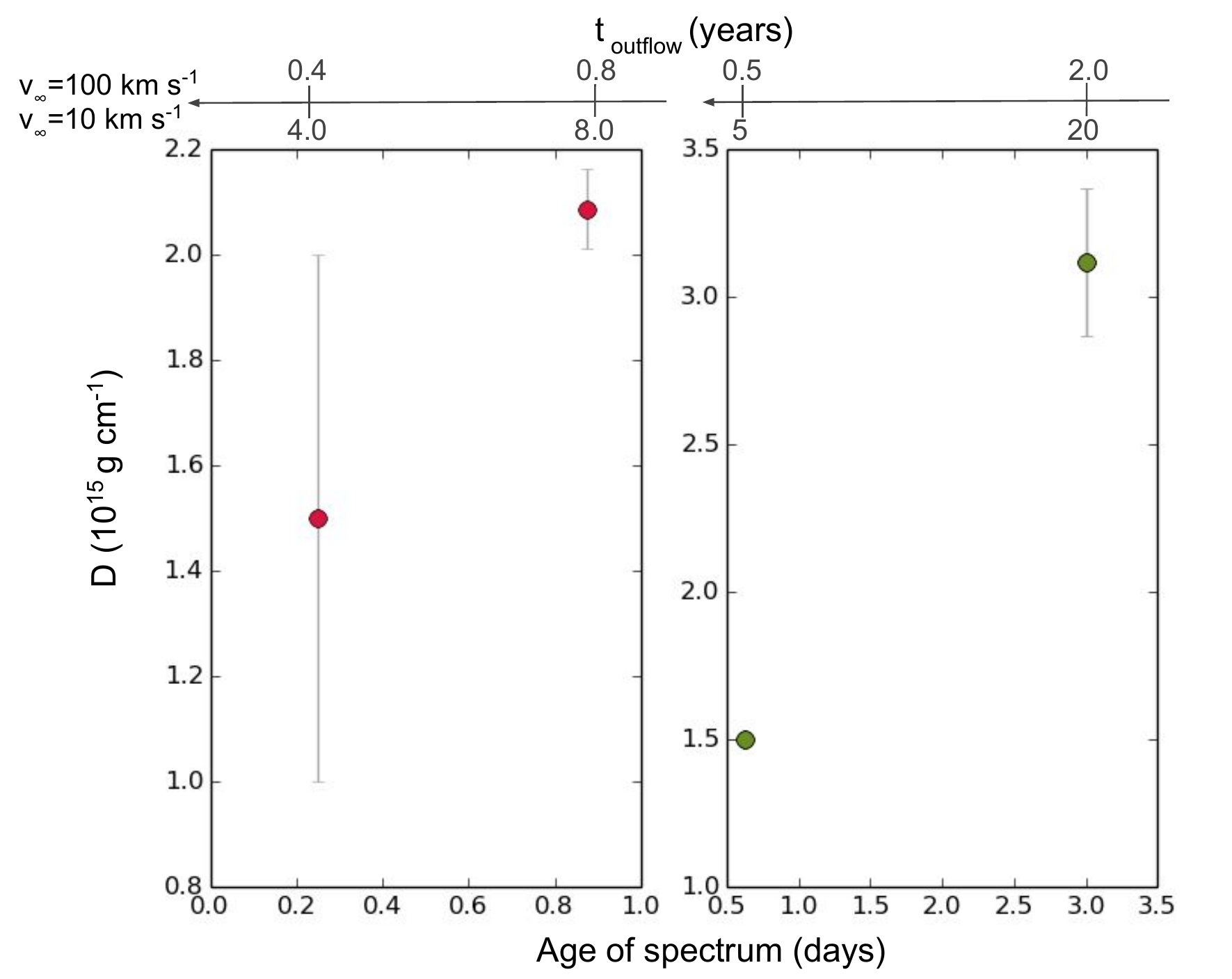}
    \caption{Temporal evolution of the wind density parameter with time for SN 2013fs (left)  and SN 2013cu (right). The bottom axis is the post-explosion time at which the spectrum was obtained, while the top axes represent the times at which the material would have been ejected, with respect to the explosion epoch, assuming the two limits of the wind velocity.}
    \label{fig:tempevol}
\end{figure}

We can also infer how these events evolve over the first few days by comparing the pre-existing results for the very early spectra of SN 2013fs and SN 2013cu with our properties obtained from fitting slightly older spectra (15 h later for 2013fs, and 2 days later for 2013cu). We can see the events evolve rapidly, decreasing in T, due to the expanding radius, but increasing in \lsn\ since all the spectra were taken pre-photometric-peak in both cases. Surprisingly, the density of the CSM increases slightly with time in both cases as well (Fig. \ref{fig:tempevol}). Assuming $\vinf =100 ~\kms$, the material seen in the two observations would have been ejected 0.5 years and 2 years prior to collapse for 2013cu, and 0.4 years and 0.8 respectively, for SN 2013fs. Depending on the mass of the star the two timescales might correspond to different burning stages. The timescales calculated here are reminiscent of the envelope instabilities induced via gravity waves by the turbulent Ne and O burning, as derived by \cite{fuller17}. 

There are also a number of events that stand out from our analysis. The low-ionisation events (SN 2010mc, SN 2018cvk, SN 2016eso) seem to be isolated from the other events, clustering in the high \mdot\, low-to-medium \lsn\ corner of Fig. \ref{fig:TLMD}b. This shows that their low temperatures are not a result of very low luminosities, but rather of their high mass-loss rates. They may be part of the classical type IIn class, and they may show interaction for a longer period of time than their medium and high ionisation counterparts. Follow-up spectra of these events should be obtained in order to test this hypothesis. They also have in common a linearly declining LC. 

PTF 10abyy is a very unusual event. First of all its spectrum is incomplete, missing the red side containing the \ion{He}{i} line, which could have had an effect on our results. However, the available part of the spectrum is very similar to other medium-to-high ionisation events, as can be seen from the T determination in Fig. \ref{fig:TLMD}a as well. Not only does PTF 10abyy seem to have a much higher \lsn\ and \mdot\ than the other events, but it does so while having one of the oldest spectra in our sample, i.e. at a similar age to the other events it could have been even more extreme. The main reason for the unusually high \lsn\ is the scaling factor that had to be applied in order to match the absolute flux. Therefore we theorise that the distance quoted in literature might be inaccurate. PTF 10abyy also has the highest \vinf\ in our sample, of $800 ~\kms$. Even more puzzling are the properties that point to the progenitor of PTF 10abyy as having been a RSG star, i.e. its IIP-like LC, its extended radius \citep{rubin16}, and its possible solar surface abundances.

For PTF 10gva, there is a discrepancy between our derived \tstar , which is in the $33\,600$ to $47\,500$ (or $\teff \simeq 30\,000 - 40\,000$) and the $20\,000$ K temperature obtained by \cite{chenko10} using a black-body fit to UV measurements obtained one day after the spectrum, which may be explained by the fast T evolution at early times post-explosion.

Due to clumps in the CSM or asymmetries, some interacting supernovae may exhibit intermediate components in their emission lines in addition to the narrow components, however we see no evidence of this in the spectra in our sample.

\section{Connecting supernovae to their progenitors}
\label{sect:disc}

Since the SN properties depend on those of the progenitor, connecting the two stages sheds light on the evolution of massive stars. In this section, we compare the SN types of the events in our sample to the pre-explosion properties derived from our analysis.

\begin{figure}
    \centering
    \includegraphics[width=1.0\columnwidth]{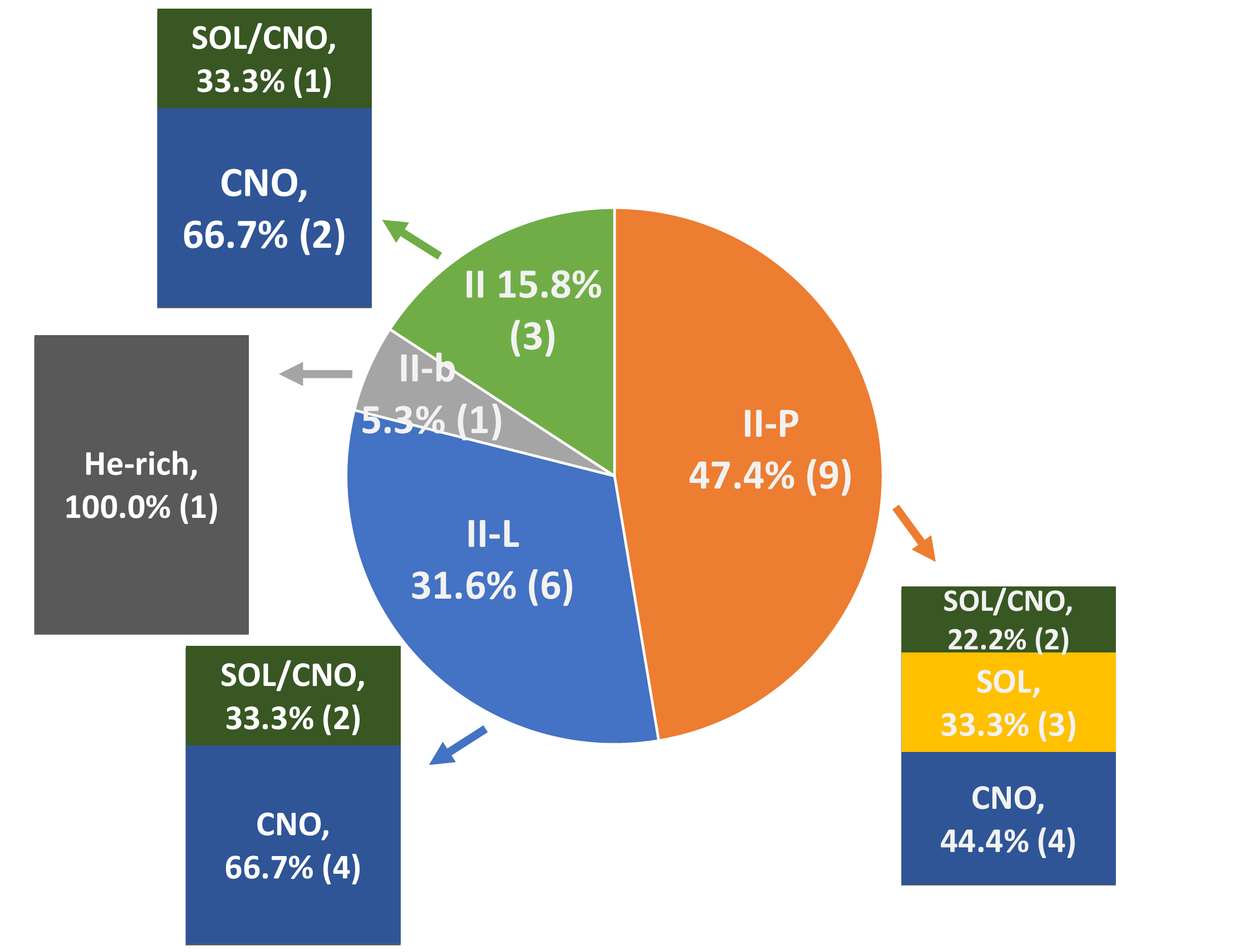}
    \caption{Relative rates of SN types in this sample and the surface abundances of their progenitors derived from modelling their post-explosion spectra. The absolute number of events for each type is included in brackets.}
    \label{fig:piechart}
\end{figure}

Our sample is dominated by type II-P SNe (47.4\%), followed by type II-Ls (31.6 \%), and 1 type IIb. 15.8\% of our events did not have enough photometric points for a LC classification, and are generically classified as type II (Fig. \ref{fig:piechart}). These rates qualitatively match previously observed and modelled rates of type II SNe \citep{li11,smith11sn}, with the exception that when considering all type II SNe the type II-b SNe are more common than the type II-L. This could have several causes, as for example, some of the events in our sample could evolve into type II-b SNe but there are no publicly available spectra to confirm this, we have selected a sample of SNe that show relatively strong \ion{H}{} lines thus disfavouring type II-b SNe in our sample, and CSM interaction can produce faster-declining lightcurves \citep{hillier19}, leading to more interacting events being classified as II-L SNe.

The mass-loss history of a SN progenitor is expected to be connected to the SN class. Depending on the mass of the H envelope available at the time of the explosion, SNe range from type II-P SNe, where most of the envelope is retained, to type II-b, which have very little H left, all the way to type I core-collapse SNe, whose progenitors have lost all of their H envelope. Modelling the spectra of interacting supernovae can place good constraints on \mdot\ and \vinf\ at the very late evolutionary stages, but linking these values to a progenitor type would not give an accurate representation of the SN-progenitor connections. Comparing our results of the mass-loss rates and wind velocities to those of typical evolved massive stars it is obvious that most of the interacting SN progenitors have much stronger winds, clearly supporting the scenario of strongly enhanced mass-loss at the pre-SN stage. 

The surface abundances can be used to link SNe to their progenitors with more accuracy. Not only does the \ion{H}{} envelope mass vary between the different SN types, but the amount of CNO-processed material at the surface of a massive star is expected to increase with ZAMS, if the star followed single star evolution \citep{ekstrom12}. While type II-P supernovae are the most common in our sample, the progenitors prefer CNO-processed surface abundances. If all type II-P SNe originate from RSGs, and only the more massive RSGs ($\gtrapprox 15 \msun$) exhibit CNO-processed material at the surface \citep{gmg13}, then our sample shows shows a clear preference for more massive progenitors. Our results also show that all the events that definitely had solar-like surface abundances are type II-P SNe, which follows the RSG - type II-P scenario, but a number of events for which the CNO fractions could not be constrained are also present in type II-L and type II SNe. The only SN with a significantly low \ion{H}{} abundances is SN 2013cu which exploded as type II-b SNe, confirming the aforementioned scenario. These results should be interpreted carefully, keeping in mind the impacts of binarity, uncertainties in the modelling, and uncertainties in the observed masses of SN progenitors. For example, $1/3$ to $1/2$ of type II SN progenitors (including type II-P) have interacted with a binary companion \citep{zapartas19}. Binary interaction can lead either to mass exchange or mergers, thus modifying the observed surface abundances of the resulting SN progenitors and the mapping to their initial masses. Additionally the mixing of chemical elements in post-main sequence stars may not be fully captured in the stellar evolution models. \cite{farrell20} have also suggested that, due to the weak dependence of RSG luminosities on the envelope mass, the masses of the observed type II-P progenitors have much larger errors than previously estimated.

The mass-loss rates however do not seem to correlate to the surface abundances, and hence to the stellar mass in the single star scenario, as for example, SN 2018zd is a type II-P SN with solar-like surface abundances, but is placed in the higher end of the \mdot\ or D parameter space (Figs. \ref{fig:TLMD} b and d). SN 2013cu, which has a \ion{H}{}-depleted CSM and therefore is expected to have had high \mdot, exhibits lower \mdot\ or similar \mdot\ to many \ion{H}{}-rich events. This may mean that the mechanism responsible for the enhanced mass-loss at the pre-explosion stage does not depend on the stellar mass. 

We recognise the limitations of our results due to the small sample of events and uncertain classification of the lightcurves and we are hopeful that a similar analysis of the upcoming larger number of early-time interacting supernovae will further improve our understanding of the final stages of massive star evolution and the SN-progenitor connections. Better photometric coverage and higher resolution spectroscopy would also be beneficial in reducing uncertainties, such as the LC classification and in particular the derived mass-loss rates, respectively.

\section{Conclusions and Summary}
\label{sect:summary}

In this paper we have compiled a set of 17 observed SNe showing interaction with a CSM at early-times and using a library of synthetic spectra computed with the radiative transfer code, CMFGEN, we have constrained the progenitor and explosion properties of these events. We summarise our main findings below.

\begin{enumerate}
    \item We devised empirical relations based on the relative strengths of the \ion{H}{} and \ion{He}{} lines, that allow us to produce a phase diagram of interacting SNe in order to classify these events. 
    
    \item Due to the diversity of explosion and CSM properties, interacting SNe span a wide range of properties. The sample of events analysed in this paper cover luminosities from $10^{8}$ to $10^{12} ~\lsun$, mass-loss rates from a few $10^{-4}$ to $1 ~\msunyr$, wind velocities from less than $ 100 ~\kms$ up to $800 ~\kms$, temperatures from $10\,000$ to nearly $60\,000$ K, and solar-like, CNO-processed, and \ion{H}{}-depleted surface abundances.
    
    \item The relative strengths of certain emission lines, e.g. \ion{H}{} and \ion{He}{} can be successfully employed in estimating SN temperatures and CSM densities. 
    
    \item The wind densities derived from our modelling support the recent hypothesis that many SN progenitors will have significantly increased mass-loss rates right before explosion. Additionally there seems to be a lack of correlation between the pre-explosion mass-loss rates and the mass of the progenitor if it evolved as a single star.
    
    \item We find that the majority of these events have CNO-processed surface abundances. Considering that most of the SNe are type II-P, this points to a preference towards high-mass RSGs ($\gtrapprox 15 \msun$) when it comes to SNe that interact with CSM in the single star scenario. The mapping between surface abundances and initial masses can be modified by binary interaction, which could affect the conclusion above.
    
\end{enumerate}

This work showcases the importance of early-time spectroscopic observations of core-collapse supernovae in studying the late-time evolution of massive stars and their explosive deaths, and supports the continued efforts of current and future time-domain surveys such as the ZTF, ASAS-SN, and LSST.

\section*{Acknowledgements}

I.B. acknowledges funding from a Trinity College Postgraduate Award through the School of Physics, and J.H.G. acknowledges support from the Irish Research Council New Foundations Award 206086.14414 "Physics of Supernovae and Stars".




\bibliographystyle{mnras}
\bibliography{refs} 

\begin{thebibliography}{}
\makeatletter
\relax
\def\mn@urlcharsother{\let\do\@makeother \do\$\do\&\do\#\do\^\do\_\do\%\do\~}
\def\mn@doi{\begingroup\mn@urlcharsother \@ifnextchar [ {\mn@doi@}
  {\mn@doi@[]}}
\def\mn@doi@[#1]#2{\def\@tempa{#1}\ifx\@tempa\@empty \href
  {http://dx.doi.org/#2} {doi:#2}\else \href {http://dx.doi.org/#2} {#1}\fi
  \endgroup}
\def\mn@eprint#1#2{\mn@eprint@#1:#2::\@nil}
\def\mn@eprint@arXiv#1{\href {http://arxiv.org/abs/#1} {{\tt arXiv:#1}}}
\def\mn@eprint@dblp#1{\href {http://dblp.uni-trier.de/rec/bibtex/#1.xml}
  {dblp:#1}}
\def\mn@eprint@#1:#2:#3:#4\@nil{\def\@tempa {#1}\def\@tempb {#2}\def\@tempc
  {#3}\ifx \@tempc \@empty \let \@tempc \@tempb \let \@tempb \@tempa \fi \ifx
  \@tempb \@empty \def\@tempb {arXiv}\fi \@ifundefined
  {mn@eprint@\@tempb}{\@tempb:\@tempc}{\expandafter \expandafter \csname
  mn@eprint@\@tempb\endcsname \expandafter{\@tempc}}}

\bibitem[\protect\citeauthoryear{{ALFOSC User Manual}}{{ALFOSC User
  Manual}}{2018}]{ALFOSC}
{ALFOSC User Manual} 2018, \url {http://www.not.iac.es/instruments/alfosc/}

\bibitem[\protect\citeauthoryear{{Arnett} \& {Meakin}}{{Arnett} \&
  {Meakin}}{2011}]{arnett11}
{Arnett} W.~D.,  {Meakin} C.,  2011, \mn@doi [\apj]
  {10.1088/0004-637X/741/1/33}, \href
  {https://ui.adsabs.harvard.edu/abs/2011ApJ...741...33A} {741, 33}

\bibitem[\protect\citeauthoryear{{Bellm} et~al.,}{{Bellm}
  et~al.}{2019}]{bellm19}
{Bellm} E.~C.,  et~al., 2019, \mn@doi [\pasp] {10.1088/1538-3873/aaecbe}, \href
  {https://ui.adsabs.harvard.edu/abs/2019PASP..131a8002B} {131, 018002}

\bibitem[\protect\citeauthoryear{{Boian} \& {Groh}}{{Boian} \&
  {Groh}}{2018}]{boian18}
{Boian} I.,  {Groh} J.~H.,  2018, \mn@doi [\aap] {10.1051/0004-6361/201731794},
  \href {https://ui.adsabs.harvard.edu/abs/2018A&A...617A.115B} {617, A115}

\bibitem[\protect\citeauthoryear{{Boian} \& {Groh}}{{Boian} \&
  {Groh}}{2019}]{boian19}
{Boian} I.,  {Groh} J.~H.,  2019, \mn@doi [\aap] {10.1051/0004-6361/201833779},
  \href {http://adsabs.harvard.edu/abs/2019A%26A...621A.109B} {621, A109}

\bibitem[\protect\citeauthoryear{{Bose}, {Kumar}, {Misra}, {Matsumoto},
  {Kumar}, {Singh}, {Fukushima}  \& {Kawabata}}{{Bose} et~al.}{2016}]{bose16}
{Bose} S.,  {Kumar} B.,  {Misra} K.,  {Matsumoto} K.,  {Kumar} B.,  {Singh} M.,
   {Fukushima} D.,   {Kawabata} M.,  2016, \mn@doi [\mnras]
  {10.1093/mnras/stv2351}, \href
  {http://adsabs.harvard.edu/abs/2016MNRAS.455.2712B} {455, 2712}

\bibitem[\protect\citeauthoryear{{Brimacombe} \& {Stanek}}{{Brimacombe} \&
  {Stanek}}{2018}]{brimacombe18b}
{Brimacombe} J.,  {Stanek} K.~Z.,  2018, Transient Name Server Discovery
  Report, \href {https://ui.adsabs.harvard.edu/abs/2018TNSTR1954....1B}
  {2018-1954, 1}

\bibitem[\protect\citeauthoryear{{Brimacombe} et~al.,}{{Brimacombe}
  et~al.}{2016}]{brimacombe16}
{Brimacombe} J.,  et~al., 2016, The Astronomer's Telegram, \href
  {http://adsabs.harvard.edu/abs/2016ATel.9344....1B} {9344}

\bibitem[\protect\citeauthoryear{{Brimacombe} et~al.,}{{Brimacombe}
  et~al.}{2018}]{brimacombe18}
{Brimacombe} J.,  et~al., 2018, The Astronomer's Telegram, \href
  {https://ui.adsabs.harvard.edu/abs/2018ATel12031....1B} {12031}

\bibitem[\protect\citeauthoryear{{Bullivant} et~al.,}{{Bullivant}
  et~al.}{2018}]{bullivant18}
{Bullivant} C.,  et~al., 2018, \mn@doi [\mnras] {10.1093/mnras/sty045}, \href
  {http://adsabs.harvard.edu/abs/2018MNRAS.tmp...35B} {}

\bibitem[\protect\citeauthoryear{{Cenko} et~al.,}{{Cenko}
  et~al.}{2010}]{chenko10}
{Cenko} S.~B.,  et~al., 2010, The Astronomer's Telegram, \href
  {http://adsabs.harvard.edu/abs/2010ATel.2606....1C} {2606}

\bibitem[\protect\citeauthoryear{{Chevalier} \& {Fransson}}{{Chevalier} \&
  {Fransson}}{1994}]{chevalier94}
{Chevalier} R.~A.,  {Fransson} C.,  1994, \mn@doi [\apj] {10.1086/173557},
  \href {http://adsabs.harvard.edu/abs/1994ApJ...420..268C} {420, 268}

\bibitem[\protect\citeauthoryear{{Chevalier} \& {Irwin}}{{Chevalier} \&
  {Irwin}}{2011}]{chevalier11}
{Chevalier} R.~A.,  {Irwin} C.~M.,  2011, \mn@doi [\apjl]
  {10.1088/2041-8205/729/1/L6}, \href
  {http://adsabs.harvard.edu/abs/2011ApJ...729L...6C} {729, L6}

\bibitem[\protect\citeauthoryear{{Chugai}}{{Chugai}}{2001}]{chugai01}
{Chugai} N.~N.,  2001, \mn@doi [\mnras] {10.1111/j.1365-2966.2001.04717.x},
  \href {http://adsabs.harvard.edu/abs/2001MNRAS.326.1448C} {326, 1448}

\bibitem[\protect\citeauthoryear{{Clemens}, {Crain}  \& {Anderson}}{{Clemens}
  et~al.}{2004}]{clemens04}
{Clemens} J.~C.,  {Crain} J.~A.,   {Anderson} R.,  2004, in {Moorwood}
  A.~F.~M.,  {Iye} M.,  eds,  \procspie Vol. 5492, Ground-based Instrumentation
  for Astronomy. pp 331--340, \mn@doi{10.1117/12.550069}

\bibitem[\protect\citeauthoryear{{Crowther}}{{Crowther}}{2007}]{crowther07}
{Crowther} P.~A.,  2007, \mn@doi [\araa]
  {10.1146/annurev.astro.45.051806.110615}, \href
  {http://adsabs.harvard.edu/abs/2007ARA%26A..45..177C} {45, 177}

\bibitem[\protect\citeauthoryear{{DIS User Manual}}{{DIS User
  Manual}}{1995}]{DIS}
{DIS User Manual} 1995, {DIS: The Double Imaging Spectrograph}, \url
  {https://www.astro.princeton.edu/~rhl/dis.html}

\bibitem[\protect\citeauthoryear{{Davies} \& {Dessart}}{{Davies} \&
  {Dessart}}{2019}]{davies19}
{Davies} B.,  {Dessart} L.,  2019, \mn@doi [\mnras] {10.1093/mnras/sty3138},
  \href {https://ui.adsabs.harvard.edu/abs/2019MNRAS.483..887D} {483, 887}

\bibitem[\protect\citeauthoryear{{De Veny}}{{De Veny}}{1992}]{mayallRC}
{De Veny} J.,  1992, {The R.C. spectrograph for the Mayall 4-meter telescope:
  instrument reference manual}, \url
  {https://www.noao.edu/kpno/KPManuals/oldrcsp.pdf}

\bibitem[\protect\citeauthoryear{Deckers, Groh  \& Boian}{Deckers
  et~al.}{2019}]{deckers19}
Deckers M.,  Groh J.~H.,   Boian I.,  2019, The origins of low-luminosity
  supernovae: the case of SN2016bkv, in prep.

\bibitem[\protect\citeauthoryear{{Dessart}, {Hillier}, {Gezari}, {Basa}  \&
  {Matheson}}{{Dessart} et~al.}{2009}]{dessart09}
{Dessart} L.,  {Hillier} D.~J.,  {Gezari} S.,  {Basa} S.,   {Matheson} T.,
  2009, \mn@doi [\mnras] {10.1111/j.1365-2966.2008.14042.x}, \href
  {http://adsabs.harvard.edu/abs/2009MNRAS.394...21D} {394, 21}

\bibitem[\protect\citeauthoryear{{Dessart}, {Audit}  \& {Hillier}}{{Dessart}
  et~al.}{2015}]{dessart15}
{Dessart} L.,  {Audit} E.,   {Hillier} D.~J.,  2015, \mn@doi [\mnras]
  {10.1093/mnras/stv609}, \href
  {http://adsabs.harvard.edu/abs/2015MNRAS.449.4304D} {449, 4304}

\bibitem[\protect\citeauthoryear{{Dessart}, {John Hillier}, {Yoon}, {Waldman}
  \& {Livne}}{{Dessart} et~al.}{2017}]{dessart17}
{Dessart} L.,  {John Hillier} D.,  {Yoon} S.-C.,  {Waldman} R.,   {Livne} E.,
  2017, \mn@doi [\aap] {10.1051/0004-6361/201730873}, \href
  {http://adsabs.harvard.edu/abs/2017A%26A...603A..51D} {603, A51}

\bibitem[\protect\citeauthoryear{{Ekstr{\"o}m} et~al.,}{{Ekstr{\"o}m}
  et~al.}{2012}]{ekstrom12}
{Ekstr{\"o}m} S.,  et~al., 2012, \mn@doi [\aap] {10.1051/0004-6361/201117751},
  \href {http://adsabs.harvard.edu/abs/2012A%26A...537A.146E} {537, A146}

\bibitem[\protect\citeauthoryear{{Farrell}, {Groh}, {Meynet}  \&
  {Eldridge}}{{Farrell} et~al.}{2020}]{farrell20}
{Farrell} E.,  {Groh} J.,  {Meynet} G.,   {Eldridge} J.,  2020, arXiv e-prints,
  \href {https://ui.adsabs.harvard.edu/abs/2020arXiv200108711F} {p.
  arXiv:2001.08711}

\bibitem[\protect\citeauthoryear{{Fitzpatrick}}{{Fitzpatrick}}{1999}]{fitzpatrick99}
{Fitzpatrick} E.~L.,  1999, \pasp, \href
  {http://adsabs.harvard.edu/cgi-bin/nph-bib_query?bibcode=1999PASP..111...63F&db_key=AST}
  {111, 63}

\bibitem[\protect\citeauthoryear{{Foley}, {Smith}, {Ganeshalingam}, {Li},
  {Chornock}  \& {Filippenko}}{{Foley} et~al.}{2007}]{foley07}
{Foley} R.~J.,  {Smith} N.,  {Ganeshalingam} M.,  {Li} W.,  {Chornock} R.,
  {Filippenko} A.~V.,  2007, \mn@doi [\apjl] {10.1086/513145}, \href
  {https://ui.adsabs.harvard.edu/abs/2007ApJ...657L.105F} {657, L105}

\bibitem[\protect\citeauthoryear{{F{\"o}rster} et~al.,}{{F{\"o}rster}
  et~al.}{2019}]{forster19}
{F{\"o}rster} F.,  et~al., 2019, \mn@doi [Nature Astronomy]
  {10.1038/s41550-018-0641-7}, \href
  {https://ui.adsabs.harvard.edu/abs/2019NatAs...3..107F} {3, 107}

\bibitem[\protect\citeauthoryear{{Fuller}}{{Fuller}}{2017}]{fuller17}
{Fuller} J.,  2017, \mn@doi [\mnras] {10.1093/mnras/stx1314}, \href
  {http://adsabs.harvard.edu/abs/2017MNRAS.470.1642F} {470, 1642}

\bibitem[\protect\citeauthoryear{{Gal-Yam}}{{Gal-Yam}}{2012}]{galyam12}
{Gal-Yam} A.,  2012, \mn@doi [Science] {10.1126/science.1203601}, \href
  {http://adsabs.harvard.edu/abs/2012Sci...337..927G} {337, 927}

\bibitem[\protect\citeauthoryear{{Gal-Yam} \& {Leonard}}{{Gal-Yam} \&
  {Leonard}}{2009}]{galyam09}
{Gal-Yam} A.,  {Leonard} D.~C.,  2009, \mn@doi [\nat] {10.1038/nature07934},
  \href {http://adsabs.harvard.edu/abs/2009Natur.458..865G} {458, 865}

\bibitem[\protect\citeauthoryear{{Gal-Yam} et~al.,}{{Gal-Yam}
  et~al.}{2010}]{galyam10}
{Gal-Yam} A.,  et~al., 2010, The Astronomer's Telegram, \href
  {http://adsabs.harvard.edu/abs/2010ATel.2603....1G} {2603}

\bibitem[\protect\citeauthoryear{{Gal-Yam} et~al.,}{{Gal-Yam}
  et~al.}{2014}]{galyam14}
{Gal-Yam} A.,  et~al., 2014, \mn@doi [\nat] {10.1038/nature13304}, \href
  {http://adsabs.harvard.edu/abs/2014Natur.509..471G} {509, 471}

\bibitem[\protect\citeauthoryear{{Gr{\"a}fener} \& {Vink}}{{Gr{\"a}fener} \&
  {Vink}}{2016}]{grafener16}
{Gr{\"a}fener} G.,  {Vink} J.~S.,  2016, \mn@doi [\mnras]
  {10.1093/mnras/stv2283}, \href
  {http://adsabs.harvard.edu/abs/2016MNRAS.455..112G} {455, 112}

\bibitem[\protect\citeauthoryear{{Groh}}{{Groh}}{2014a}]{groh14}
{Groh} J.~H.,  2014a, \mn@doi [\aap] {10.1051/0004-6361/201424852}, 572, L11

\bibitem[\protect\citeauthoryear{{Groh}}{{Groh}}{2014b}]{groh14a}
{Groh} J.~H.,  2014b, \mn@doi [\aap] {10.1051/0004-6361/201424852}, \href
  {http://adsabs.harvard.edu/abs/2014A%26A...572L..11G} {572, L11}

\bibitem[\protect\citeauthoryear{{Groh}, {Meynet}, {Georgy}  \&
  {Ekstr{\"o}m}}{{Groh} et~al.}{2013}]{gmg13}
{Groh} J.~H.,  {Meynet} G.,  {Georgy} C.,   {Ekstr{\"o}m} S.,  2013, \mn@doi
  [\aap] {10.1051/0004-6361/201321906}, \href
  {http://adsabs.harvard.edu/abs/2013A%26A...558A.131G} {558, A131}

\bibitem[\protect\citeauthoryear{{Guillochon}, {Parrent}, {Kelley}  \&
  {Margutti}}{{Guillochon} et~al.}{2017}]{guillochon17}
{Guillochon} J.,  {Parrent} J.,  {Kelley} L.~Z.,   {Margutti} R.,  2017,
  \mn@doi [\apj] {10.3847/1538-4357/835/1/64}, \href
  {https://ui.adsabs.harvard.edu/abs/2017ApJ...835...64G} {835, 64}

\bibitem[\protect\citeauthoryear{{Hillier} \& {Dessart}}{{Hillier} \&
  {Dessart}}{2019}]{hillier19}
{Hillier} D.~J.,  {Dessart} L.,  2019, \mn@doi [\aap]
  {10.1051/0004-6361/201935100}, \href
  {https://ui.adsabs.harvard.edu/abs/2019A&A...631A...8H} {631, A8}

\bibitem[\protect\citeauthoryear{{Hillier} \& {Miller}}{{Hillier} \&
  {Miller}}{1998}]{hm98}
{Hillier} D.~J.,  {Miller} D.~L.,  1998, \mn@doi [\apj] {10.1086/305350}, \href
  {http://adsabs.harvard.edu/cgi-bin/nph-bib_query?bibcode=1998ApJ...496..407H&db_key=AST}
  {496, 407}

\bibitem[\protect\citeauthoryear{{Hook}, {J{\o}rgensen}, {Allington-Smith},
  {Davies}, {Metcalfe}, {Murowinski}  \& {Crampton}}{{Hook}
  et~al.}{2004}]{hook04}
{Hook} I.~M.,  {J{\o}rgensen} I.,  {Allington-Smith} J.~R.,  {Davies} R.~L.,
  {Metcalfe} N.,  {Murowinski} R.~G.,   {Crampton} D.,  2004, \mn@doi [\pasp]
  {10.1086/383624}, \href
  {https://ui.adsabs.harvard.edu/abs/2004PASP..116..425H} {116, 425}

\bibitem[\protect\citeauthoryear{{Hosseinzadeh} et~al.,}{{Hosseinzadeh}
  et~al.}{2018}]{hosseinzadeh18}
{Hosseinzadeh} G.,  et~al., 2018, \mn@doi [\apj] {10.3847/1538-4357/aac5f6},
  \href {https://ui.adsabs.harvard.edu/abs/2018ApJ...861...63H} {861, 63}

\bibitem[\protect\citeauthoryear{{Itagaki}}{{Itagaki}}{2018}]{itagaki18}
{Itagaki} K.,  2018, Transient Name Server Discovery Report, \href
  {https://ui.adsabs.harvard.edu/abs/2018TNSTR.285....1I} {2018-285, 1}

\bibitem[\protect\citeauthoryear{{Kasliwal} et~al.,}{{Kasliwal}
  et~al.}{2009}]{kasliwal09}
{Kasliwal} M.~M.,  et~al., 2009, Central Bureau Electronic Telegrams, \href
  {http://adsabs.harvard.edu/abs/2009CBET.1820....2K} {1820}

\bibitem[\protect\citeauthoryear{{Khazov} et~al.,}{{Khazov}
  et~al.}{2016}]{khazov16}
{Khazov} D.,  et~al., 2016, \mn@doi [\apj] {10.3847/0004-637X/818/1/3}, \href
  {http://adsabs.harvard.edu/abs/2016ApJ...818....3K} {818, 3}

\bibitem[\protect\citeauthoryear{{Kiewe} et~al.,}{{Kiewe}
  et~al.}{2012}]{kiewe12}
{Kiewe} M.,  et~al., 2012, \mn@doi [\apj] {10.1088/0004-637X/744/1/10}, \href
  {http://adsabs.harvard.edu/abs/2012ApJ...744...10K} {744, 10}

\bibitem[\protect\citeauthoryear{{Kilpatrick} et~al.,}{{Kilpatrick}
  et~al.}{2017}]{kilpatrick17}
{Kilpatrick} C.~D.,  et~al., 2017, preprint, \href
  {http://adsabs.harvard.edu/abs/2017arXiv170609962K} {} (\mn@eprint {arXiv}
  {1706.09962})

\bibitem[\protect\citeauthoryear{{Li} et~al.,}{{Li} et~al.}{2011}]{li11}
{Li} W.,  et~al., 2011, \mn@doi [\mnras] {10.1111/j.1365-2966.2011.18160.x},
  \href {https://ui.adsabs.harvard.edu/abs/2011MNRAS.412.1441L} {412, 1441}

\bibitem[\protect\citeauthoryear{{Margutti} et~al.,}{{Margutti}
  et~al.}{2017}]{margutti14b}
{Margutti} R.,  et~al., 2017, \mn@doi [\apj] {10.3847/1538-4357/835/2/140},
  \href {https://ui.adsabs.harvard.edu/abs/2017ApJ...835..140M} {835, 140}

\bibitem[\protect\citeauthoryear{{Mauron} \& {Josselin}}{{Mauron} \&
  {Josselin}}{2011}]{mauron11}
{Mauron} N.,  {Josselin} E.,  2011, \mn@doi [\aap]
  {10.1051/0004-6361/201013993}, \href
  {http://adsabs.harvard.edu/abs/2011A%26A...526A.156M} {526, A156}

\bibitem[\protect\citeauthoryear{{Mikolajczyk} \& {Wyrzykowski}}{{Mikolajczyk}
  \& {Wyrzykowski}}{2018}]{mikolajczyk18}
{Mikolajczyk} P.,  {Wyrzykowski} L.,  2018, The Astronomer's Telegram, \href
  {https://ui.adsabs.harvard.edu/abs/2018ATel12079....1M} {12079}

\bibitem[\protect\citeauthoryear{Miller}{Miller}{1994}]{miller94}
Miller J.,  1994, {The Kast Double Spectrograph}, \url
  {https://mthamilton.ucolick.org/techdocs/instruments/kast/Tech\%20Report\%2066\%20KAST\%20Miller\%20Stone.pdf}

\bibitem[\protect\citeauthoryear{{Moriya} \& {Maeda}}{{Moriya} \&
  {Maeda}}{2014}]{moriya14b}
{Moriya} T.~J.,  {Maeda} K.,  2014, \mn@doi [\apjl]
  {10.1088/2041-8205/790/2/L16}, \href
  {http://adsabs.harvard.edu/abs/2014ApJ...790L..16M} {790, L16}

\bibitem[\protect\citeauthoryear{{Moriya}, {Maeda}, {Taddia}, {Sollerman},
  {Blinnikov}  \& {Sorokina}}{{Moriya} et~al.}{2014}]{moriya14a}
{Moriya} T.~J.,  {Maeda} K.,  {Taddia} F.,  {Sollerman} J.,  {Blinnikov} S.~I.,
    {Sorokina} E.~I.,  2014, \mn@doi [\mnras] {10.1093/mnras/stu163}, \href
  {http://adsabs.harvard.edu/abs/2014MNRAS.439.2917M} {439, 2917}

\bibitem[\protect\citeauthoryear{{Moriya}, {F{\"o}rster}, {Yoon},
  {Gr{\"a}fener}  \& {Blinnikov}}{{Moriya} et~al.}{2018}]{Moriya18}
{Moriya} T.~J.,  {F{\"o}rster} F.,  {Yoon} S.-C.,  {Gr{\"a}fener} G.,
  {Blinnikov} S.~I.,  2018, \mn@doi [\mnras] {10.1093/mnras/sty475}, \href
  {http://adsabs.harvard.edu/abs/2018MNRAS.476.2840M} {476, 2840}

\bibitem[\protect\citeauthoryear{{Morozova}, {Piro}  \& {Valenti}}{{Morozova}
  et~al.}{2017}]{morozova17}
{Morozova} V.,  {Piro} A.~L.,   {Valenti} S.,  2017, \mn@doi [\apj]
  {10.3847/1538-4357/aa6251}, \href
  {http://adsabs.harvard.edu/abs/2017ApJ...838...28M} {838, 28}

\bibitem[\protect\citeauthoryear{{Nakano}}{{Nakano}}{2014}]{nakano14}
{Nakano} S.,  2014, Central Bureau Electronic Telegrams, \href
  {https://ui.adsabs.harvard.edu/abs/2014CBET.3787....1N} {3787}

\bibitem[\protect\citeauthoryear{{Nakaoka} et~al.,}{{Nakaoka}
  et~al.}{2018}]{nakaoka18}
{Nakaoka} T.,  et~al., 2018, \mn@doi [\apj] {10.3847/1538-4357/aabee7}, \href
  {https://ui.adsabs.harvard.edu/abs/2018ApJ...859...78N} {859, 78}

\bibitem[\protect\citeauthoryear{{Nicholls}, {Stanek}  \& {Dong}}{{Nicholls}
  et~al.}{2018}]{nicholls18}
{Nicholls} B.,  {Stanek} K.~Z.,   {Dong} S.,  2018, Transient Name Server
  Discovery Report, \href
  {https://ui.adsabs.harvard.edu/abs/2018TNSTR.887....1N} {2018-887, 1}

\bibitem[\protect\citeauthoryear{{Ofek} et~al.,}{{Ofek} et~al.}{2007}]{ofek07}
{Ofek} E.~O.,  et~al., 2007, \mn@doi [\apjl] {10.1086/516749}, \href
  {https://ui.adsabs.harvard.edu/abs/2007ApJ...659L..13O} {659, L13}

\bibitem[\protect\citeauthoryear{{Ofek} et~al.,}{{Ofek} et~al.}{2013}]{ofek13b}
{Ofek} E.~O.,  et~al., 2013, \mn@doi [\nat] {10.1038/nature11877}, \href
  {http://adsabs.harvard.edu/abs/2013Natur.494...65O} {494, 65}

\bibitem[\protect\citeauthoryear{{Ofek} et~al.,}{{Ofek} et~al.}{2014}]{ofek14b}
{Ofek} E.~O.,  et~al., 2014, \mn@doi [\apj] {10.1088/0004-637X/789/2/104},
  \href {http://adsabs.harvard.edu/abs/2014ApJ...789..104O} {789, 104}

\bibitem[\protect\citeauthoryear{{Oke} \& {Gunn}}{{Oke} \&
  {Gunn}}{1982}]{oke82}
{Oke} J.~B.,  {Gunn} J.~E.,  1982, \mn@doi [\pasp] {10.1086/131027}, \href
  {https://ui.adsabs.harvard.edu/abs/1982PASP...94..586O} {94, 586}

\bibitem[\protect\citeauthoryear{{Oke} et~al.,}{{Oke} et~al.}{1995}]{Oke95}
{Oke} J.~B.,  et~al., 1995, \mn@doi [\pasp] {10.1086/133562}, \href
  {https://ui.adsabs.harvard.edu/abs/1995PASP..107..375O} {107, 375}

\bibitem[\protect\citeauthoryear{{Parrent} et~al.,}{{Parrent}
  et~al.}{2011}]{Parrent11}
{Parrent} J.,  et~al., 2011, The Astronomer's Telegram, \href
  {http://adsabs.harvard.edu/abs/2011ATel.3510....1P} {3510}

\bibitem[\protect\citeauthoryear{{Pastorello} et~al.,}{{Pastorello}
  et~al.}{2008}]{pastorello08}
{Pastorello} A.,  et~al., 2008, \mn@doi [\mnras]
  {10.1111/j.1365-2966.2008.13602.x}, \href
  {https://ui.adsabs.harvard.edu/abs/2008MNRAS.389..113P} {389, 113}

\bibitem[\protect\citeauthoryear{{Pastorello} et~al.,}{{Pastorello}
  et~al.}{2018}]{pastorello18}
{Pastorello} A.,  et~al., 2018, \mn@doi [\mnras] {10.1093/mnras/stx2668}, \href
  {http://adsabs.harvard.edu/abs/2018MNRAS.474..197P} {474, 197}

\bibitem[\protect\citeauthoryear{{Pastorello} et~al.,}{{Pastorello}
  et~al.}{2019}]{pastorello19}
{Pastorello} A.,  et~al., 2019, \mn@doi [\aap] {10.1051/0004-6361/201935420},
  \href {https://ui.adsabs.harvard.edu/abs/2019A&A...628A..93P} {628, A93}

\bibitem[\protect\citeauthoryear{{Quataert}, {Fern{\'a}ndez}, {Kasen}, {Klion}
  \& {Paxton}}{{Quataert} et~al.}{2016}]{quataert16}
{Quataert} E.,  {Fern{\'a}ndez} R.,  {Kasen} D.,  {Klion} H.,   {Paxton} B.,
  2016, \mn@doi [\mnras] {10.1093/mnras/stw365}, \href
  {https://ui.adsabs.harvard.edu/abs/2016MNRAS.458.1214Q} {458, 1214}

\bibitem[\protect\citeauthoryear{{Rafanelli} \& {Siviero}}{{Rafanelli} \&
  {Siviero}}{2012}]{rafanelli12}
{Rafanelli} P.,  {Siviero} A.,  2012, \mn@doi [Baltic Astronomy]
  {10.1515/astro-2017-0351}, \href
  {https://ui.adsabs.harvard.edu/abs/2012BaltA..21....1R} {21, 1}

\bibitem[\protect\citeauthoryear{{Rubin} et~al.,}{{Rubin}
  et~al.}{2016}]{rubin16}
{Rubin} A.,  et~al., 2016, \mn@doi [\apj] {10.3847/0004-637X/820/1/33}, \href
  {http://adsabs.harvard.edu/abs/2016ApJ...820...33R} {820, 33}

\bibitem[\protect\citeauthoryear{{Sand}}{{Sand}}{2014}]{sand14}
{Sand} D.,  2014, in {Wozniak} P.~R.,  {Graham} M.~J.,  {Mahabal} A.~A.,
  {Seaman} R.,  eds, The Third Hot-wiring the Transient Universe Workshop.
  p.~187

\bibitem[\protect\citeauthoryear{{Schlafly} \& {Finkbeiner}}{{Schlafly} \&
  {Finkbeiner}}{2011}]{schlafly11}
{Schlafly} E.~F.,  {Finkbeiner} D.~P.,  2011, \mn@doi [\apj]
  {10.1088/0004-637X/737/2/103}, \href
  {http://adsabs.harvard.edu/abs/2011ApJ...737..103S} {737, 103}

\bibitem[\protect\citeauthoryear{{Schlegel}, {Finkbeiner}  \&
  {Davis}}{{Schlegel} et~al.}{1998}]{Schlegel98}
{Schlegel} D.~J.,  {Finkbeiner} D.~P.,   {Davis} M.,  1998, \mn@doi [\apj]
  {10.1086/305772}, \href {http://adsabs.harvard.edu/abs/1998ApJ...500..525S}
  {500, 525}

\bibitem[\protect\citeauthoryear{{Shivvers}, {Groh}, {Mauerhan}, {Fox},
  {Leonard}  \& {Filippenko}}{{Shivvers} et~al.}{2015}]{shivvers15}
{Shivvers} I.,  {Groh} J.~H.,  {Mauerhan} J.~C.,  {Fox} O.~D.,  {Leonard}
  D.~C.,   {Filippenko} A.~V.,  2015, \mn@doi [\apj]
  {10.1088/0004-637X/806/2/213}, \href
  {http://adsabs.harvard.edu/abs/2015ApJ...806..213S} {806, 213}

\bibitem[\protect\citeauthoryear{{Smartt} et~al.,}{{Smartt}
  et~al.}{2015}]{smartt15b}
{Smartt} S.~J.,  et~al., 2015, \mn@doi [\aap] {10.1051/0004-6361/201425237},
  \href {https://ui.adsabs.harvard.edu/abs/2015A&A...579A..40S} {579, A40}

\bibitem[\protect\citeauthoryear{{Smith}}{{Smith}}{2006}]{smith06}
{Smith} N.,  2006, \mn@doi [\apj] {10.1086/503766}, \href
  {http://adsabs.harvard.edu/cgi-bin/nph-bib_query?bibcode=2006ApJ...644.1151S&db_key=AST}
  {644, 1151}

\bibitem[\protect\citeauthoryear{{Smith}}{{Smith}}{2014}]{smith14araa}
{Smith} N.,  2014, \mn@doi [\araa] {10.1146/annurev-astro-081913-040025}, \href
  {http://adsabs.harvard.edu/abs/2014ARA%26A..52..487S} {52, 487}

\bibitem[\protect\citeauthoryear{{Smith}}{{Smith}}{2017}]{smith17hsn}
{Smith} N.,  2017, {Interacting Supernovae: Types IIn and Ibn}.
Springer International Publishing AG, p.~403,
  \mn@doi{10.1007/978-3-319-21846-5_38}

\bibitem[\protect\citeauthoryear{{Smith} \& {Arnett}}{{Smith} \&
  {Arnett}}{2014}]{smith14b}
{Smith} N.,  {Arnett} W.~D.,  2014, \mn@doi [\apj]
  {10.1088/0004-637X/785/2/82}, \href
  {https://ui.adsabs.harvard.edu/abs/2014ApJ...785...82S} {785, 82}

\bibitem[\protect\citeauthoryear{{Smith}, {Li}, {Filippenko}  \&
  {Chornock}}{{Smith} et~al.}{2011}]{smith11sn}
{Smith} N.,  {Li} W.,  {Filippenko} A.~V.,   {Chornock} R.,  2011, \mn@doi
  [\mnras] {10.1111/j.1365-2966.2011.17229.x}, \href
  {http://adsabs.harvard.edu/abs/2011MNRAS.412.1522S} {412, 1522}

\bibitem[\protect\citeauthoryear{{Smith} et~al.,}{{Smith}
  et~al.}{2015}]{smith15}
{Smith} N.,  et~al., 2015, \mn@doi [\mnras] {10.1093/mnras/stv354}, \href
  {http://adsabs.harvard.edu/abs/2015MNRAS.449.1876S} {449, 1876}

\bibitem[\protect\citeauthoryear{{Stanek}}{{Stanek}}{2016}]{stanek16}
{Stanek} K.~Z.,  2016, Transient Name Server Discovery Report, \href
  {https://ui.adsabs.harvard.edu/abs/2016TNSTR.534....1S} {2016-534, 1}

\bibitem[\protect\citeauthoryear{{Stathakis} \& {Sadler}}{{Stathakis} \&
  {Sadler}}{1991}]{stathakis91}
{Stathakis} R.~A.,  {Sadler} E.~M.,  1991, \mn@doi [\mnras]
  {10.1093/mnras/250.4.786}, \href
  {http://adsabs.harvard.edu/abs/1991MNRAS.250..786S} {250, 786}

\bibitem[\protect\citeauthoryear{{Stritzinger} et~al.,}{{Stritzinger}
  et~al.}{2012}]{stritzinger12}
{Stritzinger} M.,  et~al., 2012, \mn@doi [\apj] {10.1088/0004-637X/756/2/173},
  \href {https://ui.adsabs.harvard.edu/abs/2012ApJ...756..173S} {756, 173}

\bibitem[\protect\citeauthoryear{{Taddia} et~al.,}{{Taddia}
  et~al.}{2013}]{taddia13}
{Taddia} F.,  et~al., 2013, \mn@doi [\aap] {10.1051/0004-6361/201321180}, \href
  {http://adsabs.harvard.edu/abs/2013A%26A...555A..10T} {555, A10}

\bibitem[\protect\citeauthoryear{{Terreran} et~al.,}{{Terreran}
  et~al.}{2016}]{terreran16}
{Terreran} G.,  et~al., 2016, \mn@doi [\mnras] {10.1093/mnras/stw1591}, \href
  {http://adsabs.harvard.edu/abs/2016MNRAS.462..137T} {462, 137}

\bibitem[\protect\citeauthoryear{{Vink}, {de Koter}  \& {Lamers}}{{Vink}
  et~al.}{2001}]{vink01}
{Vink} J.~S.,  {de Koter} A.,   {Lamers} H.~J.~G.~L.~M.,  2001, \mn@doi [\aap]
  {10.1051/0004-6361:20010127}, \href
  {http://adsabs.harvard.edu/abs/2001A%26A...369..574V} {369, 574}

\bibitem[\protect\citeauthoryear{{Yaron} et~al.,}{{Yaron}
  et~al.}{2017}]{yaron17}
{Yaron} O.,  et~al., 2017, \mn@doi [Nature Physics] {10.1038/nphys4025}, \href
  {http://adsabs.harvard.edu/abs/2017NatPh..13..510Y} {13, 510}

\bibitem[\protect\citeauthoryear{{Yoon} \& {Cantiello}}{{Yoon} \&
  {Cantiello}}{2010}]{yooncantiello10}
{Yoon} S.-C.,  {Cantiello} M.,  2010, \mn@doi [\apjl]
  {10.1088/2041-8205/717/1/L62}, \href
  {http://adsabs.harvard.edu/abs/2010ApJ...717L..62Y} {717, L62}

\bibitem[\protect\citeauthoryear{{Zapartas} et~al.,}{{Zapartas}
  et~al.}{2019}]{zapartas19}
{Zapartas} E.,  et~al., 2019, \mn@doi [\aap] {10.1051/0004-6361/201935854},
  \href {https://ui.adsabs.harvard.edu/abs/2019A&A...631A...5Z} {631, A5}

\makeatother
\end{thebibliography}



\appendix

\onecolumn

\section{Best-fit models of individual supernovae}
\label{apx:bfs}
This section contains the best-fitting models for the other SNe in our sample in chronological order.
\begin{figure}
\label{fig:09ij_bf}
    \begin{subfigure}{0.99\textwidth}
        \includegraphics[width=0.94\textwidth]{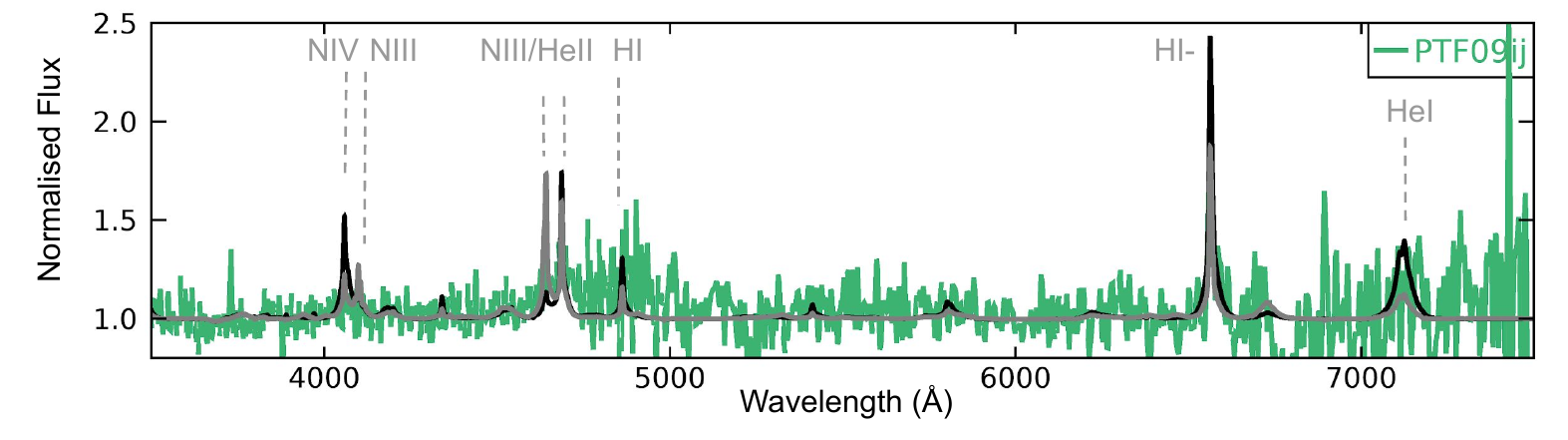}
    \end{subfigure}
    
    \begin{subfigure}{0.99\textwidth}
        \includegraphics[width=0.94\textwidth]{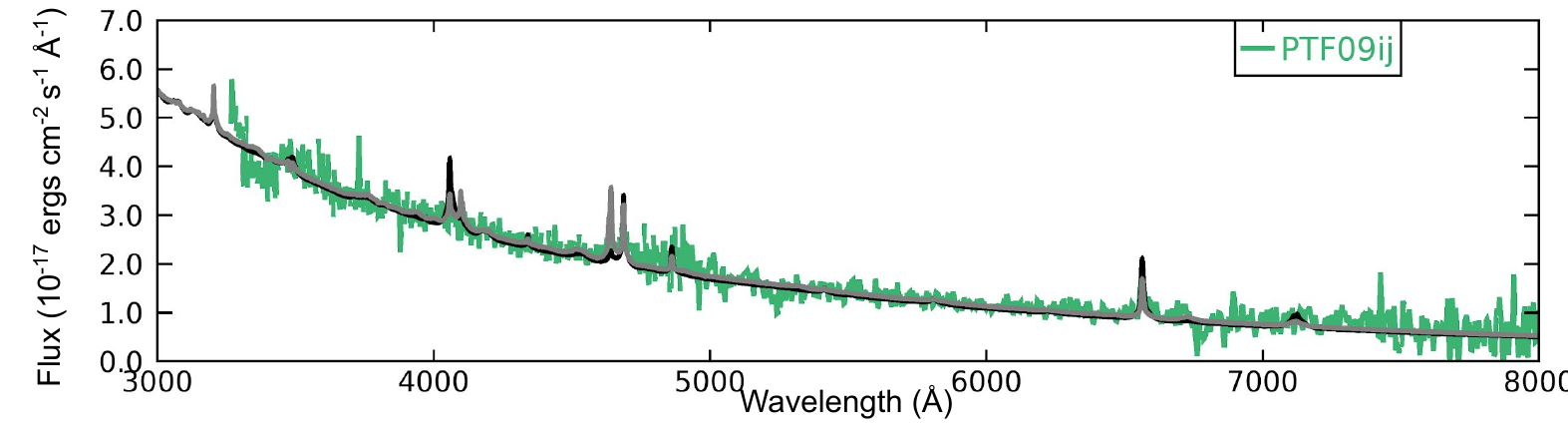}
    \end{subfigure}
    
    \caption{Best-fit models for PTF 09ij. Top: normalised spectra. Bottom: absolute flux. PTF 09ij falls in between a model with $L = 5.7 \times 10^{10} ~\lsun$, $\tstar = 29300 $ K, $\mdot = 22.7 \times 10^{-3} ~\msunyr$, $\rin = 62 \times 10^{13}$ cm, $\vinf<250~\kms$, and CNO-processed surface abundances (black) and  $L = 4.3 \times 10^{10} ~\lsun$,  $\tstar = 26700 $ K, $\mdot = 14.9 \times 10^{-3} ~\msunyr$, $\rin = 66.4 \times 10^{13}$ cm, $\vinf<250~\kms$, and CNO-processed surface abundances (grey).  The SED was fit assuming a distance of $d_L = 598.7 Mpc$, and the best-fit was given by a color excess of $E(B-V) = 0.22$, and relative visibility $R_V = 3.1$.}
\end{figure}

\begin{figure*}
\label{fig:10abyy_bf}
    \begin{subfigure}[t]{0.99\textwidth}
        \includegraphics[width=0.94\textwidth]{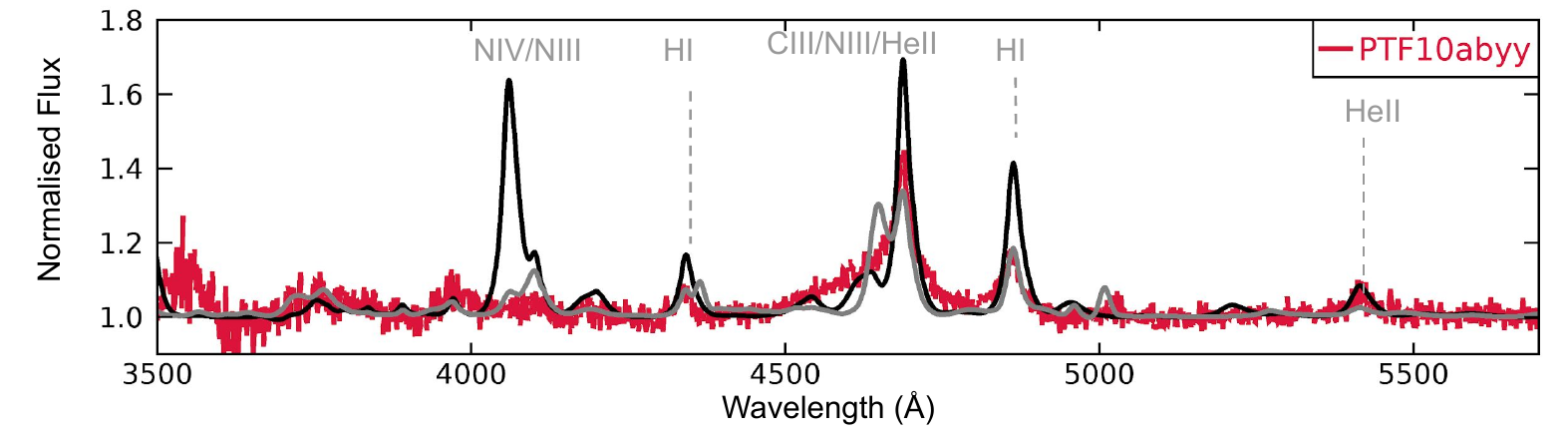}
    \end{subfigure}
    
    \begin{subfigure}[b]{0.99\textwidth}
        \includegraphics[width=0.94\textwidth]{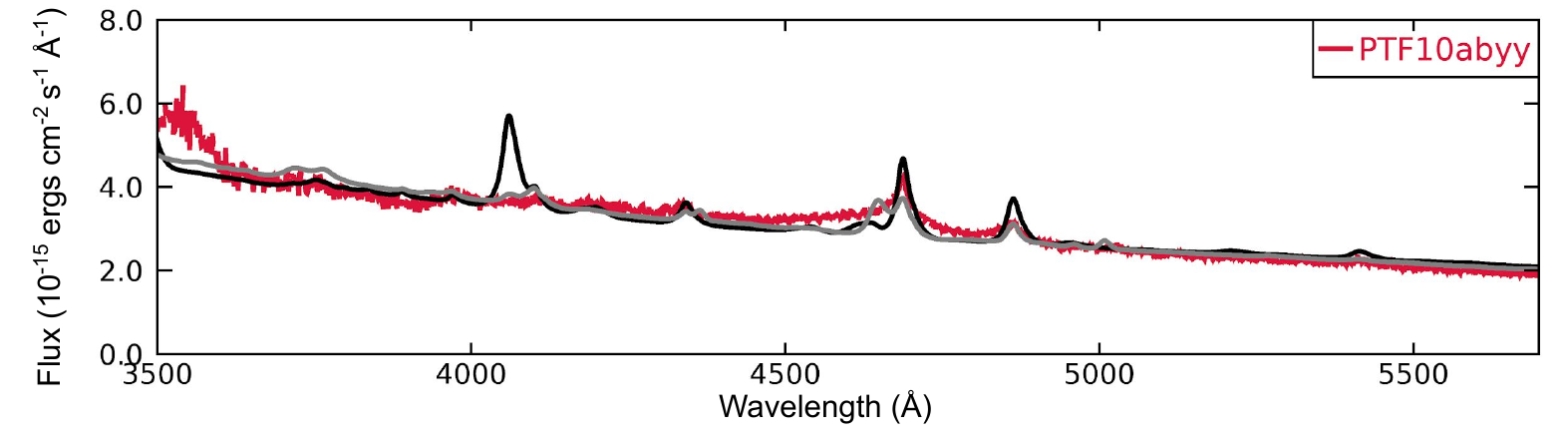}
    \end{subfigure}
    
    \caption{The $6.8 \pm 0.5$ day-old spectrum of PTF10abyy and two encompassing models. Top: normalised spectra. Bottom: absolute flux. The black model corresponds to a SN with $L = 1.37 \times 10^{12} ~\lsun$, $\mdot = 1 ~\msunyr$ ($\vinf = 800 ~\kms$), $\rin = 2.4 \times 10^{15}$ cm, and CNO-processed surface abundances. The grey model has $L = 5.1 \times 10^{11} ~\lsun$, $\mdot = 5 \times 10^{-1} ~\msunyr$ ($\vinf = 800 ~\kms$), $\rin = 2.3 \times 10^{15}$ cm, and solar surface abundances. Fitting the SED assuming $d = 134.4$ Mpc, requires $E(B-V) = 0.44$, $R_V = 3.1$ for the black model and  $E(B-V) = 0.37$, $R_V = 3.1$ for the grey model. The models also match the closest photometric measurement of $M_R = -18.190$ mag.}
\end{figure*}

\begin{figure*}
    \label{fig:10gva_bf}
    \begin{subfigure}[t]{0.99\textwidth}
        \includegraphics[width=0.94\textwidth]{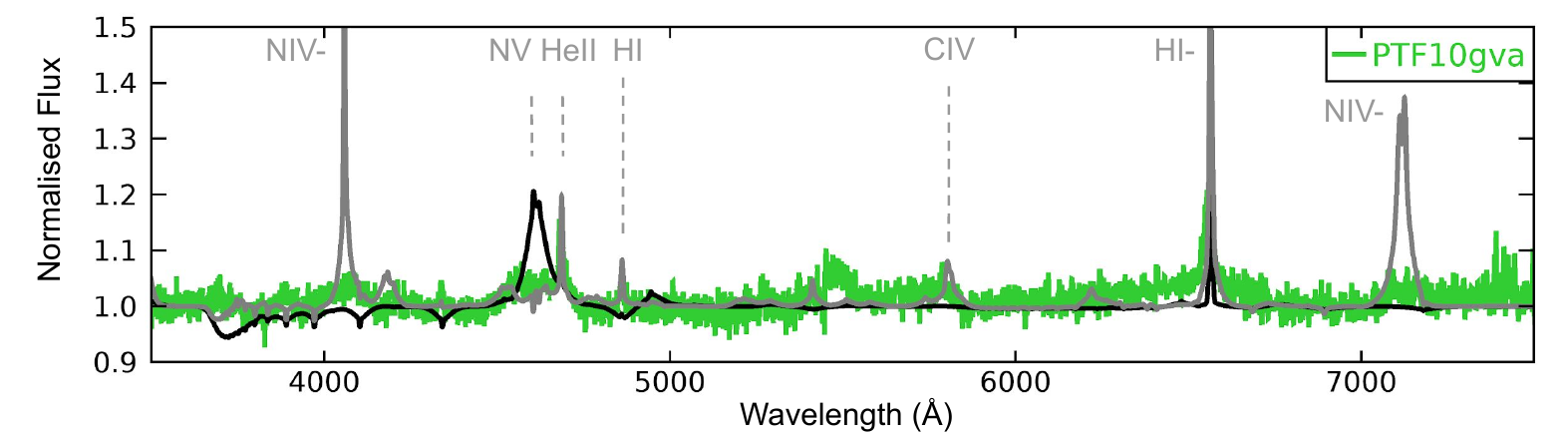}
    \end{subfigure}
    
    \begin{subfigure}[b]{0.99\textwidth}
        \includegraphics[width=0.94\textwidth]{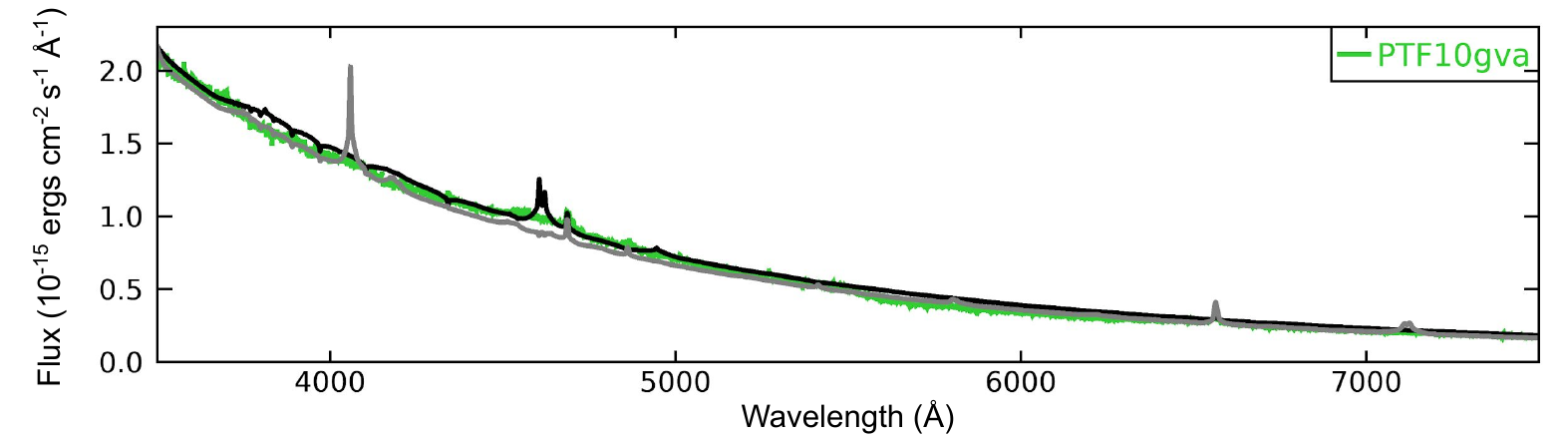}
    \end{subfigure}
    
    \caption{The spectrum of PTF10gva and its two closest fitting models. Top: normalised spectra. Bottom: absolute flux. The black spectrum has $L = 2.0 \times 10^{11} \lsun$, $\mdot = 8.708 \times 10^{-3} ~\msunyr$, $\vinf = 275 ~\kms$, $\rin = 45.25 \times 10^{13}$ cm, and CNO-processed surface abundances. The grey model has $L = 5.67 \times 10^{10} \lsun$, $\mdot = 9.526 \times 10^{-3} ~\msunyr$, $\vinf = 275 ~\kms$, $\rin = 48 \times 10^{13}$ cm, and CNO-processed surface abundances. In order to match the absolute flux a distance of $d_L = 124.3 Mpc$, a colour excess of $E(B-V) = 0.01$, and $R_V = 3.1$ was assumed for the black model and $E(B-V) = 0.02$, and $R_V = 3.1$ for the grey model.  }
\end{figure*}

\begin{figure*}
    \label{fig:10uls_bf}
    \begin{subfigure}[t]{0.99\textwidth}
        \includegraphics[width=0.94\textwidth]{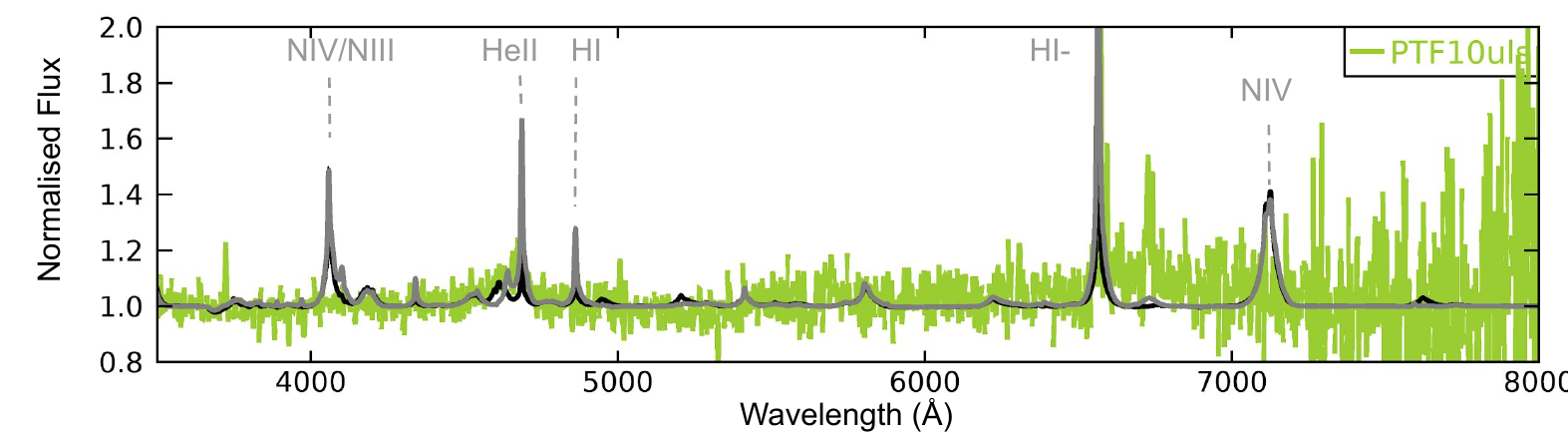}
    \end{subfigure}
    
    \begin{subfigure}[b]{0.99\textwidth}
        \includegraphics[width=0.94\textwidth]{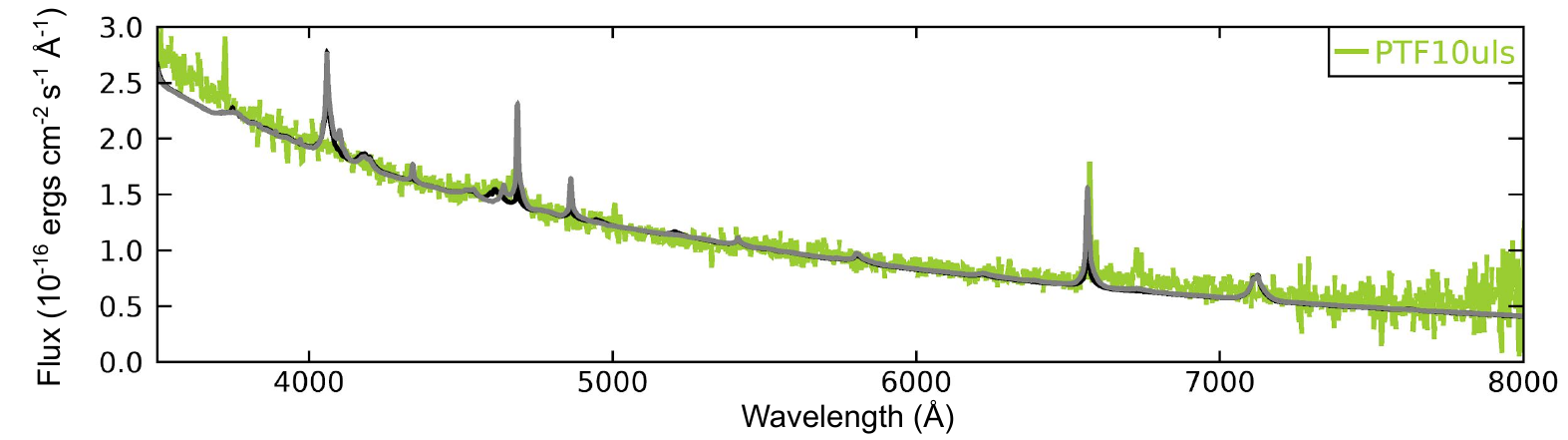}
    \end{subfigure}
    
    \caption{The spectrum of PTF10uls and its two closest fitting models. Top: normalised spectra. Bottom: absolute flux. The black spectrum has $L = 8.75 \times 10^{10} \lsun$, $\mdot = 15.4 \times 10^{-3} ~\msunyr$, $\vinf = 300 ~\kms$, $\rin = 59.9 \times 10^{13}$ cm, and CNO-processed surface abundances. The grey model has $L = 6.0 \times 10^{10} \lsun$, $\mdot = 28.3 \times 10^{-3} ~\msunyr$, $\vinf = 300 ~\kms$, $\rin = 64 \times 10^{13}$ cm, and CNO-processed surface abundances. In order to match the absolute flux a distance of $d_L = 201$ Mpc, a colour excess of $E(B-V) = 0.3$, and $R_V = 3.1$ was assumed for both models.}
\end{figure*}

\begin{figure*}
    \label{fig:10tel}
    \begin{subfigure}[t]{0.99\textwidth}
        \includegraphics[width=0.94\textwidth]{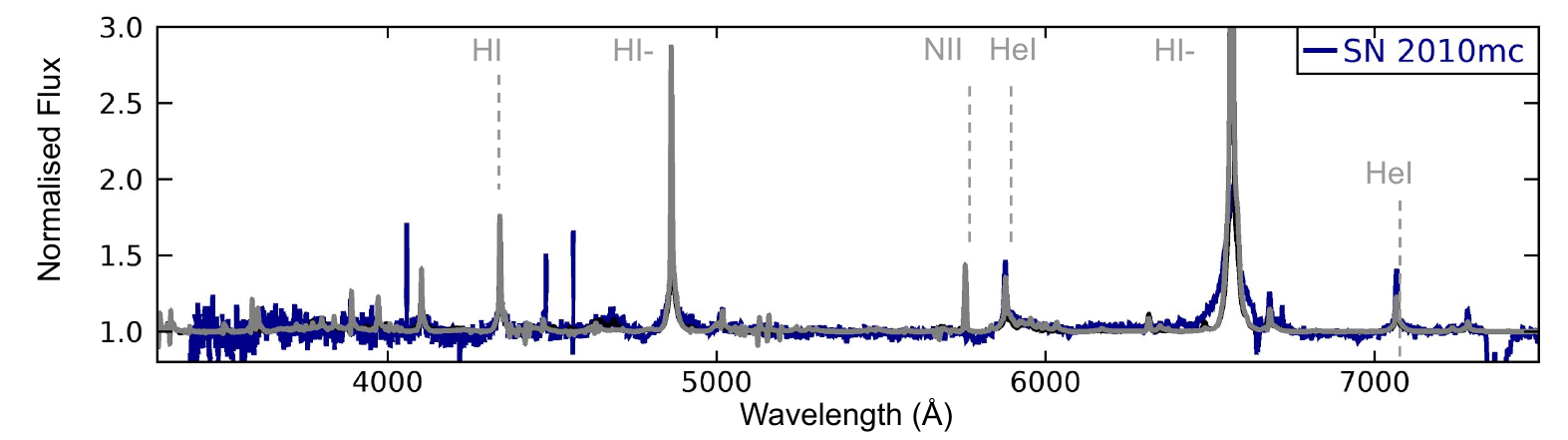}
    \end{subfigure}
    
    \begin{subfigure}[b]{0.99\textwidth}
        \includegraphics[width=0.94\textwidth]{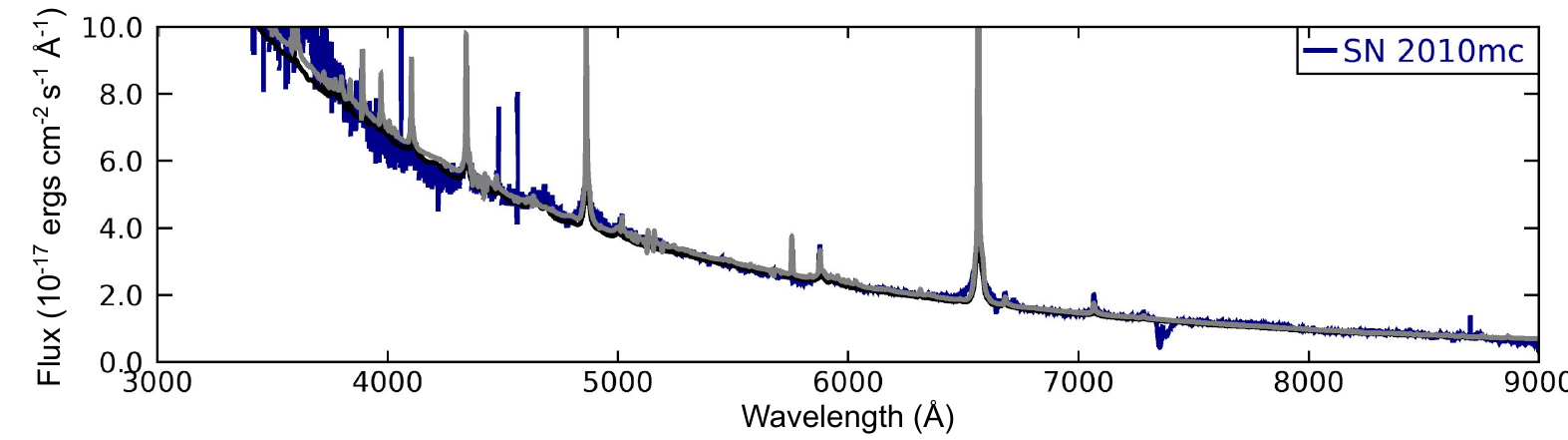}
    \end{subfigure}
    
    \caption{Closest fitting models for SN 2010mc. Top: normalised spectra. Bottom: absolute flux. The black model has $L = 1.5 \times 10^{9} ~\lsun$, $\mdot = 10.08 \times 10^{-3} ~\msunyr$, $\vinf = 300 ~\kms$, $\tstar = 19800$ K, $\rin = 22.6 \times 10^{13}$ cm and CNO-processed surface abundances. The grey model has $L = 9.8 \times 10^{8} ~\lsun$, $\mdot = 11.93 \times 10^{-3} ~\msunyr$, $\vinf = 300 ~\kms$, $\tstar = 16700$ K, $\rin = 25.3 \times 10^{13}$ cm, and CNO-processed surface abundances. The SED was matched with $E(B-V) = 0.1$, and $R_V = 3.1$ for the black model and $E(B-V) = 0.07$, and $R_V = 3.1$ for the grey, assuming a distance of 153 Mpc. }
\end{figure*}

\begin{figure*}
    \label{fig:11iqb_bf}
    \begin{subfigure}[t]{0.99\textwidth}
        \includegraphics[width=0.94\textwidth]{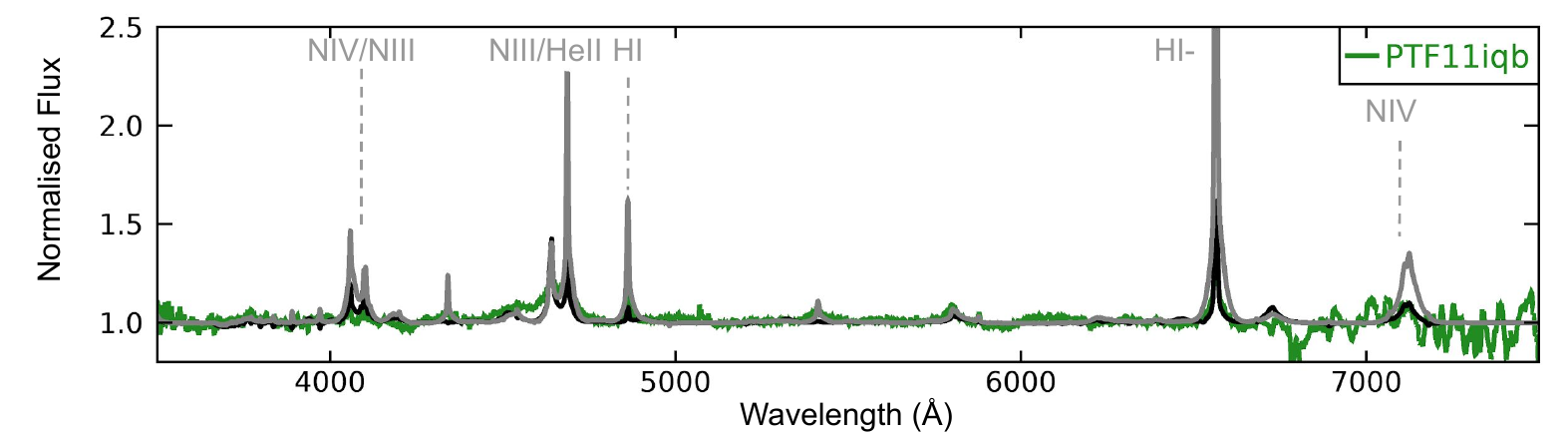}
    \end{subfigure}
    
    \begin{subfigure}[b]{0.99\textwidth}
        \includegraphics[width=0.94\textwidth]{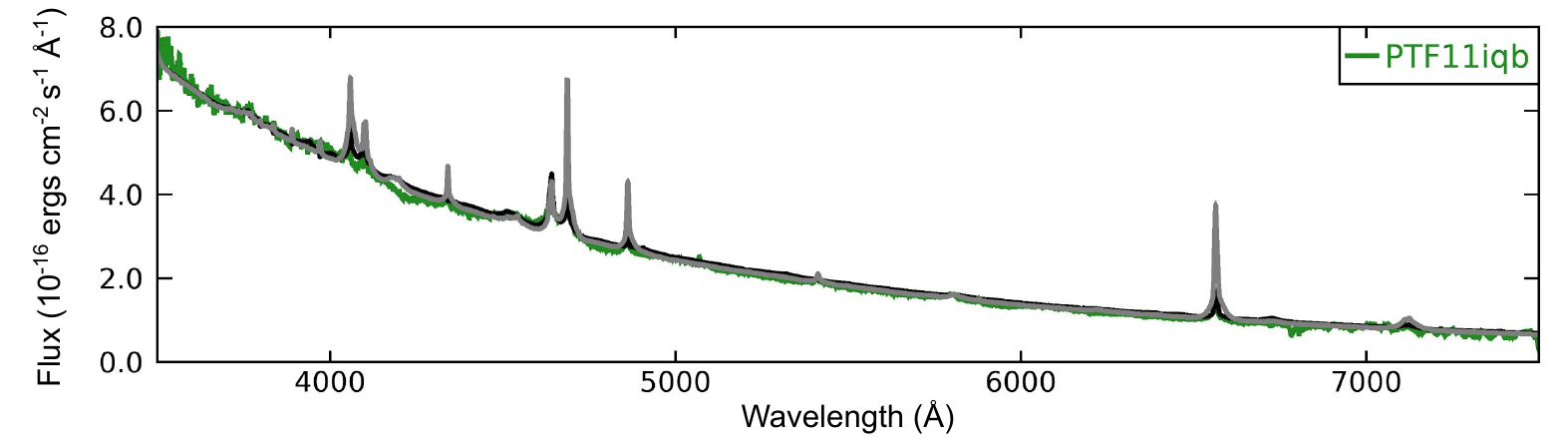}
    \end{subfigure}
    
    \caption{Best-fit models for PTF11iqb. Top: normalised spectra. Bottom: absolute flux. The black model has $L = 3.1 \times 10^{9} \lsun$, $\mdot = 0.66 \times 10^{-3} ~\msunyr$, $\vinf = 100 ~\kms$, $\rin = 16 \times 10^{13}$ cm, and CNO-processed surface abundances, while the grey model has $L = 4.0 \times 10^{9} \lsun$, $\mdot = 2.0 \times 10^{-3} ~\msunyr$, $\vinf = 100 ~\kms$, $\rin = 16 \times 10^{13}$ cm, and CNO-processed surface abundances. The models have been convolved with a Gaussian kernel having FWTH = $300 ~\kms$ in order to match the resolution of the observations. Assuming a distance of $55.85$ Mpc, $E(B-V) = 0.3$ and $R_V = 3.1$ fits both models. }
\end{figure*}

\begin{figure*}
    \label{fig:12gnn_bf}
    \begin{subfigure}[t]{0.99\textwidth}
        \includegraphics[width=0.94\textwidth]{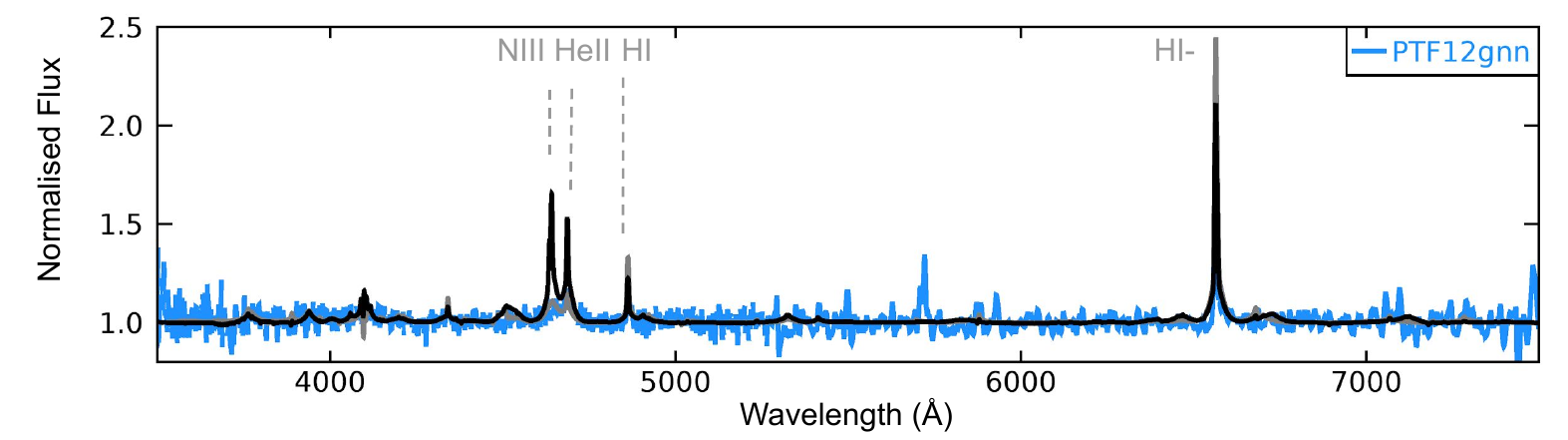}
    \end{subfigure}
    
    \begin{subfigure}[b]{0.99\textwidth}
        \includegraphics[width=0.94\textwidth]{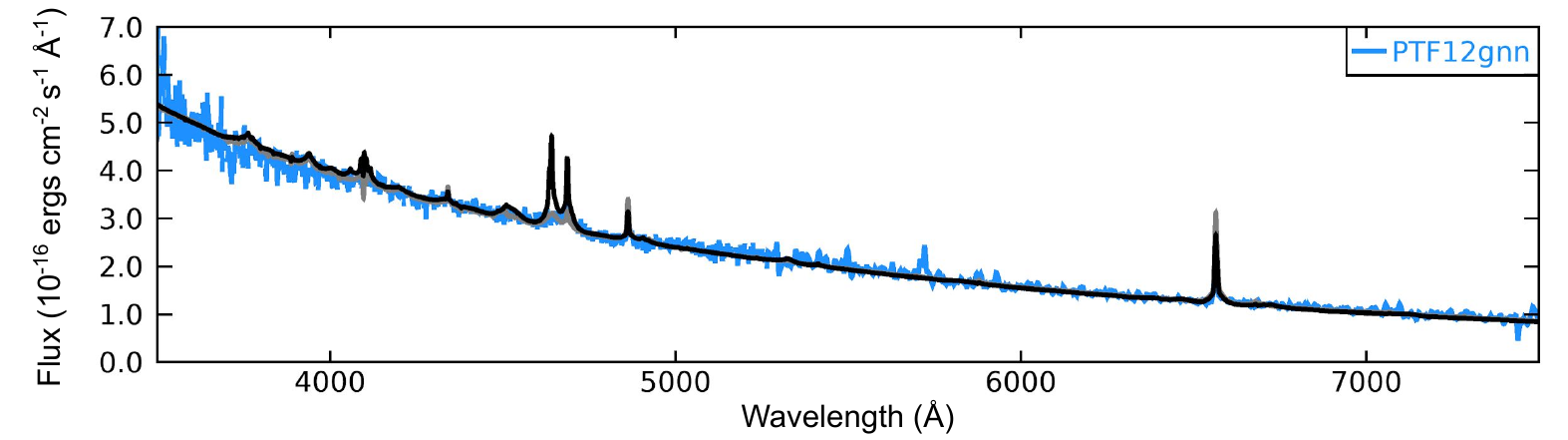}
    \end{subfigure}
    
    \caption{Best-fit models for PTF12gnn. Top: normalised spectra. Bottom: absolute flux. The black model has $L = 23.3 \times 10^{9} \lsun$, $\mdot = 13.3 \times 10^{-3} ~\msunyr$, $\vinf = 250 ~\kms$, $\rin = 61.5 \times 10^{13}$ cm, and CNO-processed surface abundances, while the grey model has $L = 13.9 \times 10^{9} \lsun$, $\mdot = 15.4 \times 10^{-3} ~\msunyr$, $\vinf = 250 ~\kms$, $\rin = 67.88 \times 10^{13}$ cm, and CNO-processed surface abundances. For the comparison to the absolute flux, assuming a distance of $d = 139.5$ Mpc, an extinction of $E(B-V) = 0.22$ ($R_V=3.1$) provided the best-fit for both models.}
\end{figure*}

\begin{figure*}
    \label{fig:12krf_bf}
    \begin{subfigure}[t]{0.99\textwidth}
        \includegraphics[width=0.94\textwidth]{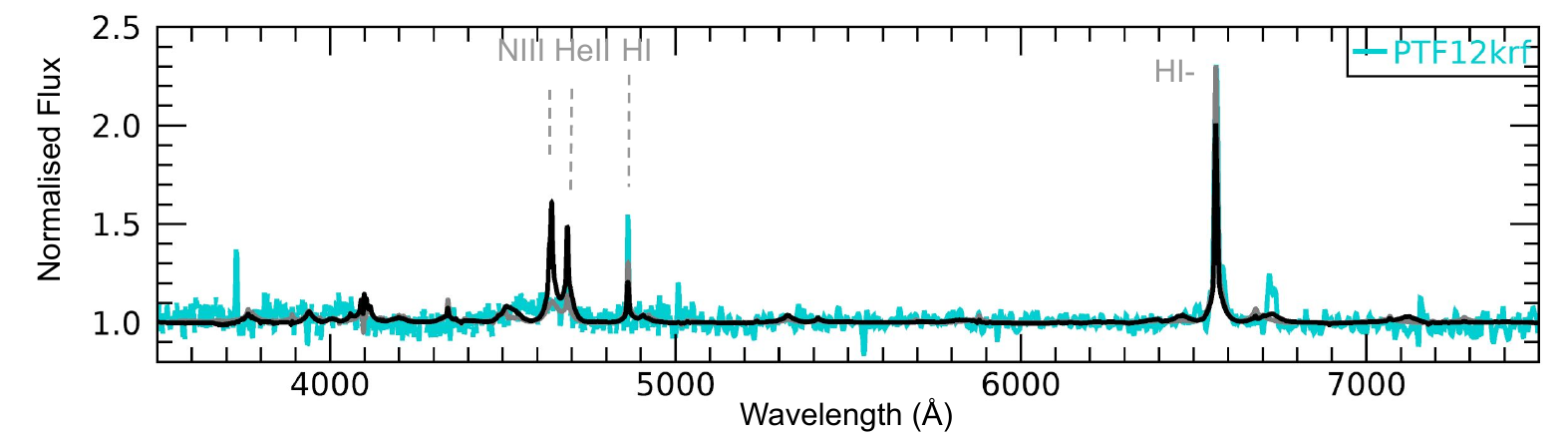}
    \end{subfigure}
    
    \begin{subfigure}[b]{0.99\textwidth}
        \includegraphics[width=0.94\textwidth]{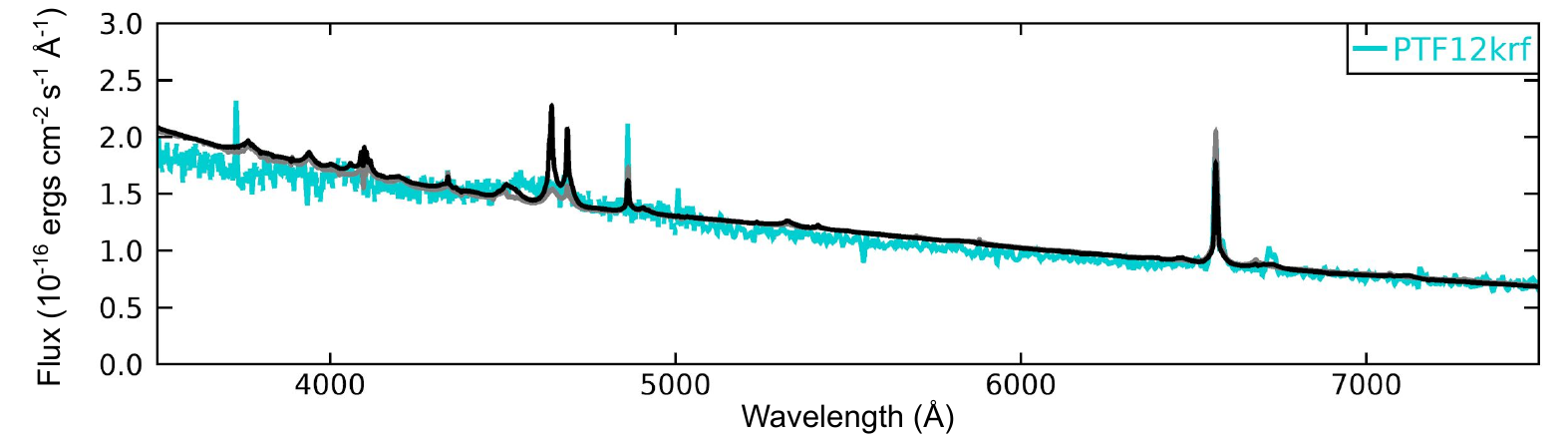}
    \end{subfigure}
    
    \caption{Best-fit models for PTF12krf. Top: normalised spectra. Bottom: absolute flux. The black model has $L = 119.7 \times 10^{9} \lsun$, $\mdot = 36.4 \times 10^{-3} ~\msunyr$, $\vinf = 200 ~\kms$, $\rin = 139.5 \times 10^{13}$ cm, and CNO-processed surface abundances, while the grey model has $L = 71.3 \times 10^{9} \lsun$, $\mdot = 42.0 \times 10^{-3} ~\msunyr$, $\vinf = 200 ~\kms$, $\rin = 153.5 \times 10^{13}$ cm, and CNO-processed surface abundances. This event has only R-band photometry, and around the time the spectrum was taken, it had $M_R = -18.44$ mag, which is well matched by the models. For the comparison to the absolute flux, a distance of $d = 289.6$ Mpc was assumed and $E(B-V) = 0.47$, $R_V=3.1$ for both models.}
\end{figure*}

\begin{figure*}
    \label{fig:13dqy_bf}
    \begin{subfigure}[t]{0.99\textwidth}
        \includegraphics[width=0.94\textwidth]{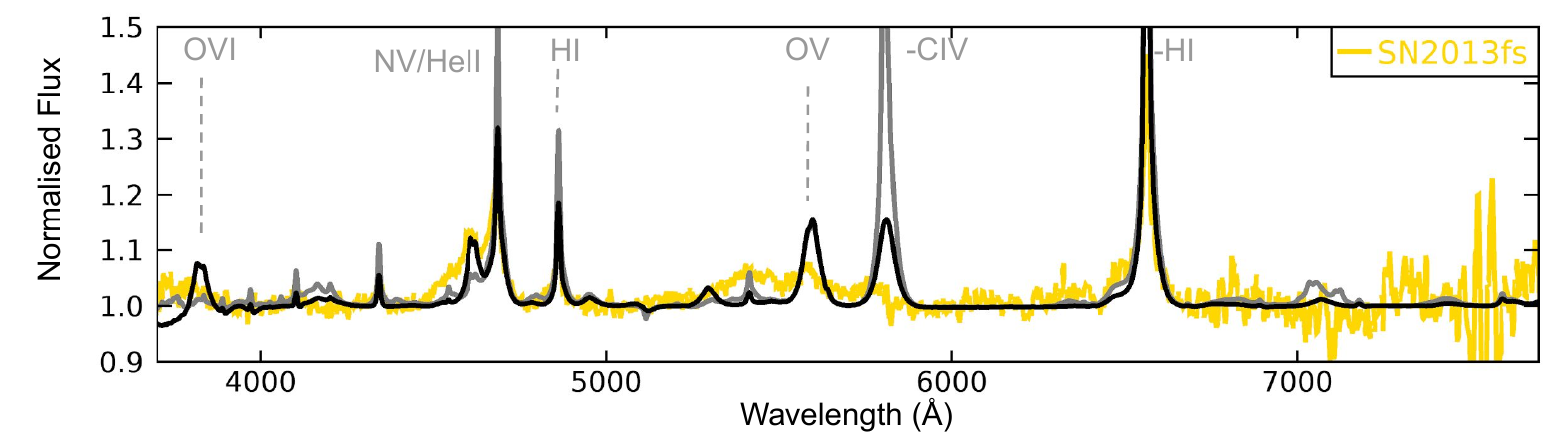}
    \end{subfigure}
    
    \begin{subfigure}[b]{0.99\textwidth}
        \includegraphics[width=0.94\textwidth]{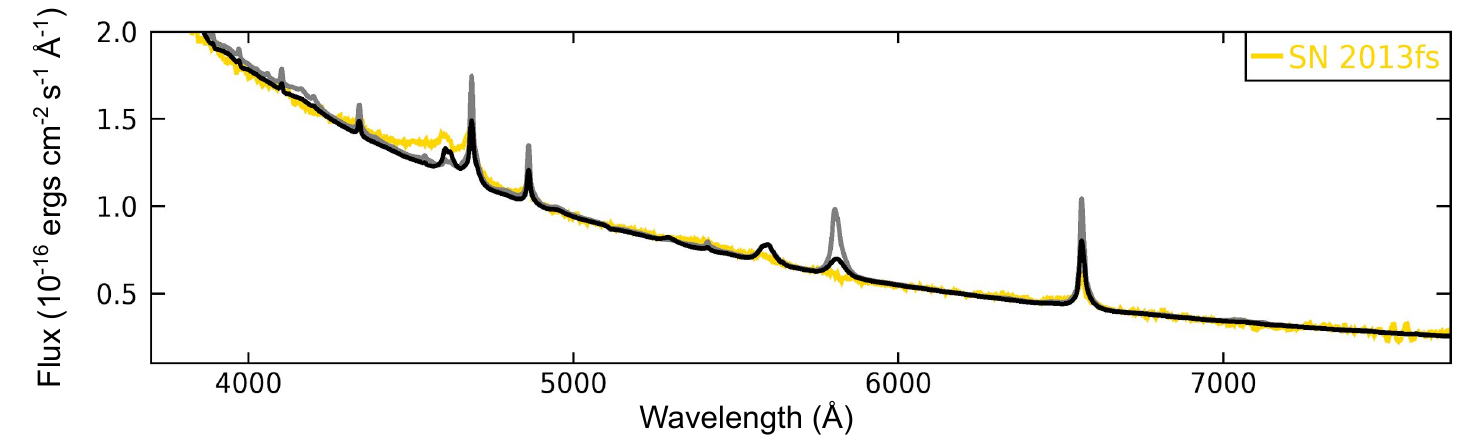}
    \end{subfigure}
    
    \caption{Best-fit models for the $21.1$ hours spectrum of SN 2013fs (iPTF13dqy). Top: normalised spectra. Bottom: absolute flux.
    This spectrum of SN 2013fs is best-fit by a model in between $L = 3.5 \times 10^{10} \lsun$, $\mdot = 4.3 \times 10^{-3} ~\msunyr$, $\vinf = 100 ~\kms$, $\rin = 26.8 \times 10^{13}$ cm, and solar surface abundances (grey) and $L = 6.3 \times 10^{10} \lsun$, $\mdot = 4.0 \times 10^{-3} ~\msunyr$, $\vinf = 100 ~\kms$, $\rin = 4.0 \times 10^{13}$ cm, and solar surface abundances (black). The grey model overestimates \ion{C}{iv} emission and underestimates \ion{N}{v} emission, pointing to the model being cooler than the observed spectrum, but the hotter black model overestimates \ion{O}{vi} lines, which are not present in the spectrum. The models have been convolved with a Gaussian kernel having FWTH = $500 ~\kms$ in order to match the resolution of the observations. To match the absolute flux a distance of $d_L = 50.95 Mpc$ was assumed, and a colour excess of $E(B-V) = 0.1$, and $R_V = 3.1$ provided the best fit for the SED.}
\end{figure*}

\begin{figure*}
    \label{fig:13fr_bf}
    \begin{subfigure}[t]{0.99\textwidth}
        \includegraphics[width=0.94\textwidth]{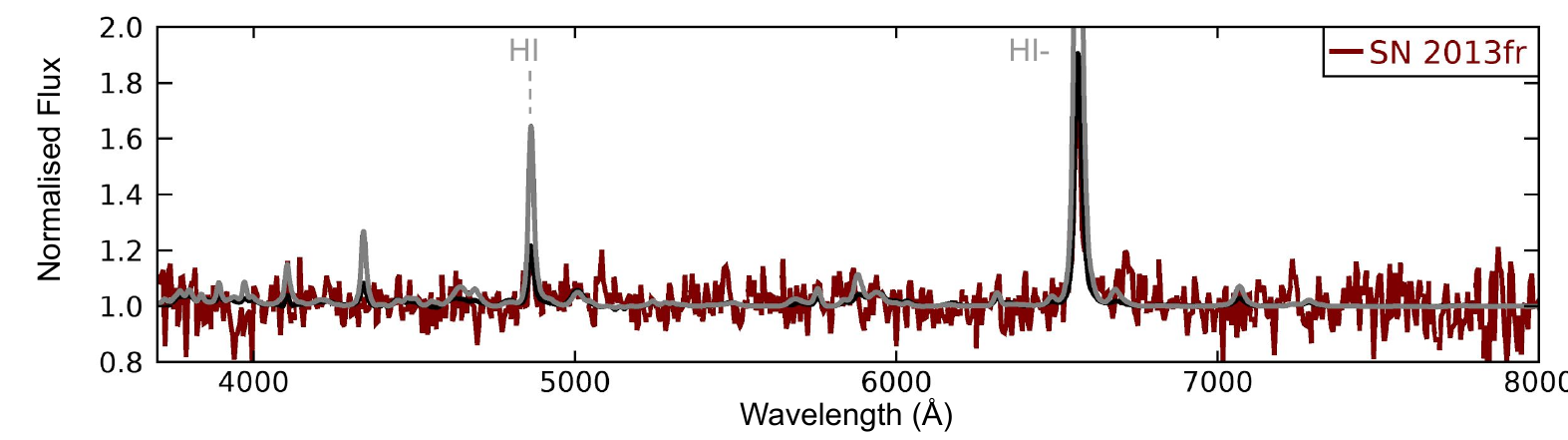}
    \end{subfigure}
    
    \begin{subfigure}[b]{0.99\textwidth}
        \includegraphics[width=0.94\textwidth]{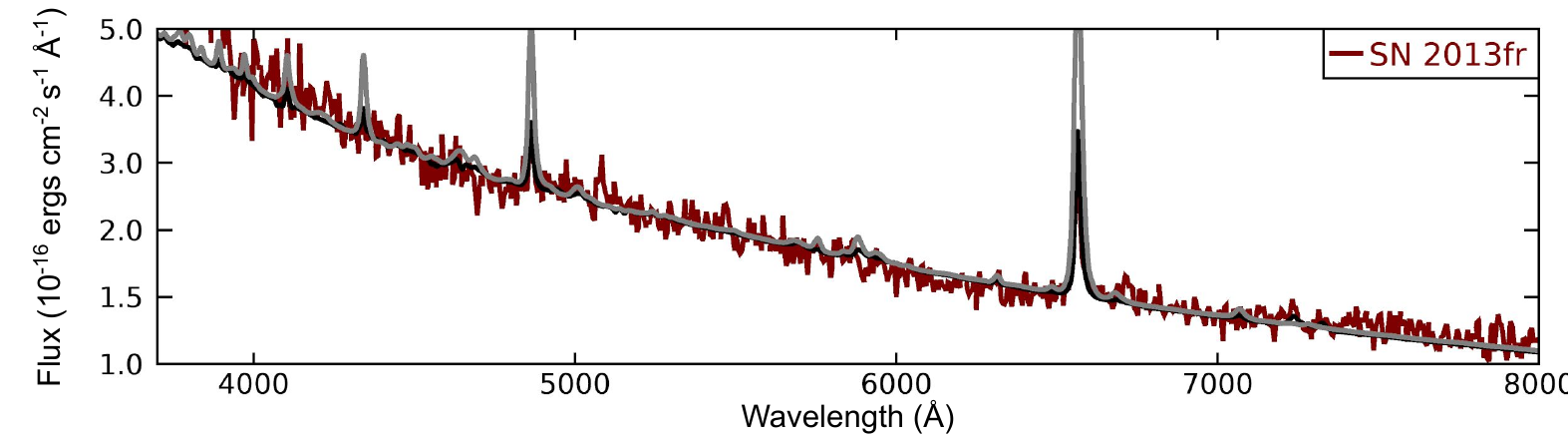}
    \end{subfigure}
    
    \caption{Best-fit models for the 4 day spectrum of SN 2013fr. Top: normalised spectra. Bottom: absolute flux. The black model has $L = 5.9 \times 10^{9} \lsun$, $\mdot = 46.9 \times 10^{-3} ~\msunyr$, $\vinf = 845 ~\kms$, $\rin = 63.2 \times 10^{13}$ cm, and CNO-processed surface abundances, while the grey model has $L = 9.4 \times 10^{9} \lsun$, $\mdot = 108.9 \times 10^{-3} ~\msunyr$, $\vinf = 845 ~\kms$, $\rin = 55.4 \times 10^{13}$ cm, and CNO-processed surface abundances. For the comparison to the absolute flux, a distance of $d = 87.0$ Mpc was assumed and $E(B-V) = 0.33$, $R_V=3.1$ for the black model and $E(B-V) = 0.36$, $R_V=3.1$ for the grey model.}
\end{figure*}

\begin{figure*}
    \label{fig:14bag_bf}
    \begin{subfigure}[t]{0.99\textwidth}
        \includegraphics[width=0.94\textwidth]{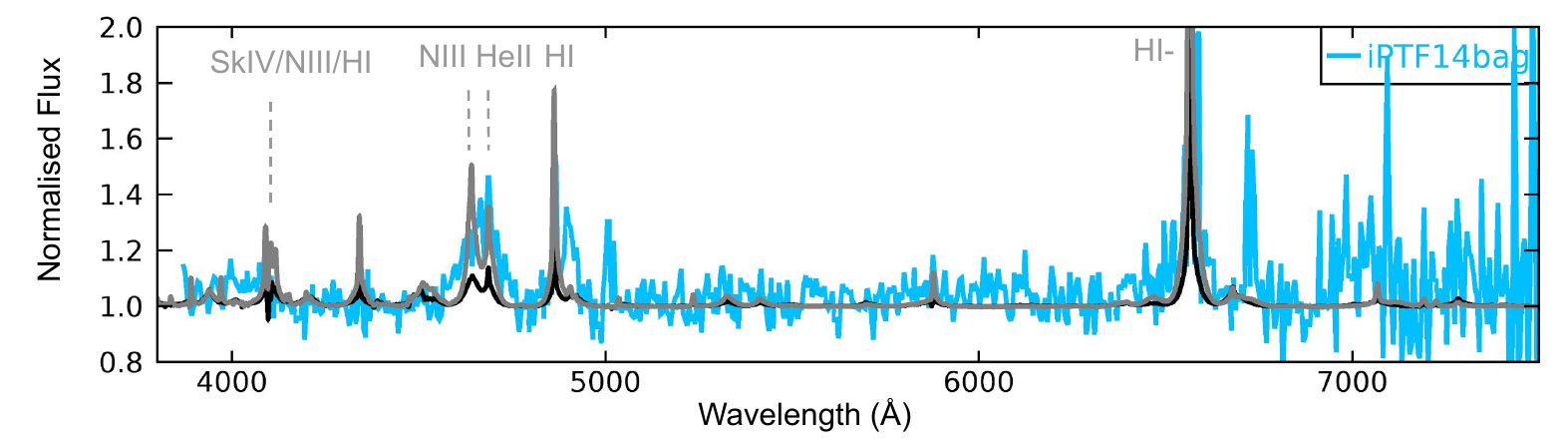}
    \end{subfigure}
    
    \begin{subfigure}[b]{0.99\textwidth}
        \includegraphics[width=0.94\textwidth]{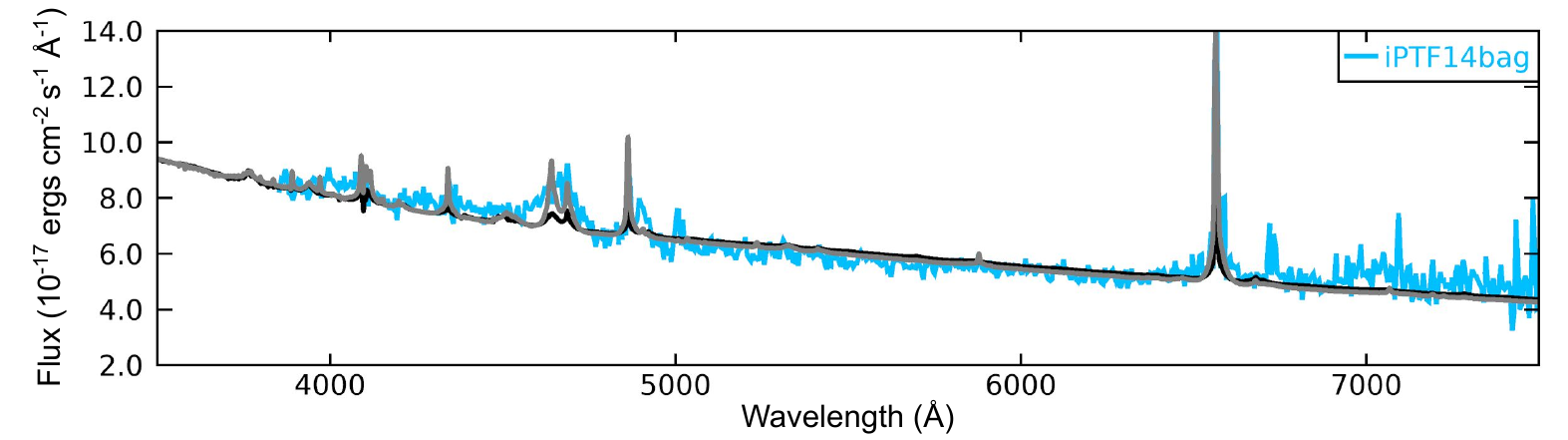}
    \end{subfigure}
    
    \caption{Best-fit models for iPTF14bag. Top: normalised spectra. Bottom: absolute flux. The best-fit models for iPTF have $L = 1.5 \times 10^{11} \lsun$, $\mdot = 189.7 \times 10^{-3} ~\msunyr$, $\vinf = 300 ~\kms$, $\rin = 160 \times 10^{13}$ cm, and CNO-processed surface abundances (grey) and $L = 1.1 \times 10^{11} \lsun$, $\mdot = 86.3 \times 10^{-3} ~\msunyr$, $\vinf = 300 ~\kms$, $\rin = 189.3 \times 10^{13}$ cm, and CNO-processed surface abundances (black). In order to match the absolute flux a distance of $d_L = 557.2 Mpc$, a colour excess of $E(B-V) = 0.45$, and $R_V = 3.1$ were assumed for both models. 
    }
\end{figure*}

\begin{figure*}
    \label{fig:16eso_bf}
    \begin{subfigure}[t]{0.99\textwidth}
        \includegraphics[width=0.94\textwidth]{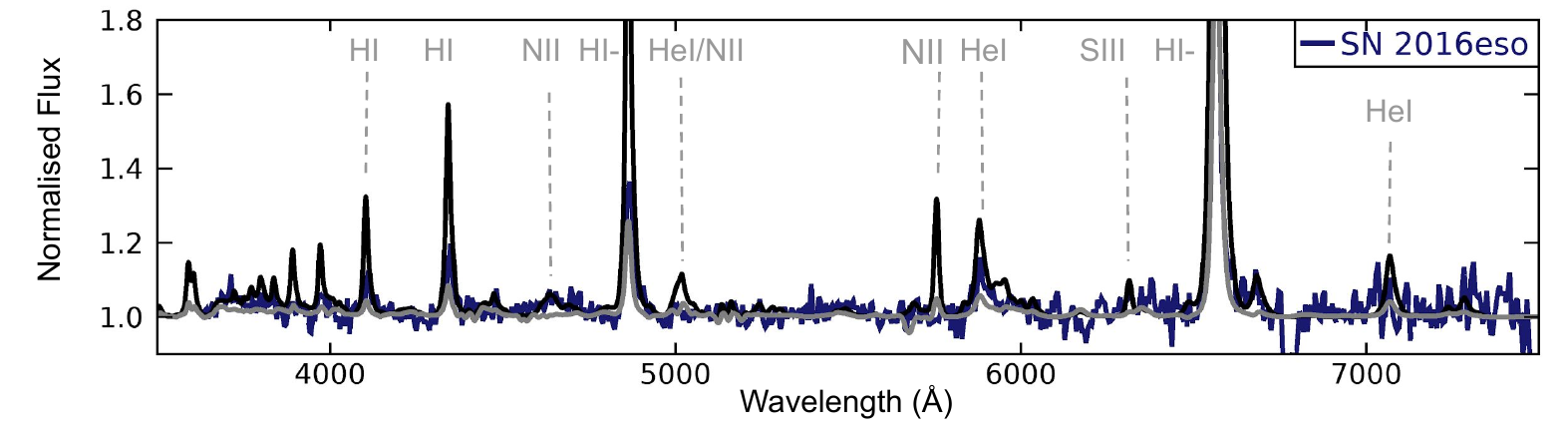}
    \end{subfigure}
    
    \begin{subfigure}[b]{0.99\textwidth}
        \includegraphics[width=0.94\textwidth]{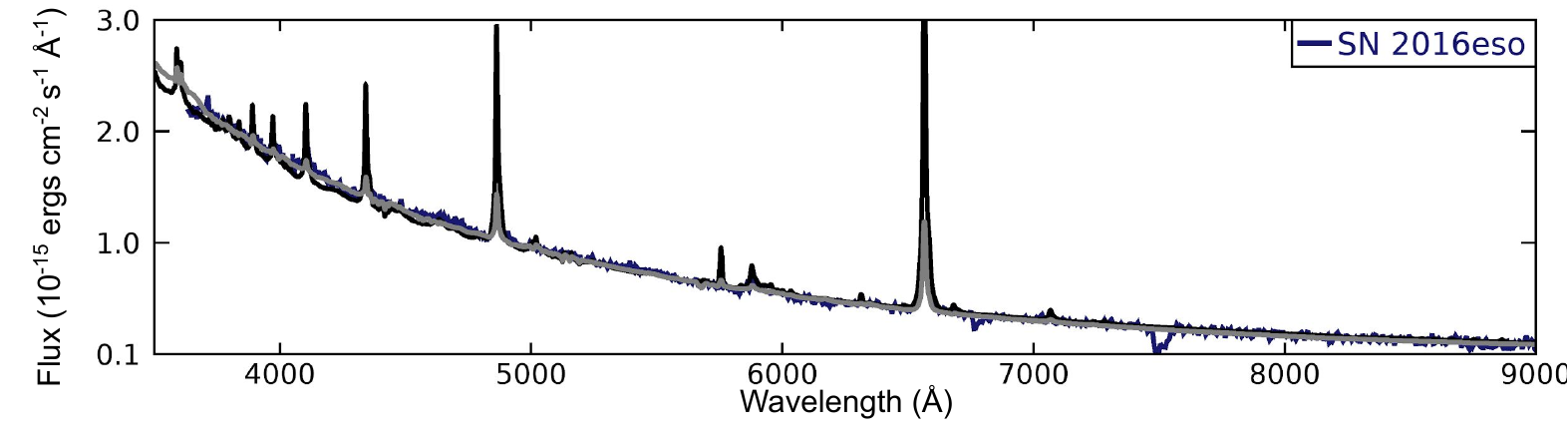}
    \end{subfigure}
    
    \caption{Best-fit models for SN 2016eso. Top: normalised spectra. Bottom: absolute flux. The black model has $L = 7.0 \times 10^{9} ~\lsun$, $R = 69.2 \times 10^{13}$ cm, $\tstar = 16.6$ kK, $\mdot = 1.8 \times 10^{-1} \msunyr (\vinf / 835 ~\kms )$, and CNO-processed surface abundances. The grey model has $L = 3.7 \times 10^{9} ~\lsun$, $R = 70.1 \times 10^{13}$ cm, $\tstar = 14.2$ kK, $\mdot = 5.4 \times 10^{-2} \msunyr (\vinf / 835 ~\kms )$, and CNO-processed surface abundances. In order to match the absolute flux, we assumed a distance of $d_L = 71.96 Mpc$, and $E(B-V) = 0.12$, $R_V = 3.1$ for the black model and $E(B-V) = 0.04$, $R_V = 3.1$ for the grey model. }
\end{figure*}

\begin{figure*}
    \label{fig:18khh_bf}
    \begin{subfigure}[t]{0.99\textwidth}
        \includegraphics[width=0.94\textwidth]{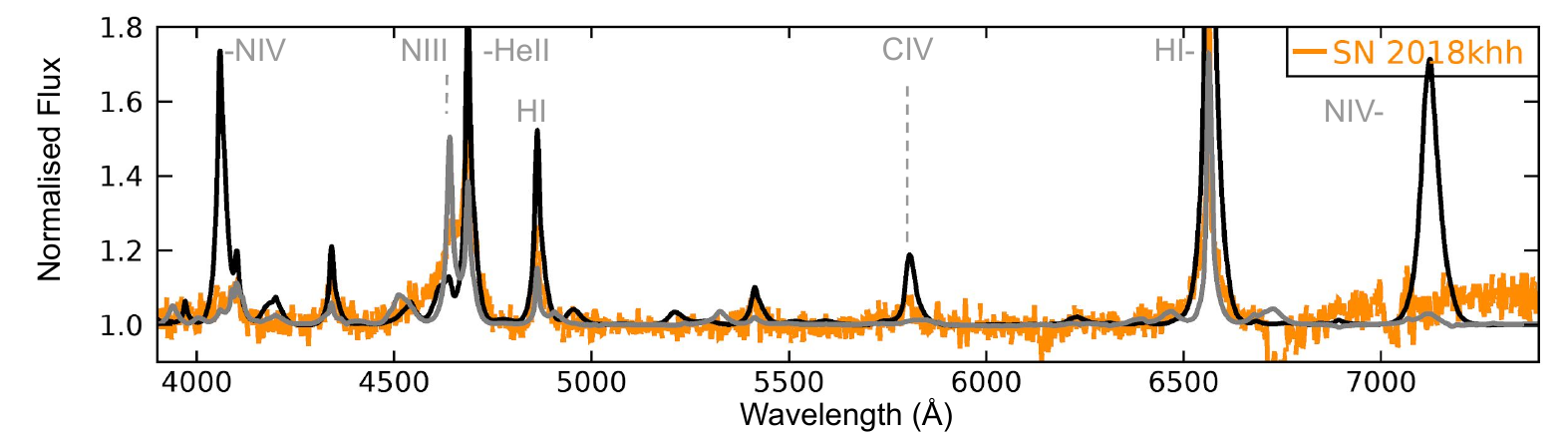}
    \end{subfigure}
    
    \begin{subfigure}[b]{0.99\textwidth}
        \includegraphics[width=0.94\textwidth]{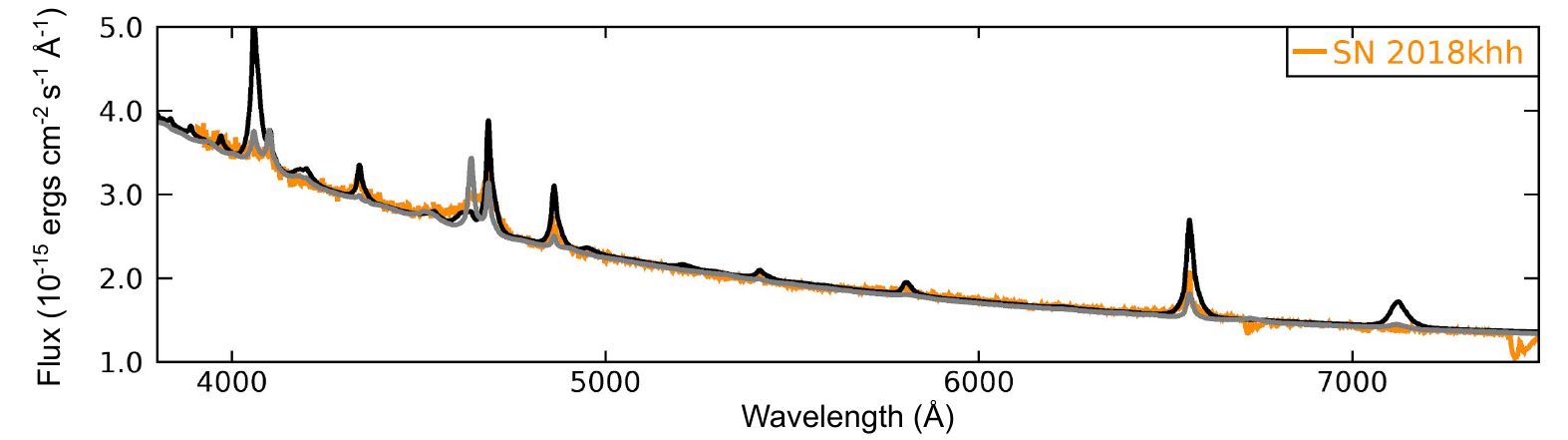}
    \end{subfigure}
    
    \caption{Observed spectrum of SN 2018khh and two encompassing models. Top: normalised spectra. Bottom: absolute flux. The black spectrum is a model with $L = 15.0 \times 10^{10} ~\lsun$, $\mdot = 1 \times 10^{-2} ~\msunyr$, $T_{\tau = 10} = 33.5$ kK, $\rin = 32 \times 10^{13}$ cm, and CNO-processed surface abundances. The grey model has $L = 4.4 \times 10^{10} ~\lsun$, $\mdot = 3 \times 10^{-3} ~\msunyr$, $T_{\tau = 10} = 23.8$ kK, $\rin = 32 \times 10^{13}$ cm, and CNO-processed surface abundances. The models have been convolved with a Gaussian of FWHM = 500 \kms. For fitting the absolute flux we assumed a distance of $d_L = 103.1 Mpc$, and color extiction $E(B-V) = 0.06$, $R_V = 3.1$ for $L = 15.0 \times 10^{10} ~\lsun$ model and $E(B-V) = 0.02$, $R_V = 3.1$ for $L = 4.4 \times 10^{10} ~\lsun$ model.}
\end{figure*}

\begin{figure*}
    \label{fig:18zd_bf}
    \begin{subfigure}[t]{0.99\textwidth}
        \includegraphics[width=0.94\textwidth]{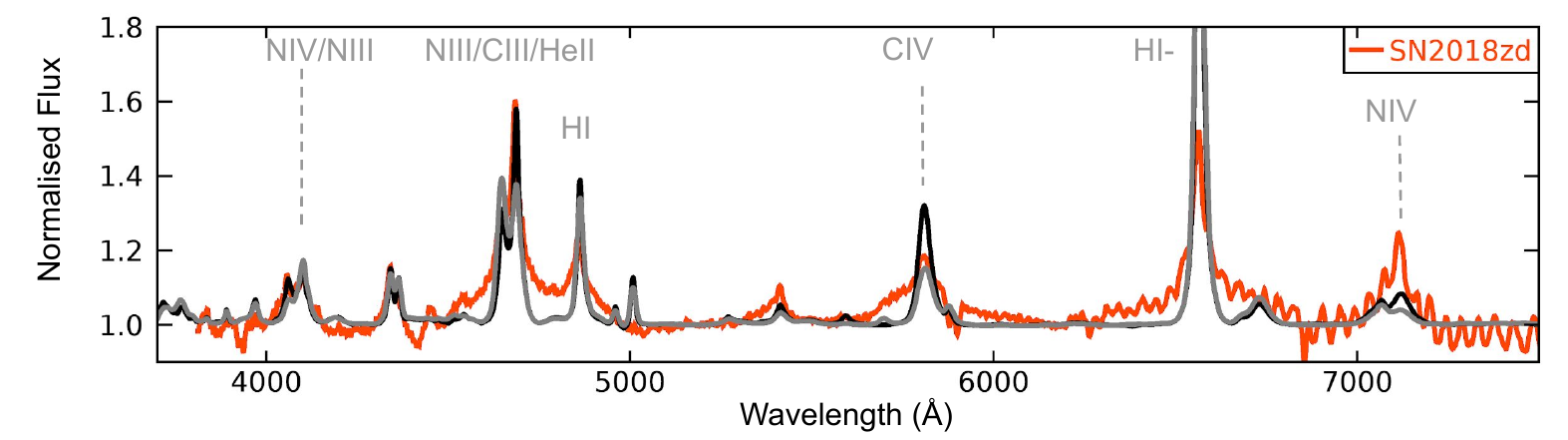}
    \end{subfigure}
    
    \begin{subfigure}[b]{0.99\textwidth}
        \includegraphics[width=0.94\textwidth]{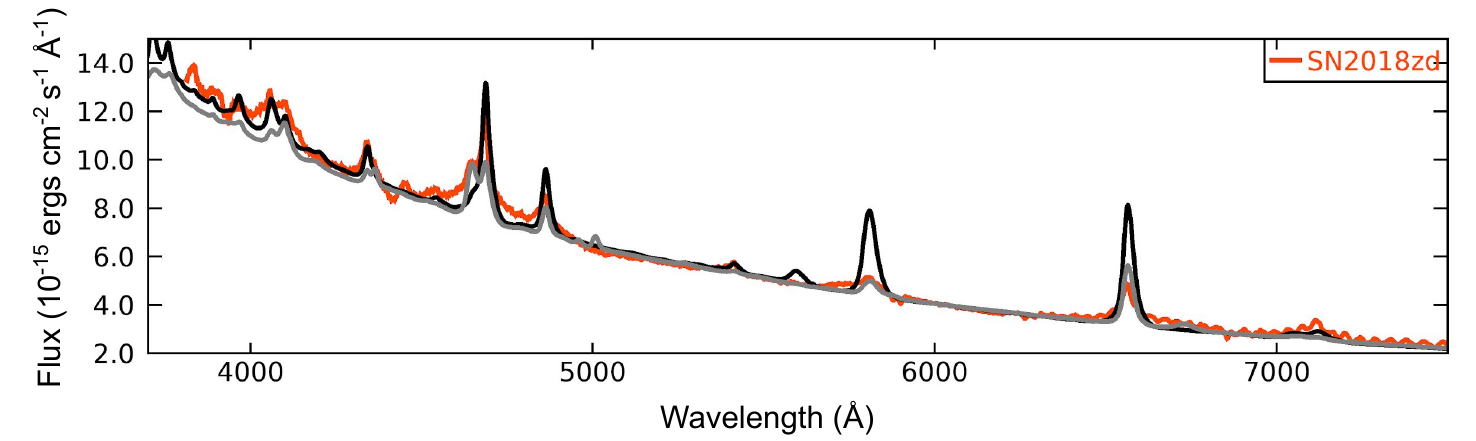}
    \end{subfigure}
    
     \caption{Best-fit models for SN 2018zd. Top: normalised spectra. The black spectrum is a model with $L = 15 \times 10^{10} ~\lsun$, $\mdot = 21.3 \times 10^{-2} ~\msun$, $\vinf = 835 ~\kms$, $T_{\tau = 10} = 33.4$ kK, $\rin = 78.4 \times 10^{13}$ cm, and solar-like surface abundances. The grey model has $L = 7 \times 10^{10} ~\lsun$, $\mdot = 11.9 \times 10^{-2} ~\msun$, $\vinf = 835 ~\kms$, $T_{\tau = 10} = 26.6$ kK, $\rin = 84.7 \times 10^{13}$ cm, and solar-like surface abundances. Bottom: absolute flux, assuming a distance of $d_L = 13.2 Mpc$, and $E(B-V) = 0.19$, $R_V = 3.1$ for both models.}
\end{figure*}

\clearpage 
\onecolumn
\section{CMFGEN models: properties and measured equivalent widths}
\label{apx:models}
\begin{longtable}{cccccccc}
\caption{ The varying properties of all the CMFGEN models included in this paper and the equivalent widths of the \ion{H}{$\beta$}, \ion{He}{i}, and \ion{He}{ii} lines of the corresponding spectra.}\\
\hline
Luminosity & Mass-loss Rate & Temperature & Inner Radius & Abundance & $ \mathrm{W_{\ion{H}{$\beta$}}} $  & $\mathrm{W_{\ion{He}{ii}}}$ & $\mathrm{W_{\ion{He}{i}}}$ \\
\lsun\ & \msunyr\ & K & cm & & & & \\
\hline
\hline
1.50E+09 & 1.00E-02 & 27250 & 8.60E+13 & SOL & -117.199 & -71.296 & -29.574 \\
1.50E+09 & 1.00E-03 & 32190 & 8.60E+13 & SOL & -4.469 & -10.755 & -0.727 \\
1.50E+09 & 3.00E-03 & 32010 & 8.60E+13 & SOL & -26.5905 & -25.44 & -4.682 \\
1.90E+08 & 1.00E-02 & 16190 & 8.60E+13 & SOL & -163.524 & -8.238 & -48.216 \\
1.90E+08 & 1.00E-03 & 19130 & 8.60E+13 & SOL & -11.5 & -0.268 & -2.546 \\
1.90E+08 & 3.00E-03 & 19020 & 8.60E+13 & SOL & -53.268 & -1.064 & -15.387 \\
2.50E+10 & 1.00E-02 & 55970 & 8.60E+13 & SOL & -40.026 & -50.356 & -0.581 \\
2.50E+10 & 1.00E-03 & 64410 & 8.60E+13 & SOL & -0.67 & -2.063 & -0.268 \\
2.50E+10 & 3.00E-03 & 64060 & 8.60E+13 & SOL & -2.928 & -5.781 & -0.274 \\
3.10E+09 & 3.00E-03 & 38070 & 8.60E+13 & SOL & -22.142 & -35.697 & -2.838 \\
3.90E+08 & 3.00E-03 & 22620 & 8.60E+13 & SOL & -39.377 & -8.209 & -10.374 \\
6.30E+09 & 3.00E-03 & 45290 & 8.60E+13 & SOL & -15.274 & -21.608 & -0.689 \\
7.80E+08 & 3.00E-03 & 26910 & 8.60E+13 & SOL & -29.85 & -16.522 & -7.464 \\
1.50E+09 & 1.00E-02 & 27280 & 8.60E+13 & CNO & -102.804 & -81.488 & -18.807 \\
1.50E+09 & 1.00E-03 & 31860 & 8.60E+13 & CNO & -4.227 & -10.575 & -0.259 \\
1.50E+09 & 3.00E-03 & 32010 & 8.60E+13 & CNO & -24.223 & -32.897 & -1.834 \\
1.90E+08 & 1.00E-02 & 16120 & 8.60E+13 & CNO & -153.295 & -9.345 & -47.395 \\
1.90E+08 & 1.00E-03 & 19130 & 8.60E+13 & CNO & -11.109 & -0.338 & -2.455 \\
1.90E+08 & 3.00E-03 & 19020 & 8.60E+13 & CNO & -51.218 & -1.702 & -14.920 \\
2.50E+10 & 1.00E-02 & 56110 & 8.60E+13 & CNO & -16.260 & -29.015 & -0.166 \\
2.50E+10 & 1.00E-03 & 64410 & 8.60E+13 & CNO & 0.692 & -2.038 & 0.115 \\
2.50E+10 & 3.00E-03 & 64060 & 8.60E+13 & CNO & -2.404 & -5.367 & 0.078 \\
3.10E+09 & 3.00E-03 & 38070 & 8.60E+13 & CNO & -21.594 & -33.797 & -0.499 \\
3.90E+08 & 3.00E-03 & 22620 & 8.60E+13 & CNO & -37.658 & -8.616 & -9.555 \\
6.30E+09 & 3.00E-03 & 45290 & 8.60E+13 & CNO & -16.455 & -20.895 & -0.027 \\
7.80E+08 & 3.00E-03 & 26910 & 8.60E+13 & CNO & -27.564 & -19.565 & -5.690 \\
1.50E+09 & 1.00E-02 & 31680 & 8.60E+13 & He & -36.865 & -86.700 & -35.469 \\
1.50E+09 & 1.00E-03 & 31910 & 8.60E+13 & He & -0.968 & -14.949 & 0.079 \\
1.50E+09 & 3.00E-03 & 32070 & 8.60E+13 & He & -9.529 & -39.336 & -2.709 \\
1.90E+08 & 1.00E-02 & 18930 & 8.60E+13 & He & -52.319 & -9.032 & -97.791 \\
1.90E+08 & 1.00E-03 & 19170 & 8.60E+13 & He & -2.145 & -0.505 & -4.175 \\
1.90E+08 & 3.00E-03 & 19080 & 8.60E+13 & He & -12.210 & -1.303 & -22.441 \\
2.50E+10 & 1.00E-02 & 63760 & 8.60E+13 & He & -6.266 & -41.193 & -0.115 \\
2.50E+10 & 1.00E-03 & 64510 & 8.60E+13 & He & 1.179 & -2.421 & 0.584 \\
2.50E+10 & 3.00E-03 & 64190 & 8.60E+13 & He & -0.325 & -8.134 & 0.462 \\
3.10E+09 & 3.00E-03 & 38150 & 8.60E+13 & He & -9.134 & -44.653 & -0.522 \\
3.90E+08 & 3.00E-03 & 22680 & 8.60E+13 & He & -7.991 & -6.378 & -14.575 \\
6.30E+09 & 3.00E-03 & 45380 & 8.60E+13 & He & -5.862 & -33.163 & 0.219 \\
7.80E+08 & 3.00E-03 & 26970 & 8.60E+13 & He & -7.121 & -20.004 & -9.279 \\
1.25E+10 & 3.00E-03 & 39770 & 1.60E+14 & SOL & -4.796 & -9.056 & -0.782 \\
1.00E+09 & 3.00E-03 & 21230 & 1.60E+14 & SOL & -12.444 & -4.859 & -2.788 \\
1.00E+10 & 1.00E-03 & 37730 & 1.60E+14 & SOL & -0.730 & -3.786 & -0.387 \\
1.90E+08 & 1.00E-02 & 14000 & 1.60E+14 & SOL & -110.783 & -1.096 & -28.822 \\
1.90E+08 & 1.00E-03 & 14080 & 1.60E+14 & SOL & -4.189 & 0.352 & -1.483 \\
1.90E+08 & 3.00E-03 & 14030 & 1.60E+14 & SOL & -21.063 & 0.474 & -6.472 \\
1.90E+08 & 4.50E-03 & 14020 & 1.60E+14 & SOL & -37.712 & 0.366 & -11.439 \\
1.90E+08 & 6.75E-03 & 14010 & 1.60E+14 & SOL & -66.851 & -0.075 & -19.515 \\
1.50E+09 & 1.00E-02 & 23390 & 1.60E+14 & SOL & -61.234 & -23.344 & -15.734 \\
1.50E+09 & 1.00E-03 & 23710 & 1.60E+14 & SOL & -1.690 & -6.267 & -0.197 \\
1.50E+09 & 3.00E-03 & 23590 & 1.60E+14 & SOL & -9.876 & -7.740 & -1.973 \\
2.50E+10 & 1.00E-03 & 47460 & 1.60E+14 & SOL & 0.536 & -3.130 & -0.484 \\
2.50E+10 & 3.00E-03 & 47450 & 1.60E+14 & SOL & -3.099 & -6.461 & -0.451 \\
3.10E+09 & 1.00E-03 & 28200 & 1.60E+14 & SOL & -0.946 & -7.140 & -0.463 \\
3.10E+09 & 3.00E-03 & 28010 & 1.60E+14 & SOL & -8.289 & -11.477 & -1.670 \\
3.90E+08 & 3.00E-03 & 16710 & 1.60E+14 & SOL & -18.593 & -0.342 & -5.118 \\
4.00E+09 & 3.00E-03 & 29850 & 1.60E+14 & SOL & -7.886 & -13.819 & -1.302 \\
5.20E+09 & 3.00E-03 & 31890 & 1.60E+14 & SOL & -7.412 & -15.264 & -0.988 \\
6.30E+09 & 1.00E-03 & 33550 & 1.60E+14 & SOL & -0.912 & -3.860 & -0.316 \\
6.30E+09 & 3.00E-03 & 33380 & 1.60E+14 & SOL & -6.880 & -14.750 & -0.841 \\
7.80E+08 & 3.00E-03 & 19900 & 1.60E+14 & SOL & -14.523 & -2.882 & -3.505 \\
1.90E+08 & 1.00E-02 & 13940 & 1.60E+14 & CNO & -101.140 & -0.745 & -31.211 \\
1.90E+08 & 1.00E-03 & 14090 & 1.60E+14 & CNO & -4.437 & -0.096 & -1.251 \\
1.90E+08 & 3.00E-03 & 14030 & 1.60E+14 & CNO & -21.289 & -0.110 & -6.366 \\
1.50E+09 & 1.00E-02 & 23390 & 1.60E+14 & CNO & -56.636 & -24.932 & -13.274 \\
1.50E+09 & 1.00E-03 & 23460 & 1.60E+14 & CNO & -1.749 & -5.678 & -0.064 \\
1.50E+09 & 3.00E-03 & 23590 & 1.60E+14 & CNO & -9.513 & -7.792 & -1.573 \\
2.50E+10 & 1.00E-02 & 47110 & 1.60E+14 & CNO & -19.905 & -27.957 & -0.092 \\
2.50E+10 & 1.00E-03 & 47460 & 1.60E+14 & CNO & 0.761 & -2.131 & 0.131 \\
2.50E+10 & 3.00E-03 & 47230 & 1.60E+14 & CNO & -1.364 & -3.907 & 0.122 \\
3.10E+09 & 1.00E-03 & 28200 & 1.60E+14 & CNO & -0.883 & -6.475 & -0.186 \\
3.10E+09 & 3.00E-03 & 28070 & 1.60E+14 & CNO & -7.491 & -13.982 & -0.846 \\
3.90E+08 & 3.00E-03 & 16680 & 1.60E+14 & CNO & -18.845 & -0.412 & -5.024 \\
4.00E+09 & 3.00E-03 & 29850 & 1.60E+14 & CNO & -7.366 & -15.395 & -0.473 \\
5.00E+09 & 3.00E-03 & 31570 & 1.60E+14 & CNO & -7.069 & -14.532 & -0.303 \\
6.30E+09 & 1.00E-03 & 33550 & 1.60E+14 & CNO & -0.942 & -2.316 & -0.052 \\
6.30E+09 & 3.00E-03 & 33380 & 1.60E+14 & CNO & -6.686 & -12.981 & -0.184 \\
7.80E+08 & 3.00E-03 & 19840 & 1.60E+14 & CNO & -14.305 & -2.031 & -3.347 \\
8.00E+09 & 3.00E-03 & 35510 & 1.60E+14 & CNO & -6.292 & -10.382 & -0.090 \\
1.90E+08 & 1.00E-02 & 13980 & 1.60E+14 & He & -20.265 & -0.640 & -44.201 \\
1.90E+08 & 1.00E-03 & 14220 & 1.60E+14 & He & -1.050 & -0.227 & -2.219 \\
1.90E+08 & 3.00E-03 & 14150 & 1.60E+14 & He & -4.345 & -0.248 & -8.821 \\
1.50E+09 & 1.00E-02 & 23490 & 1.60E+14 & He & -13.698 & -22.272 & -20.596 \\
1.50E+09 & 1.00E-03 & 23500 & 1.60E+14 & He & -0.540 & -6.654 & -0.190 \\
1.50E+09 & 3.00E-03 & 23700 & 1.60E+14 & He & -2.013 & -8.183 & -2.390 \\
2.50E+10 & 1.00E-02 & 47280 & 1.60E+14 & He & -8.367 & -49.157 & 0.073 \\
2.50E+10 & 1.00E-03 & 47530 & 1.60E+14 & He & 1.042 & -2.935 & 0.632 \\
2.50E+10 & 3.00E-03 & 47240 & 1.60E+14 & He & -0.325 & -7.071 & 0.533 \\
3.10E+09 & 3.00E-03 & 28110 & 1.60E+14 & He & -1.939 & -16.069 & -0.831 \\
3.90E+08 & 3.00E-03 & 16730 & 1.60E+14 & He & -4.557 & -0.277 & -7.995 \\
6.30E+09 & 3.00E-03 & 33440 & 1.60E+14 & He & -2.341 & -19.637 & 0.112 \\
7.80E+08 & 3.00E-03 & 19880 & 1.60E+14 & He & -2.889 & -2.336 & -5.509 \\
1.90E+08 & 1.00E-02 & 9919 & 3.20E+14 & SOL & -54.334 & -0.071 & -7.535 \\
1.90E+08 & 1.00E-03 & 9906 & 3.20E+14 & SOL & -1.925 & 0.026 & 0.061 \\
1.90E+08 & 3.00E-03 & 9976 & 3.20E+14 & SOL & -8.686 & 0.036 & -1.267 \\
1.50E+09 & 1.00E-02 & 16620 & 3.20E+14 & SOL & -27.018 & -1.738 & -7.157 \\
1.50E+09 & 1.00E-03 & 16790 & 3.20E+14 & SOL & -1.548 & -1.127 & -1.095 \\
1.50E+09 & 3.00E-03 & 16790 & 3.20E+14 & SOL & -4.668 & -1.024 & -1.439 \\
2.50E+10 & 1.00E-02 & 33430 & 3.20E+14 & SOL & -10.219 & -19.515 & -1.381 \\
2.50E+10 & 1.00E-03 & 33550 & 3.20E+14 & SOL & -0.857 & -3.093 & -0.369 \\
2.50E+10 & 3.00E-03 & 33400 & 3.20E+14 & SOL & -1.631 & -4.239 & -0.444 \\
2.50E+10 & 7.00E-03 & 33500 & 3.20E+14 & SOL & -5.624 & -10.848 & -0.805 \\
3.10E+09 & 3.00E-03 & 19930 & 3.20E+14 & SOL & -3.449 & -3.447 & -0.798 \\
3.90E+08 & 3.00E-03 & 11880 & 3.20E+14 & SOL & -6.608 & 0.136 & -2.227 \\
6.30E+09 & 3.00E-03 & 23710 & 3.20E+14 & SOL & -2.506 & -8.735 & -0.516 \\
7.80E+08 & 3.00E-03 & 14120 & 3.20E+14 & SOL & -5.330 & 0.169 & -1.889 \\
1.90E+08 & 1.00E-02 & 9919 & 3.20E+14 & CNO & -54.855 & -0.081 & -10.060 \\
1.90E+08 & 1.00E-03 & 9907 & 3.20E+14 & CNO & -1.985 & 0.003 & 0.015 \\
1.90E+08 & 3.00E-03 & 9985 & 3.20E+14 & CNO & -8.913 & 0.003 & -1.530 \\
1.50E+09 & 1.00E-02 & 16620 & 3.20E+14 & CNO & -27.488 & -1.153 & -7.019 \\
1.50E+09 & 1.00E-03 & 16590 & 3.20E+14 & CNO & -2.040 & -0.341 & -1.015 \\
1.50E+09 & 3.00E-03 & 16800 & 3.20E+14 & CNO & -4.702 & -0.730 & -1.284 \\
2.50E+10 & 1.00E-02 & 33500 & 3.20E+14 & CNO & -10.764 & -17.134 & -0.235 \\
2.50E+10 & 1.00E-03 & 33550 & 3.20E+14 & CNO & -0.865 & -1.411 & -0.041 \\
2.50E+10 & 3.00E-03 & 33400 & 3.20E+14 & CNO & -1.787 & -2.686 & -0.058 \\
3.10E+09 & 3.00E-03 & 19990 & 3.20E+14 & CNO & -3.354 & -2.818 & -0.569 \\
3.90E+08 & 3.00E-03 & 11880 & 3.20E+14 & CNO & -6.784 & -0.069 & -2.056 \\
6.30E+09 & 3.00E-03 & 23770 & 3.20E+14 & CNO & -2.386 & -7.970 & -0.243 \\
7.80E+08 & 3.00E-03 & 14120 & 3.20E+14 & CNO & -5.652 & -0.271 & -1.659 \\
1.90E+08 & 1.00E-02 & 9933 & 3.20E+14 & He & -8.552 & -0.035 & -12.597 \\
1.90E+08 & 1.00E-03 & 9985 & 3.20E+14 & He & -0.835 & -0.014 & -0.368 \\
1.90E+08 & 3.00E-03 & 9982 & 3.20E+14 & He & -1.392 & -0.004 & -1.904 \\
1.50E+09 & 1.00E-02 & 16550 & 3.20E+14 & He & -5.814 & -1.258 & -10.549 \\
1.50E+09 & 1.00E-03 & 16780 & 3.20E+14 & He & -0.950 & -1.211 & -2.021 \\
1.50E+09 & 3.00E-03 & 16780 & 3.20E+14 & He & -1.471 & -1.150 & -2.634 \\
2.50E+10 & 1.00E-02 & 33360 & 3.20E+14 & He & -4.071 & -27.505 & 0.031 \\
2.50E+10 & 1.00E-03 & 33540 & 3.20E+14 & He & -0.833 & -2.018 & 0.394 \\
2.50E+10 & 3.00E-03 & 33450 & 3.20E+14 & He & -0.974 & -4.821 & 0.357 \\
3.10E+09 & 3.00E-03 & 19970 & 3.20E+14 & He & -1.376 & -3.184 & -1.554 \\
3.90E+08 & 3.00E-03 & 11880 & 3.20E+14 & He & -1.553 & -0.150 & -3.294 \\
6.30E+09 & 3.00E-03 & 23710 & 3.20E+14 & He & -0.809 & -10.239 & -0.062 \\
7.80E+08 & 3.00E-03 & 14130 & 3.20E+14 & He & -1.708 & -0.499 & -3.124 \\ 
\caption{Your caption here}
\label{tab:myfirstlongtable}
\end{longtable}

\bsp	
\label{lastpage}
\end{document}